%% file: FSQ-12-035_temp.tex
\pdfoutput=1

\documentclass[11pt,twoside,a4paper,cmspaper,final,collab]{cms-tdr}
\def\svnVersion{277448}\def\cmsCernNo{2013-037}\def\cmsCernDate{\today}\def\cmsMessage{Published in the European Physical Journal C as \href{http://dx.doi.org/10.1140/epjc/s10052-014-3232-5}{\doi{10.1140/epjc/s10052-014-3232-5.}}}

\begin{document}\cmsNoteHeader{FSQ-12-035}

\hyphenation{had-ron-i-za-tion}
\hyphenation{cal-or-i-me-ter}
\hyphenation{de-vices}
\RCS$Revision: 268549 $
\RCS$HeadURL: svn+ssh://svn.cern.ch/reps/tdr2/papers/FSQ-12-035/trunk/FSQ-12-035.tex $
\RCS$Id: FSQ-12-035.tex 268549 2014-11-20 16:26:16Z alverson $
\newlength\cmsFigWidth
\ifthenelse{\boolean{cms@external}}{\setlength\cmsFigWidth{0.85\columnwidth}}{\setlength\cmsFigWidth{0.4\textwidth}}
\ifthenelse{\boolean{cms@external}}{\providecommand{\cmsLeft}{top}}{\providecommand{\cmsLeft}{left}}
\ifthenelse{\boolean{cms@external}}{\providecommand{\cmsRight}{bottom}}{\providecommand{\cmsRight}{right}}
\ifthenelse{\boolean{cms@external}}{\providecommand{\reBox}{\relax}}{\providecommand{\reBox}{\resizebox{\textwidth}{!}}}
\providecommand{\lljj}{\ensuremath{\ell\ell\mathrm{jj}}\xspace}
\providecommand{\Zjj}{\ensuremath{\cPZ\mathrm{jj}}\xspace}
\providecommand{\ewklljj}{\ensuremath{\mathrm{EW}\,\ell\ell\mathrm{jj}}\xspace}
\providecommand{\ewkzjj}{\ensuremath{\mathrm{EW}\,\cPZ\mathrm{jj}}\xspace}
\providecommand{\ewkgjj}{\ensuremath{\mathrm{EW}\,\gamma\mathrm{jj}}\xspace}
\providecommand{\gjj}{\ensuremath{\gamma\mathrm{jj}}\xspace}
\providecommand{\dyzjj}{\ensuremath{\mathrm{DY}\,\cPZ\mathrm{jj}}\xspace}
\providecommand{\ptj}{\ensuremath{p_\mathrm{T j}}\xspace}
\providecommand{\usedSign}{\ensuremath{\bullet}}
\providecommand{\hyph}{-\penalty0\hskip0pt\relax}
\ifthenelse{\boolean{cms@external}}{\providecommand{\breakhere}{\linebreak[4]}}{\providecommand{\breakhere}}{\relax}
\cmsNoteHeader{FSQ-12-035} 
\title{Measurement of electroweak production of two jets in association with a Z boson
in proton-proton collisions at $\sqrt{s}=8\TeV$}
\titlerunning{Electroweak production of two jets in association with a Z boson}

\date{\today}

\abstract{
The purely electroweak (EW) cross section for the production of two jets in association with a Z boson,
in proton-proton collisions at $\sqrt{s}=8\TeV$, is measured using data recorded
by the CMS experiment at the CERN LHC, corresponding to an integrated luminosity of 19.7\fbinv.
The electroweak cross section for the \lljj final state (with $\ell = \Pe$ or $\mu$
and j representing the quarks produced in the hard interaction)
in the kinematic region defined by $M_{\ell\ell} >50$\GeV,
$M_\mathrm{jj} >120$\GeV, transverse momentum $\ptj  > 25$\GeV,
and pseudorapidity $\abs{\eta_\mathrm{j}}< 5$,
is found to be $\sigma_\mathrm{EW}(\lljj)=174 \pm 15\stat\pm 40\syst\unit{fb}$,
in agreement with the standard model prediction.
The associated jet activity of the selected events is studied,
in particular in a signal-enriched region of phase space,
and the measurements are found to be in agreement with QCD predictions.
}

\hypersetup{%
pdfauthor={CMS Collaboration},%
pdftitle={Measurement of electroweak production of two jets in association with a Z boson in proton-proton collisions at sqrt(s)= 8 TeV},%
pdfsubject={CMS},%
pdfkeywords={CMS, physics, Z, vector boson fusion, electroweak interaction, cross section}}

\maketitle

\section{Introduction}
\label{sec:intro}

The production of a $\cPZ$ boson  in association with two jets
in proton-proton (pp) collisions
is dominated by a mixture of electroweak (EW)
and strong processes of order
$\alpha_\mathrm{EW}^2\alpha_\mathrm{S}^2$.
For $\cPZ\to\ell\ell$ leptonic decays,
such events are referred to as
``Drell--Yan (DY) + jets'' or \dyzjj events.

\begin{figure*}[htb] {
\centering
\includegraphics[width=0.315\textwidth]{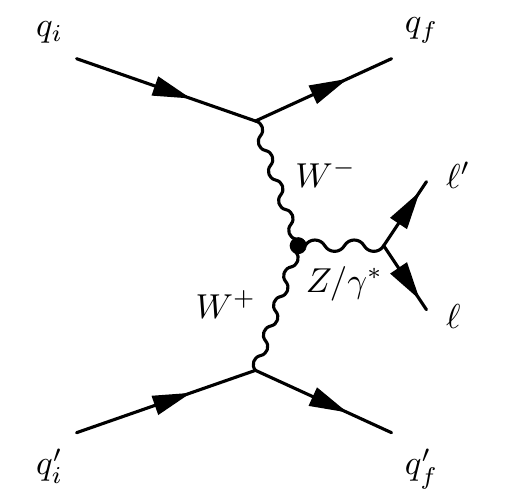}
\includegraphics[width=0.35\textwidth]{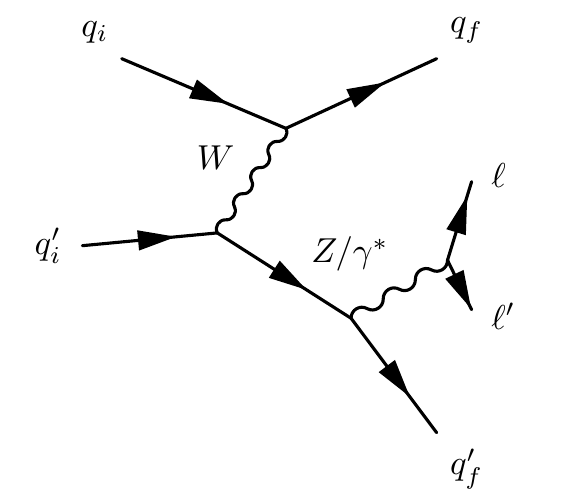}
\includegraphics[width=0.315\textwidth]{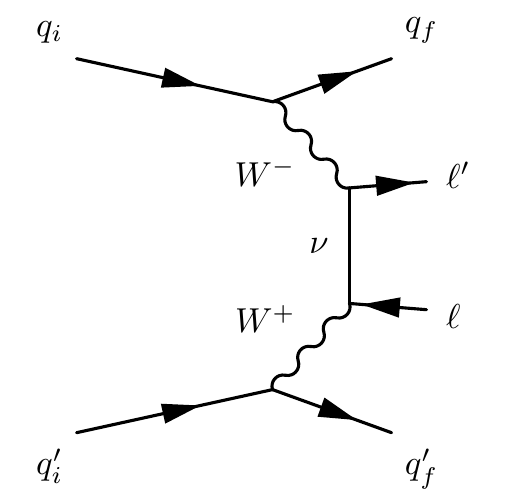}
\caption{
Representative Feynman diagrams for dilepton production in association
with two jets from purely electroweak contributions:
(left) vector boson fusion,
(middle) bremsstrahlung-like,
and (right) multiperipheral production.
\label{fig:sigdiagram}}

}
\end{figure*}

Purely electroweak \lljj\ production contributing to the same final
state is expected at order $\alpha_\mathrm{EW}^4$,
resulting in a comparatively small cross section~\cite{Oleari:2003tc}.
This process is however predicted to have a distinctive signature of
two jets of very high energy and large jj invariant mass, $M_\mathrm{jj}$,
separated by a large rapidity interval that can be occupied by the two
charged leptons and where extra gluon emission is suppressed~\cite{Rainwater:1996ud,Khoze:2002fa}.
We refer to jets produced through the fragmentation of the outgoing quarks
in pure EW processes as ``tagging jets'', and to the process from
which they originate as ``\ewkzjj''.
Figure~\ref{fig:sigdiagram} shows representative Feynman diagrams for the  \ewkzjj\
processes, namely (left) vector boson fusion (VBF),
(middle) bremsstrahlung-like, and
(right) multiperipheral production.
Detailed calculations reveal the presence of a
large negative interference between the pure VBF process and the
two other categories~\cite{Khoze:2002fa,Oleari:2003tc}.
These diagrams represent the signal (S) in the data.

\begin{figure*}[htb] {
\centering
\includegraphics[width=0.315\textwidth]{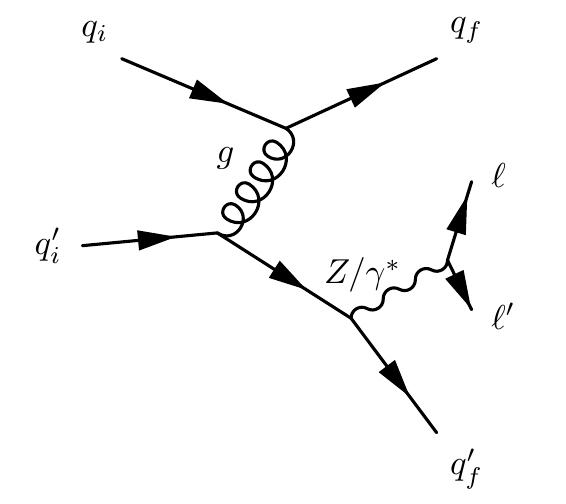}
\includegraphics[width=0.315\textwidth]{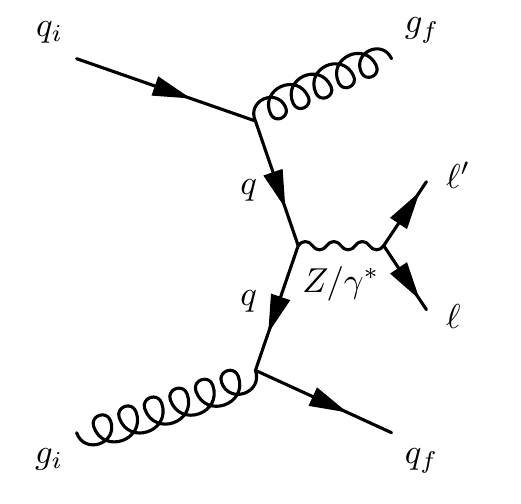}
\caption{Representative diagrams for order $\alpha_\mathrm{S}^2$
  corrections to DY production that comprise the main background (B)
  in this study.}
\label{fig:bkgdiagram}}
\end{figure*}

For inclusive \lljj\ final states,
some of the diagrams with same initial- and final-state
particles and quantum numbers
can interfere, even if they do not involve  exclusively EW interactions.
Figure~\ref{fig:bkgdiagram}~(left) shows
one example of order $\alpha_\mathrm{S}^2$ corrections to DY production
that have the same initial and final state as those in
Fig.~\ref{fig:sigdiagram}.
A different order $\alpha_\mathrm{S}^2$ correction that
does not interfere with the EW signal, is shown in
Fig.~\ref{fig:bkgdiagram}~(right).

The study of \ewkzjj\ processes is part of a more general
investigation of standard model (SM) vector boson fusion and scattering processes that include
the Higgs boson~\cite{Aad:2012tfa,Chatrchyan:2012ufa,Chatrchyan:2013lba} and
searches for physics beyond the standard model~\cite{Cho:2006sx,Dutta:2012xe}.
When isolated from the backgrounds, the properties
of \ewkzjj\ events can be compared with SM predictions.
Probing the jet activity in the selected events in particular
can shed light  on
the selection (or vetoing) of additional parton radiation to
the tagging jets~\cite{Bjorken:1992er,Schissler:2013nga}.

At the CERN LHC, the \ewkzjj\ process was first measured by the CMS
experiment using pp collisions at
$\sqrt{s}=7\TeV$~\cite{Chatrchyan:2013jya},
and more recently by the ATLAS experiment
at $\sqrt{s}=8\TeV$~\cite{Aad:2014dta}.
Both results have been found to agree with the expectations of the SM.
Our present work reflects the measurement
at CMS using pp collision data collected at $\sqrt{s}=8$\TeV
during 2012 that correspond to an integrated luminosity of 19.7\fbinv.
As the signal-to-background ratio for the measurement is small,
different methods
are used to enhance the signal fraction,
to confirm the presence of the signal, and to measure the cross section.
Besides the two multivariate analyses,
based on the methods developed for the 7\TeV
analysis~\cite{Chatrchyan:2013jya},
a new method is presented,
using a model of the main background based on real pp collisions.
The analysis of the 8\TeV data,
 offers the opportunity of
reducing the uncertainties of the 7\TeV measurements,
given the larger integrated luminosity,
and to add robustness to the results with the new data-based method.

This paper is organised as follows:
Section~\ref{sec:cmsexperiment} describes the experimental
apparatus
and Section~\ref{sec:simulation} the simulations.
Event selection procedures are described in Section~\ref{sec:evsel}, and
Section~\ref{sec:controlreg} discusses the
selection efficiencies and background models in control regions.
Section~\ref{sec:sigdisc} details the strategies adopted in our
analysis to extract the signal from the data, and the corresponding
systematic uncertainties are summarised in Section~\ref{sec:systunc}.
The results obtained are presented in Section~\ref{sec:results},
and we conclude with a study of jet properties in a
\dyzjj-dominated control region, as well as in a high-purity,
\ewkzjj-enriched region
in Section~\ref{sec:hadactivity}.
Finally, a brief summary of the results is given in Section~\ref{sec:summary}.

\section{The CMS detector}
\label{sec:cmsexperiment}

The central feature of the CMS apparatus is a
superconducting solenoid of 6\unit{m} internal diameter,
providing a magnetic field of 3.8\unit{T}. The
solenoid volume contains a silicon pixel and strip tracker,
a lead tungstate crystal electromagnetic calorimeter (ECAL),
and a brass/scintillator hadron calorimeter (HCAL),
each composed of a barrel and two endcap sections.
Muons are measured in gas-ionisation tracking detectors embedded in the steel
flux-return yoke outside the solenoid.
Extensive forward cal\-or\-im\-e\-try complements the coverage provided by the barrel and endcap detectors.

The silicon tracker consists of 1440 silicon pixel modules and 15\:148 silicon strip
detector modules, located in the
field of the superconducting solenoid.
It measures charged particles within
 $\abs{\eta}< 2.5$,
providing an impact parameter
resolution of ${\approx}15\mum$
and a transverse momentum (\pt) resolution of about 1.5\% for $\pt=100\GeV$ particles.

The energy of electrons is measured after combining the information from
the ECAL and the tracker, whereas their direction is
measured by the tracker.
The invariant mass resolution for $\cPZ \to \Pe \Pe$ decays is 1.6\%
when both electrons are in the
ECAL barrel, and 2.6\% when both electrons are in the ECAL endcap~\cite{Chatrchyan:2013dga}.
Matching muons to tracks measured in the silicon tracker yields
a \pt resolution between $1$ and 10\%, for \pt values up to 1\TeV.
The jet energy resolution (JER) is typically ${\approx}15\%$ at 10\GeV,
8\% at 100\GeV, and 4\% at 1\TeV~\cite{Chatrchyan:2011ds}.

\section{Simulation of signal and background events}
\label{sec:simulation}

Signal events are simulated at leading order (LO) using
the \MADGRAPH (v5.1.3.30) Monte Carlo (MC) generator~\cite{Alwall:2011uj,Alwall:2014hca},
interfaced to \PYTHIA (v6.4.26)~\cite{Sjostrand:2006za} for parton showering (PS) and hadronisation.
The CTEQ6L1~\cite{Pumplin:2002vw} parton distribution functions (PDF) are
used to generate the event,
the factorisation ($\mu_F$) and renormalisation ($\mu_R$) scales
being both fixed to be equal to the \cPZ-boson
mass~\cite{Beringer:1900zz}.
The underlying event is modelled with the so-called $Z2^{*}$ tune~\cite{Chatrchyan:2011id}.
The simulation does not include the generation of extra partons at
matrix-element level.
In the kinematic region defined by dilepton mass $M_{\ell\ell} >50\GeV$,
parton transverse momentum $\ptj > 25\GeV$,
parton pseudorapidity $\vert \eta_\mathrm{j}\vert< 5$,
diparton mass $M_\mathrm{jj} >120\GeV$,
and angular separation
$\Delta R_\mathrm{jj}=\sqrt{\smash[b]{(\Delta\eta_\mathrm{jj})^2+(\Delta\phi_\mathrm{jj})^2}}>0.5$,
where $\Delta\eta_\mathrm{jj}$ and $\Delta\phi_\mathrm{jj}$
are the differences in pseudorapidity and azimuthal angle
between the tagging partons,
the cross section in the $\ell\ell$jj final state (with $\ell$ = e or $\mu$) is expected to be
$\sigma_\mathrm{LO}(\mathrm{EW}~\ell\ell\mathrm{jj})=208^{+8}_{-9}\,\text{(scale)}\pm7\,\text{(PDF)}\unit{fb}$,
where the first uncertainty is obtained by changing simultaneously
$\mu_F$ and $\mu_R$ by factors of $2$ and $1/2$, and the second
from the uncertainties in the PDFs
which has been estimated following the \textsc{pdf4lhc} prescription~\cite{Alekhin:2011sk,Botje:2011sn,Ball:2010de,Pumplin:2002vw,Martin:2009iq}.
The LO signal cross section and kinematic distributions
estimated with \MADGRAPH are found to be in good agreement with the LO predictions of the
\textsc{vbfnlo} generator (v.2.6.3)~\cite{Arnold:2008rz,Arnold:2011wj,Arnold:2012xn}.

Background DY events are also generated
with \MADGRAPH using a LO matrix element (ME) calculation that includes up to four partons
generated from quantum chromodynamics (QCD) interactions. The ME-PS matching is
performed following the ktMLM prescription~\cite{Mangano:2006rw,Alwall:2007fs}.
The dilepton DY production for $M_{\ell\ell}>50\GeV$ is normalised
to $\sigma_\text{th}(\mathrm{DY})=3.504\unit{nb}$, as computed
at next-to-next-leading order (NNLO)
with \textsc{fewz}~\cite{Melnikov:2006kv}.

The evaluation of the interference between \ewkzjj\ and \dyzjj\ processes,
relies on the predictions obtained with \MADGRAPH.
Three samples, one of pure signal, one pure background,
and one including both
$\alpha_\mathrm{EW}^4$ and
$\alpha_\mathrm{EW}^2\alpha_\mathrm{S}^2$ contributions
are generated for this purpose.
The differential cross sections are compared
and used to estimate
the expected interference contributions at the parton level.

Other residual background is expected from events with
two leptons of same flavour with accompanying jets in the final state.
Production of \ttbar events
is generated with \MADGRAPH, including up to three extra partons,
and normalised to the NNLO with next\hyph{}to\hyph{}next\hyph{}to\hyph{}leading\hyph{}logarithmic
corrections to an inclusive cross section of 245.8\unit{pb} \cite{Czakon:2013goa}.
Single-top-quark processes are modelled at next-to-leading order (NLO)
with \POWHEG~\cite{Alioli:2010xd,Nason:2004rx,Frixione:2007vw,Alioli:2009je,Re:2010bp}
and normalised, respectively, to cross sections of $22\pm2\unit{pb}$,
$86\pm3\unit{pb}$, and $5.6\pm0.2\unit{pb}$ for
the tW, $t$-, and $s$- channel production~\cite{Kidonakis:2012db,Kidonakis:2013zqa}.
Diboson production processes $\PW\PW$,
$\PW\cPZ$, and $\cPZ\cPZ$ are generated with \MADGRAPH
and normalised, respectively, to the cross sections
of 59.8\unit{pb}, 33.2\unit{pb}, and 17.7\unit{pb}, computed at NNLO~\cite{Gehrmann:2014fva} and with
\MCFM~\cite{Campbell:2010ff}.
Throughout this paper we use the abbreviation VV when
referring to the sum of the processes which yield two vector bosons.

The production of a $\PW$ boson in association with jets,
where the $\PW$ decays to a charged lepton and a neutrino,
is generated with \MADGRAPH,
and normalised to a total cross section of 36.3~nb, computed at NNLO with \textsc{Fewz}.
Multijet QCD processes are also studied in simulation, but are found to yield negligible
contributions to the selected events.

A detector simulation based on \GEANTfour (v.9.4p03)~\cite{Allison:2006ve,Agostinelli:2002hh}
is applied to all the generated signal and background samples.
The presence of multiple pp interactions in the same beam crossing  (pileup)
is incorporated by simulating additional interactions
(both in-time and out-of-time with the collision) with a multiplicity
that matches the one observed in data.
The average number of pileup events is estimated as $\approx$21
interactions per bunch crossing.

\section{Reconstruction and selection of events}
\label{sec:evsel}

The event selection is optimised to identify dilepton final states
with two isolated, high-\pt leptons, and at least two high-\pt jets.
Dilepton triggers are used to acquire the data,
where one lepton is required to have $\pt>17 \GeV$ and the other to have $\pt>8 \GeV$.
Electron-based triggers include additional isolation requirements,
both in the tracker
detectors and in the calorimeters.
A single-isolated-muon trigger, with a requirement of $\pt>24 \GeV$,
is used to complement the dimuon trigger and increase the efficiency of the selection.

Electrons are reconstructed from clusters of energy
depositions in the ECAL that match tracks extrapolated from the
silicon tracker~\cite{CMS-PAS-EGM-10-004}.
Muons are reconstructed by fitting trajectories based on hits in the silicon
tracker and in the outer muon system~\cite{muon}.
Reconstructed electron or muon candidates
are required to have $\pt>20\GeV$.
Electron candidates are required to be reconstructed
within $\abs{\eta}\leq 2.5$, excluding the CMS barrel-to-endcap
transition region of the ECAL~\cite{Chatrchyan:2008zzk},
and muon candidates are required to be reconstructed in the fiducial
region $\abs{\eta}\leq2.4$ of the tracker system.
The track associated to a lepton candidate is required
to have both its transverse and longitudinal impact parameters
compatible with the position of the primary vertex (PV) of the event.
The PV for each event is defined as the one with
the largest $\sum \pt^2$, where the sum runs over
all the tracks used to fit the vertex.
A particle-based relative isolation
parameter is computed for each lepton, and
corrected on an event-by-event basis for contributions from pileup.
The particle candidates used to compute the isolation variable
are reconstructed with the particle flow algorithm which will be
detailed later.
We require that the sum of the scalar \pt of all particle candidates
reconstructed in an isolation cone
with radius $R=\sqrt{\smash[b]{(\Delta \eta)^{2}+(\Delta \phi)^{2}}}<0.4$
around the lepton's momentum vector is
$<$10\% or $<$12\% of the electron or muon \pt value, respectively.
The two leptons with opposite electric charge
and with highest \pt are chosen to form the dilepton pair.
Same-flavour dileptons (ee or $\mu\mu$)
compatible with $\cPZ\to\ell\ell$ decays are then selected by
requiring $|M_\cPZ-M_{\ell\ell}|<15 \GeV$,
where $M_\cPZ$ is the mass  of the $\cPZ$ boson~\cite{Beringer:1900zz}.

Two types of jets are used in the analysis: ``jet-plus-track'' (JPT)~\cite{CMS-PAS-JME-09-002}
and particle-flow (PF)~\cite{Chatrchyan:2011ds} jets. Both cases use the
anti-\kt algorithm~\cite{Cacciari:2008gp,Cacciari:2011ma} with
a distance parameter of 0.5 to define jets.
The information from the ECAL, HCAL and tracker are used by both algorithms in distinct ways.
The JPT algorithm improves the energy response and resolution of
calorimeter jets by incorporating additional tracking information.
For JPT jets the associated tracks are classified as in-cone or out-of-cone
if they point to within or outside the jet cone around the jet axis at
the surface of the calorimeter.
The momenta of both in-cone and out-of-cone tracks
are then added to the energy of the associated calorimeter jet
and for in-cone tracks the expected
average energy deposition in the calorimeters is subtracted based on
the momentum of the track.
The direction of the jet axis is also corrected by the algorithm.
As a result, the JPT algorithm improves both the energy and the direction of the jet.
The PF algorithm~\cite{CMS-PAS-PFT-09-001,CMS-PAS-PFT-10-001} combines the information
from all relevant CMS sub-detectors to identify and reconstruct particle candidates in the event:
muons, electrons, photons, charged hadrons, and neutral hadrons.
The PF jets are constructed by clustering these particle
candidates and the jet momentum is defined as the vectorial sum of the momenta of all particle candidates.
An area-based correction is applied
to both JPT and PF jets, to account for the
extra energy that is clustered through in-time pileup~\cite{Cacciari:2007fd,Cacciari:2008gn}.
Jet energy scale (JES) and resolution (JER) for JPT and PF jets are derived from simulation
and confirmed with in situ measurements of the \pt balance observed
in exclusive dijet and \cPZ/photon+jet events.
The simulation is corrected so that it describes the JER from real data.
Additional selection criteria are applied to each event to remove
spurious jet-like features originating from isolated noise patterns in
certain HCAL regions.
Jet identification criteria are furthermore applied to remove
contributions from jets clustered from pileup events. These criteria are described in more detail
in Ref.~\cite{CMS-PAS-JME-13-005}.
As will be detailed in Section~\ref{subsec:zp1j},
the efficiency of these algorithms has been measured in data and it is
observed to be compatible with the expectations from simulation across the full
pseudorapidity range used in the analysis.

In the preselection of events we require at least
two jets with $\pt>30\GeV$ and ${\abs{\eta}\leq4.7}$.
The two jets of highest \pt jets are defined as the tagging jets.
For the measurement of the cross section, we require
the leading jet to have $\pt>50\GeV$ and the dijet invariant mass $M_\mathrm{jj}>200\GeV$.
Other selection requirements will be described below, as they depend on the analysis.

\section{Control regions for jets and  modelling of background}
\label{sec:controlreg}

In our analysis, we select control regions for different purposes:
to validate the calibrated jet energy response and efficiencies of jet-identification
criteria, to estimate the backgrounds and
to verify the agreement between data and estimates of background.
The following details the result of these cross-checks.

\subsection{Jet identification and response}
\label{subsec:zp1j}

Events with either a $\cPZ\to\mu\mu$ or a
photon candidate, produced in association with a single jet with
\pt$>30\GeV$, are used as one of the control samples in this analysis.
The $\cPZ$ candidate or the photon, and the associated jet are required
to have
$\abs{\Delta\phi(\text{jet},\cPZ\text{ or }\gamma)}>2.7\unit{rad}$.
These events enable a measure of
the efficiency of the algorithms used to reject calorimeter noise and
pileup-induced jets,
and to check the jet energy response.

The jet identification criteria are based on the fractions of the jet energy
deposited in different calorimeter elements~\cite{Chatrchyan:2011ds}.
Besides calorimetric noise, pileup events result in additional
reconstructed jets.
Such pileup jets can be rejected through
a multivariate analysis based on the kinematics of the jet,
on the topological configuration of its constituents,
and on the fraction of tracks
in the jet, associated
to other reconstructed PVs in the same event~\cite{CMS-PAS-JME-13-005}.
The efficiency of both jet identification and pileup
rejection is measured in the control sample, and
determined to be $>98\%$ for both JPT and PF jets.
The dependence of this efficiency on $\eta$ agrees with that
predicted in MC simulation.
The residual $\eta$-dependent difference is used to assign a systematic uncertainty in the
selected signal.

{\tolerance=1000
The same control sample is also used to verify the jet
energy response~\cite{Chatrchyan:2011ds}, which is defined from the ratio
$\left[\pt(\text{jet})/\pt(\cPZ\text{ or }\gamma)\right]$.
The double ratio of the response in data and in simulation, \ie
$\big[\pt(\text{jet})/\pt(\cPZ\text{ or }\gamma)\big]_\text{data}/\breakhere
\big[\pt(\text{jet})/\pt(\cPZ\text{ or }\gamma)\big]_\mathrm{MC}$,
provides
a residual uncertainty that is assigned as a systematic source of uncertainty
to the measurement.
Although partially covered by the JES
uncertainties, this procedure considers
possible residual uncertainties
in the particular phase-space regions selected in our analysis.
This evaluation is crucial for the most forward region of $\eta$,
where the uncertainties in response are large.
The double ratio defined above is observed to be close to
unity except for a small loss in response
($\approx$5\%) observed in the region where the tracker
has no acceptance and where there is a transition from the endcap to
the forward hadron calorimeters of CMS ($2.7<\abs{\eta}<3.2$).
}

\subsection{Discriminating gluons from quarks}
\label{sec:gtag}

Jets in signal events are expected to originate from quarks
while for background events it is more probable that jets are
initiated by a gluon emitted from a radiative QCD process.
A quark-gluon (q/g) discriminant~\cite{Chatrchyan:2013jya}
is
evaluated for the two tagging jets with the intent of distinguishing
the nature of each jet.

The q/g discriminant exploits differences in the showering and fragmentation of gluons and quarks,
making use of the internal jet-composition and structure observables.
The jet particle multiplicity and the maximum energy
fraction carried by a particle inside the jet are used.
In addition
the q/g discriminant makes use of the following variables, computed
using the weighted $\pt^2$-sum of the particles inside a jet:
the jet constituents' major root-mean-square (RMS)  distance in the $\eta$-$\phi$
plane, the jet constituents' minor RMS distance in the $\eta$-$\phi$ plane, and  the jet asymmetry pull.
Further details can be found in~\cite{CMS-PAS-JME-13-002,Gallicchio:2010sw}.

The variables are used as an input to a likelihood-ratio
discriminant that is trained using the \textsc{tmva} package~\cite{Hocker:2007ht}
on gluon and quark jets from simulated 
dijet events.
To improve the separation power,
all variables are corrected for their pileup contamination using
the same estimator for the average energy density from pileup
interactions~\cite{Cacciari:2007fd,Cacciari:2008gn},
as previously defined in Section~\ref{sec:evsel}.
The performance of the q/g discriminant has been
evaluated and validated using independent,
exclusive samples of \cPZ+jet and dijet data~\cite{CMS-PAS-JME-13-002}.
The use of the gluon-quark likelihood discriminator leads to a decrease
of the statistical uncertainty of the measured signal by about 5\%.

\subsection{Modeling background}
\label{subsec:bckgmodel}

Alternative background models are explored for the dominant \dyzjj\ background.
Given that the majority of the $\ell\ell \mathrm{jj}$ final states are
produced through \dyzjj\ processes it is crucial to have different handles
on the behavior of this process, in particular, in the signal phase
space region.

\textbf{Simulation-based prediction for background}

The effect of virtual corrections to the \MADGRAPH-based
(Born-level) description of  \dyzjj\ is studied using \MCFM.
Comparisons are made between the predictions of \MCFM parton-level distributions
with NLO and LO calculations and
these studies provide a dynamic NLO to LO scale factor (K-factor)
as a function of $M_\mathrm{jj}$ and of the difference between the rapidity of the
$\cPZ$ boson and the average rapidity of the two tagging jets, \ie
\begin{equation}
y^*=y_{\cPZ}-\frac{1}{2}(y_{\mathrm{j}_1}+y_{\mathrm{j}_2}).
\end{equation}

The K-factor is observed to have a minor dependence on $M_\mathrm{jj}$,
but to increase steeply with $\abs{y^*}$, and a correction greater than
10\%, relative to the signal,  is obtained for $\abs{y^*}>1.2$.
As a consequence, an event selection of $\abs{y^*}<1.2$ is introduced in the
\dyzjj\ simulation-based analyses.
Finally, the difference between the nominal \MADGRAPH prediction
and the one obtained after reweighting it  with the dynamic
K-factor,  on an event-by-event basis,
is assigned as a systematic uncertainty for the \dyzjj\ background prediction from simulation.

\begin{figure*}[htp]
\centering
\includegraphics[width=0.32\textwidth]{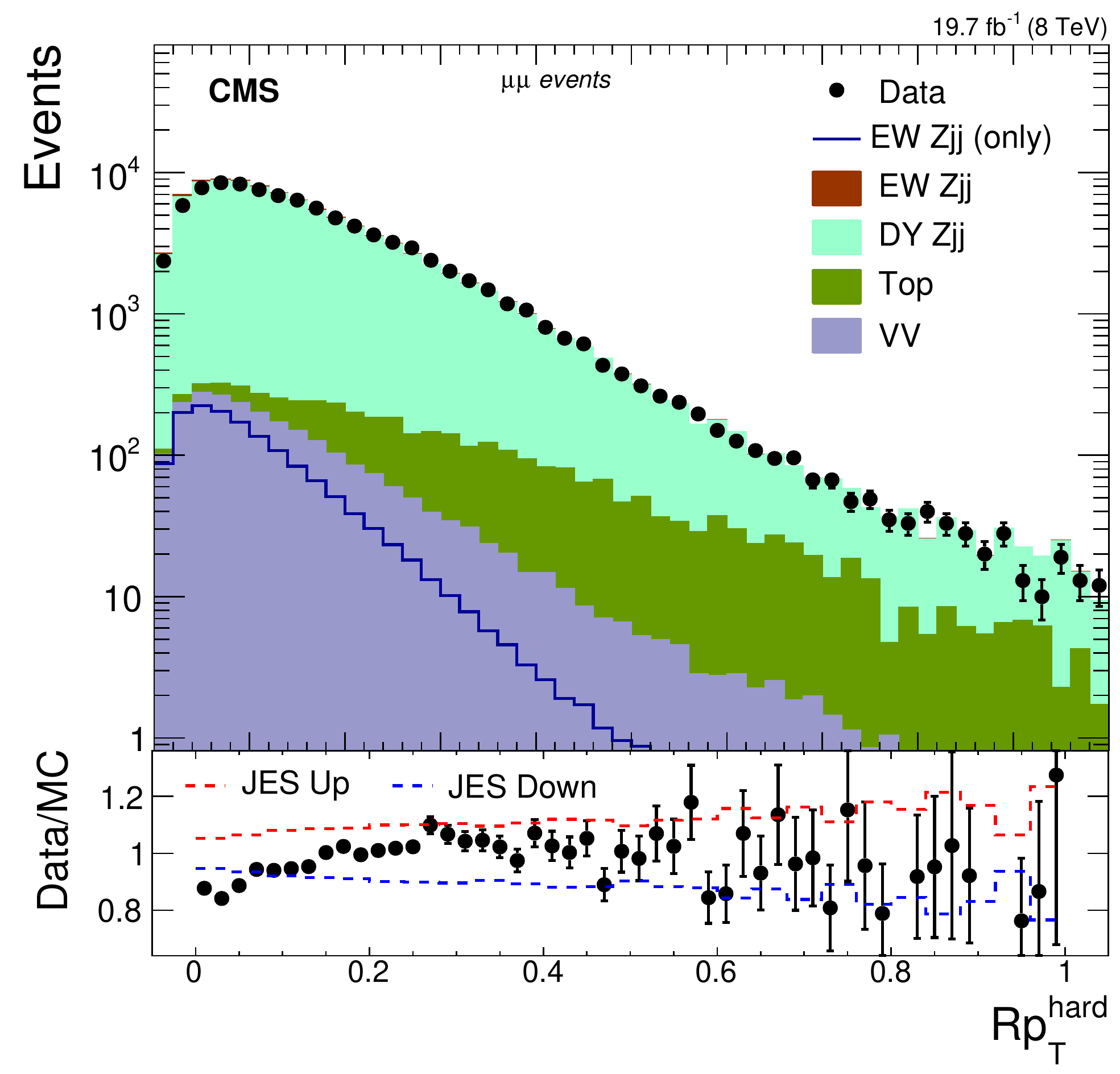}
\includegraphics[width=0.32\textwidth]{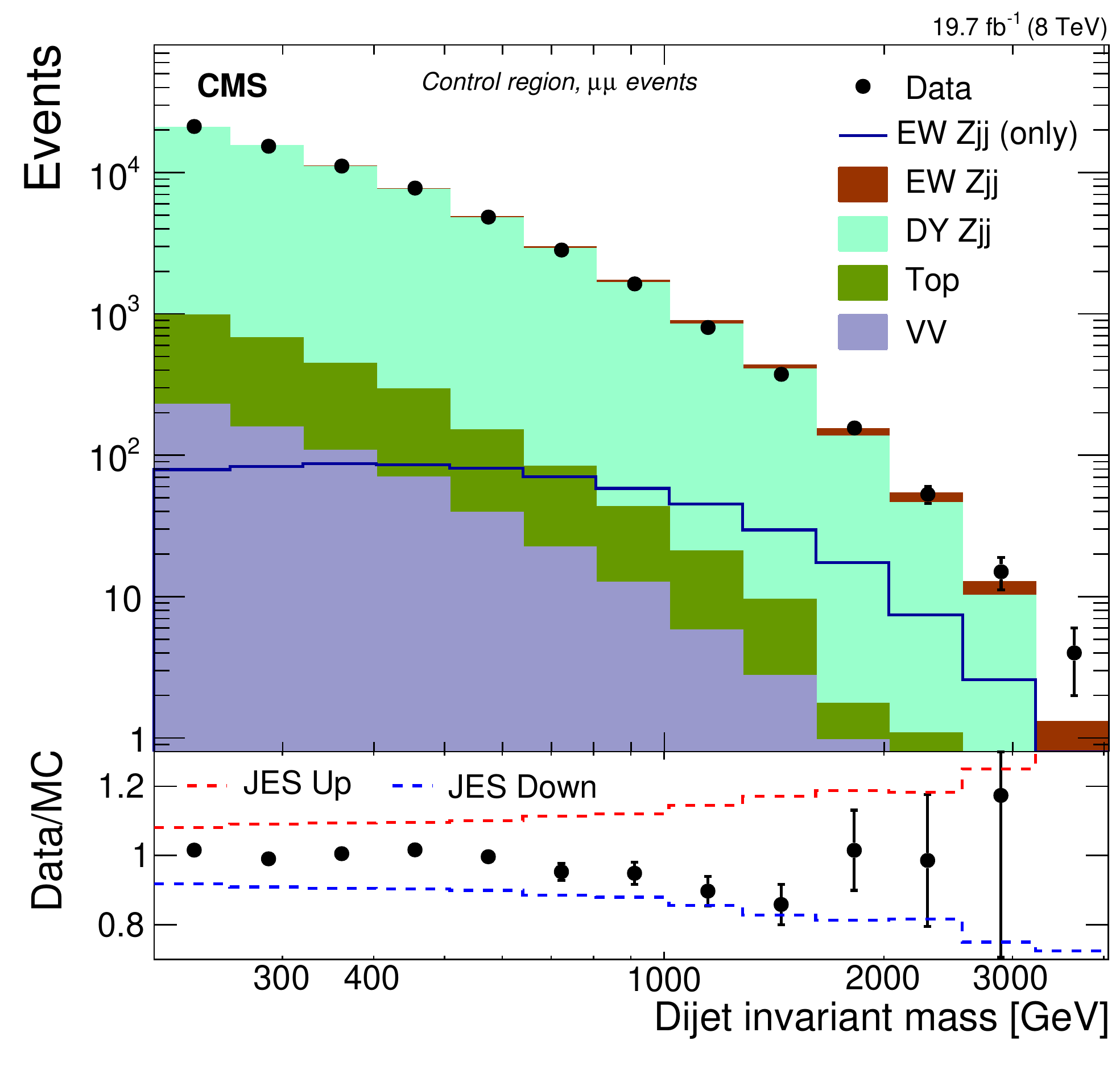}
\includegraphics[width=0.32\textwidth]{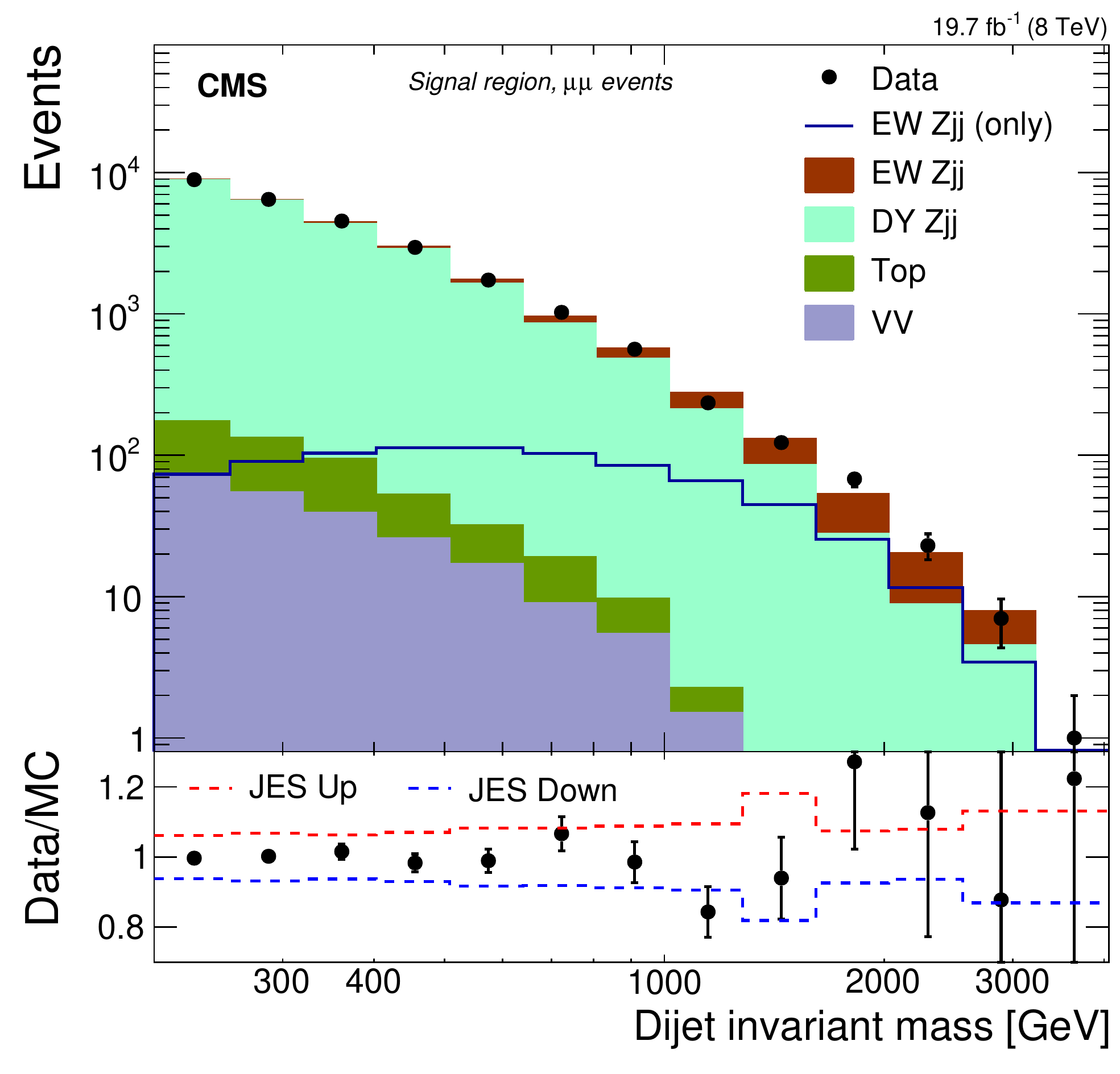}
\caption{
Distribution for (left) $R\pt^\text{hard}$
and $M_\mathrm{jj}$ for $\mu\mu$ events with
(middle) $R\pt^\text{hard}\geq0.14$ (control region)
and (right) $R\pt^\text{hard}<0.14$ (signal region).
The contributions from the different background sources and the signal are shown stacked,
with data points superimposed.
The panels below the distributions show
the ratio between the data and expectations
as well as the uncertainty
envelope for the impact of the uncertainty of the JES. \label{fig:mjj}
}
\end{figure*}

For the selection of the signal-region in the analysis where \dyzjj\
is based on simulation we make use of an event balance variable,
$R\pt^\text{hard}$, defined as\begin{equation}
R\pt^\text{hard}=
\frac
{\abs{ \vec{p}_{\mathrm{T} \mathrm{j}_1}+\vec{p}_{\mathrm{T} \mathrm{j}_2}+\vec{p}_{\mathrm{T} \cPZ}}}
{ \abs{\vec{p}_{\mathrm{T} \mathrm{j}_1}} +\abs{\vec{p}_{\mathrm{T} \mathrm{j}_2}} + \abs{\vec{p}_{\mathrm{T} \cPZ}} }
=
\frac
{ \abs{\vec{p}_{\mathrm{T}}^\text{hard}}}
{ \abs{\vec{p}_{\mathrm{T} \mathrm{j}_1}} +\abs{\vec{p}_{\mathrm{T} \mathrm{j}_2}} + \abs{\vec{p}_{\mathrm{T} \cPZ}} },
\end{equation}
where the numerator is the estimator of the \pt for the
hard process, \ie $\pt^\text{hard}$.
The distribution of the $R\pt^\text{hard}$ variable is shown in
Fig.~\ref{fig:mjj}~(left),
where data and simulation are found to be in agreement with each other.
It can be seen, from the same figure, that the variable is robust
against the variation of JES according to its uncertainty.
We apply a requirement of $R\pt^\text{hard}<0.14$
to select the signal region and the events failing this requirement
are used as a control region for the analyses.
The cut is motivated by the fact that the signal is expected to
have the \cPZ\ boson
balanced with respect to the dijet system in the transverse plane.
The events which fail this requirement are used as control region for
the modelling of the background.
The $M_\mathrm{jj}$ distribution in dimuon events
for the signal and control regions is shown in Fig.~\ref{fig:mjj},
(middle) and (right), correspondingly.
The reweighting of the \dyzjj\ background is applied to the simulation, as described above.
Data and predictions are found to be in agreement with each other.

\begin{figure*}[htp]
\centering
\includegraphics[width=0.32\textwidth]{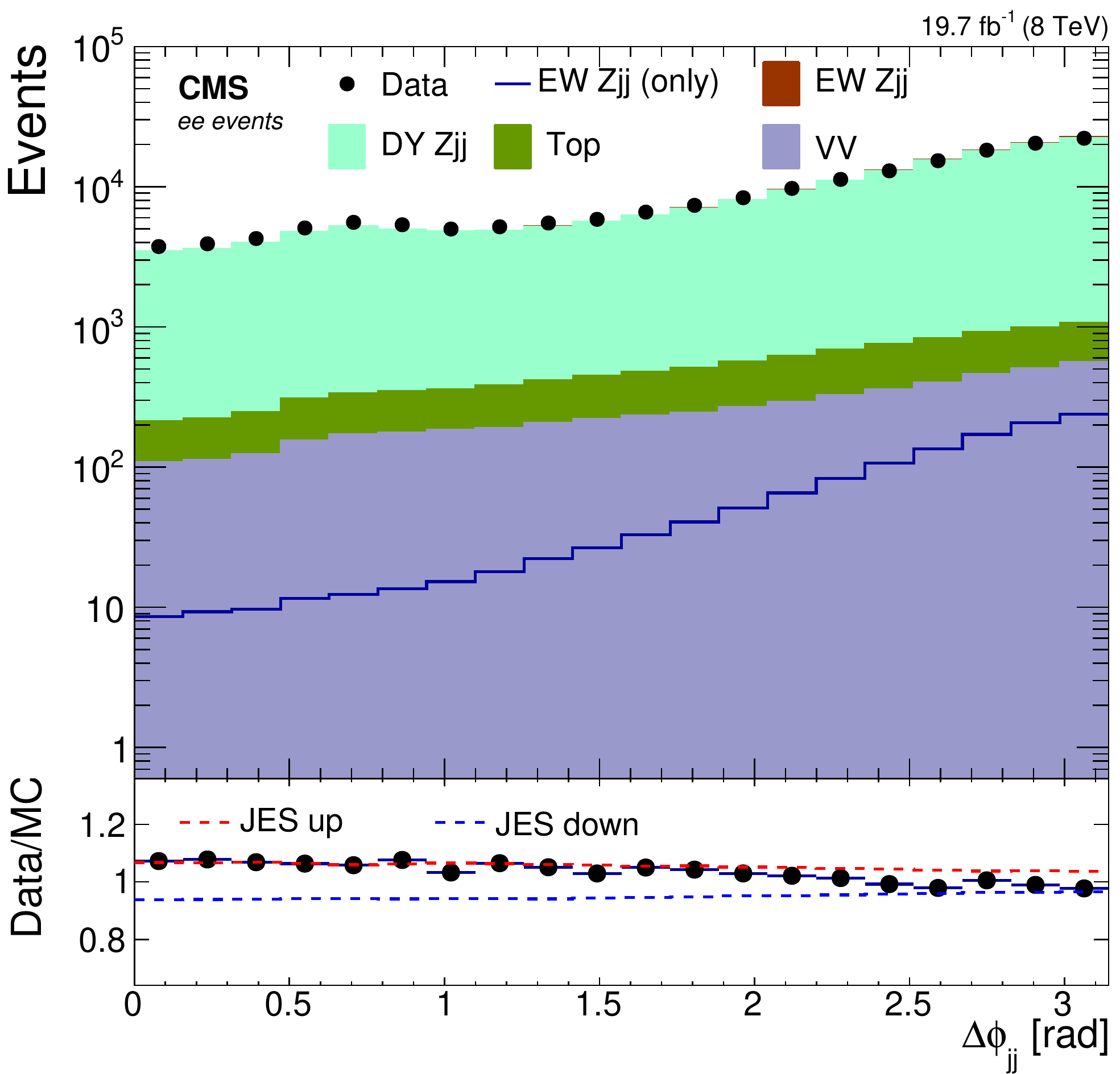}
\includegraphics[width=0.32\textwidth]{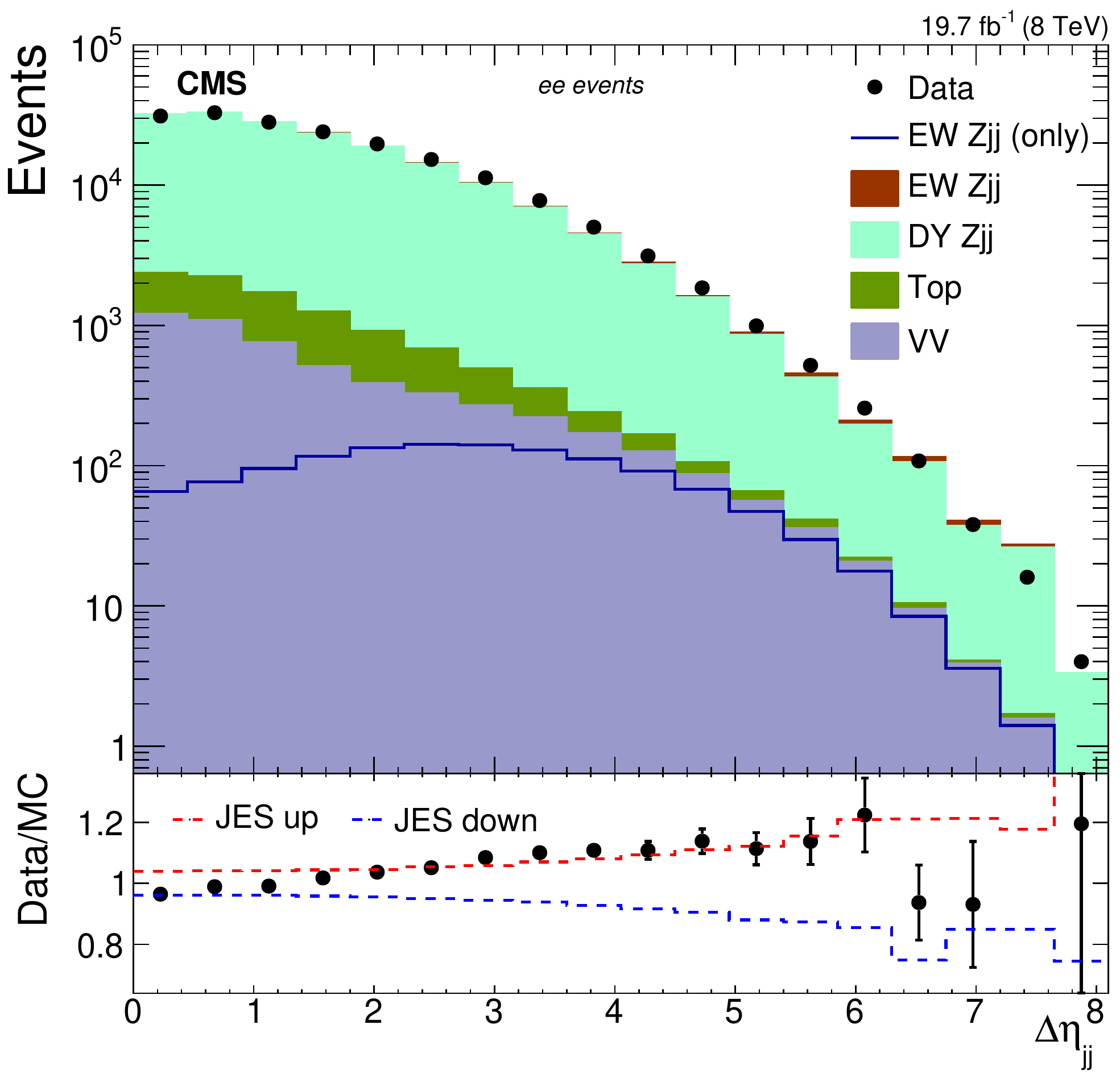}
\includegraphics[width=0.32\textwidth]{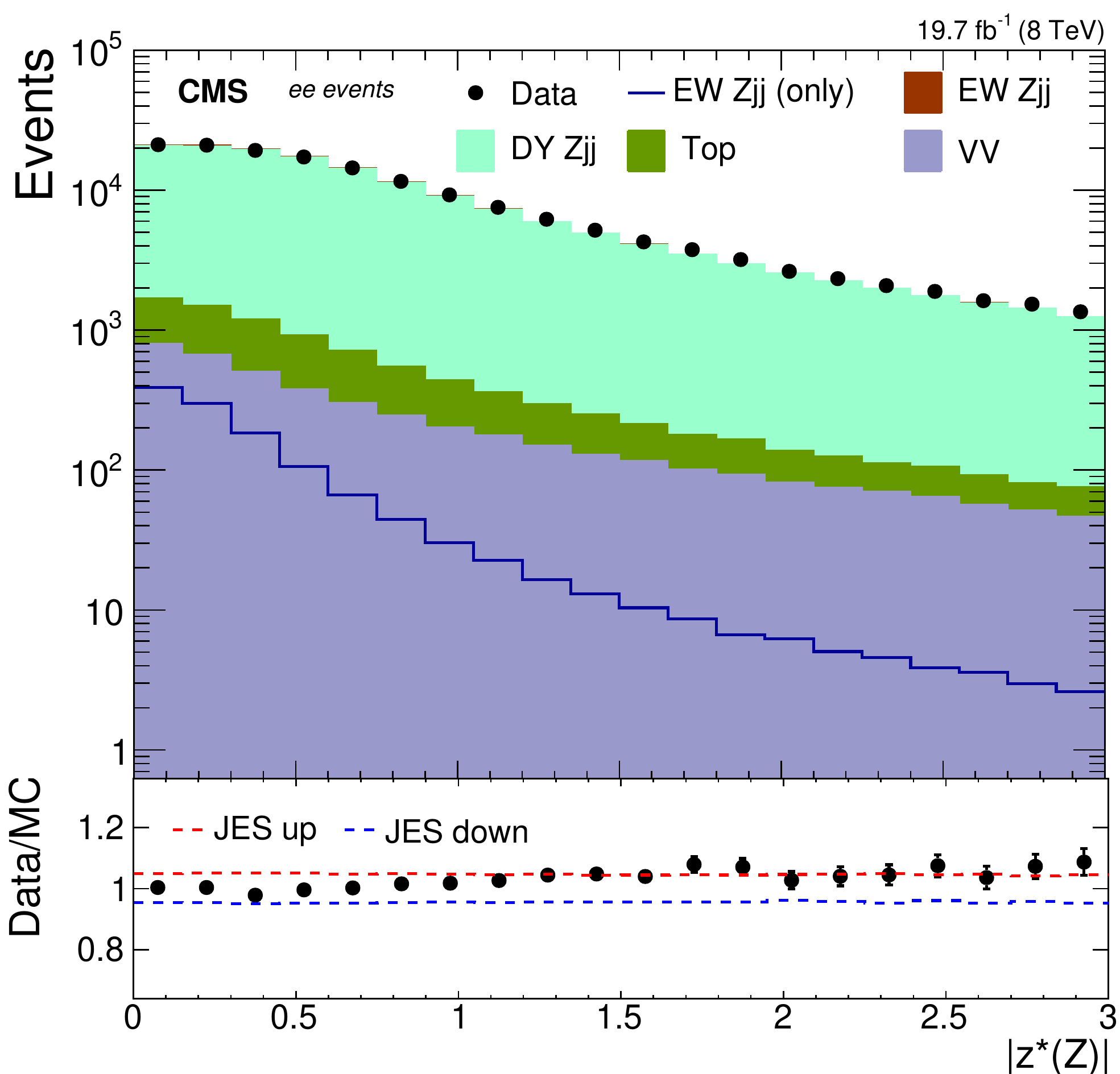} \caption{
Distribution for (left) the difference in the azimuthal angle
and (middle) difference in the pseudorapidity of the tagging jets
for ee events, with  $R\pt^\text{hard}\geq0.14$.
The $z^*$ distribution (right) is shown for the same
category of events.
The panels below the distributions show
the ratio between the data and expectations
as well as the uncertainty
envelope for the impact of the uncertainty of the JES.
\label{fig:angulardists}
}
\end{figure*}

Figure~\ref{fig:angulardists} shows distributions for angle-related variables.
Fair agreement is observed for the absolute differences in the azimuthal angle
($\Delta\phi_\mathrm{jj}$) and in the pseudorapidity ($\Delta\eta_\mathrm{jj}$) of
the tagging jets which are shown on the left and middle, respectively.
The $z^*$ variable~\cite{Schissler:2013nga} is shown in
Fig.~\ref{fig:angulardists} (right), and
it is defined as \begin{equation}
z^*=\frac{ y^* } { \Delta y_\mathrm{jj} }.
\end{equation}

Data is verified to be in good agreement with the prediction for the
distribution in $ z^*$ variable.

\textbf{Data-based prediction for background}

The diagrams contributing to the production of a photon and two jets
(\gjj)
are expected to resemble those involved in the production of
\dyzjj\ (see Fig.~\ref{fig:bkgdiagram}).
Thus, we build a data-based model for the shapes of the distributions
of the kinematic observables of the tagging jets
from \gjj\ events selected in a similar way as the
\Zjj\ ones.
The differences, specific to the $\cPZ$ or photon-sample, are expected to be mitigated by reweighting the
\pt of the photons to the \pt of the $\cPZ$ candidates.
From simulation, we expect that the differences between the $\gamma$ and $\cPZ$
masses do not contribute significantly when matching the dijet kinematics between the
two samples after $M_\mathrm{jj}>2M_{\cPZ}$ is required.
Given that the photon sample is
affected by multijet production, and that the selection of the
low-\pt region in data
is also affected by very large prescaling at the trigger stages, we impose tighter
kinematic constraints on the reconstructed boson, with respect to the
ones applied at pre-selection (Section~\ref{sec:evsel}).
To match effectively the $\cPZ$ and photon kinematics,
we require $\pt(\cPZ\text{ or }\gamma)>50\GeV$ and rapidity
$\vert y(\cPZ\text{ or }\gamma)\vert<1.44$.
The rapidity requirement corresponds to the physical boundary of the
central (barrel) region of the CMS ECAL~\cite{Chatrchyan:2008zzk}.

\begin{figure*}[htbp]
\centering
\includegraphics[width=0.49\textwidth]{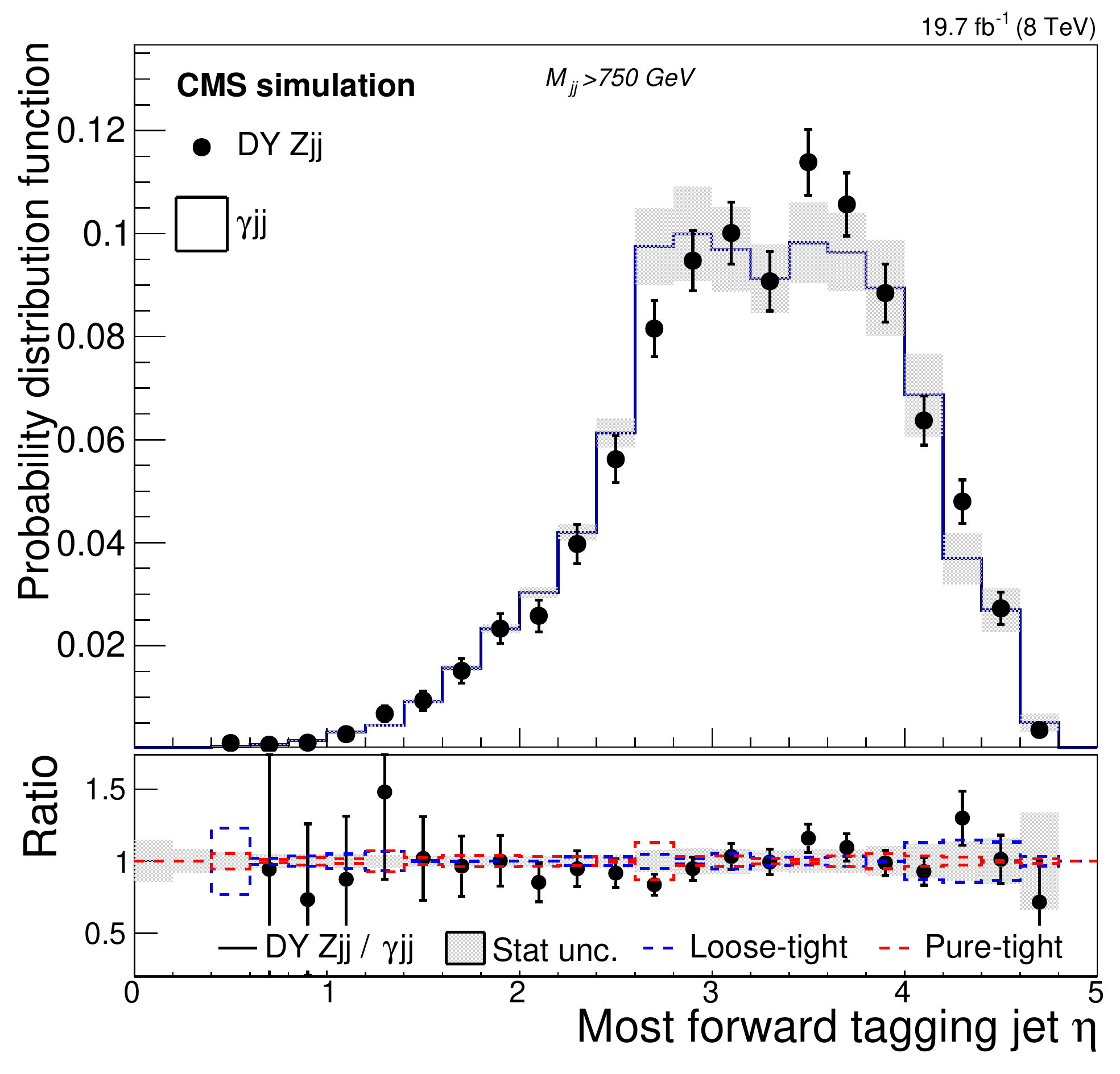}
\includegraphics[width=0.49\textwidth]{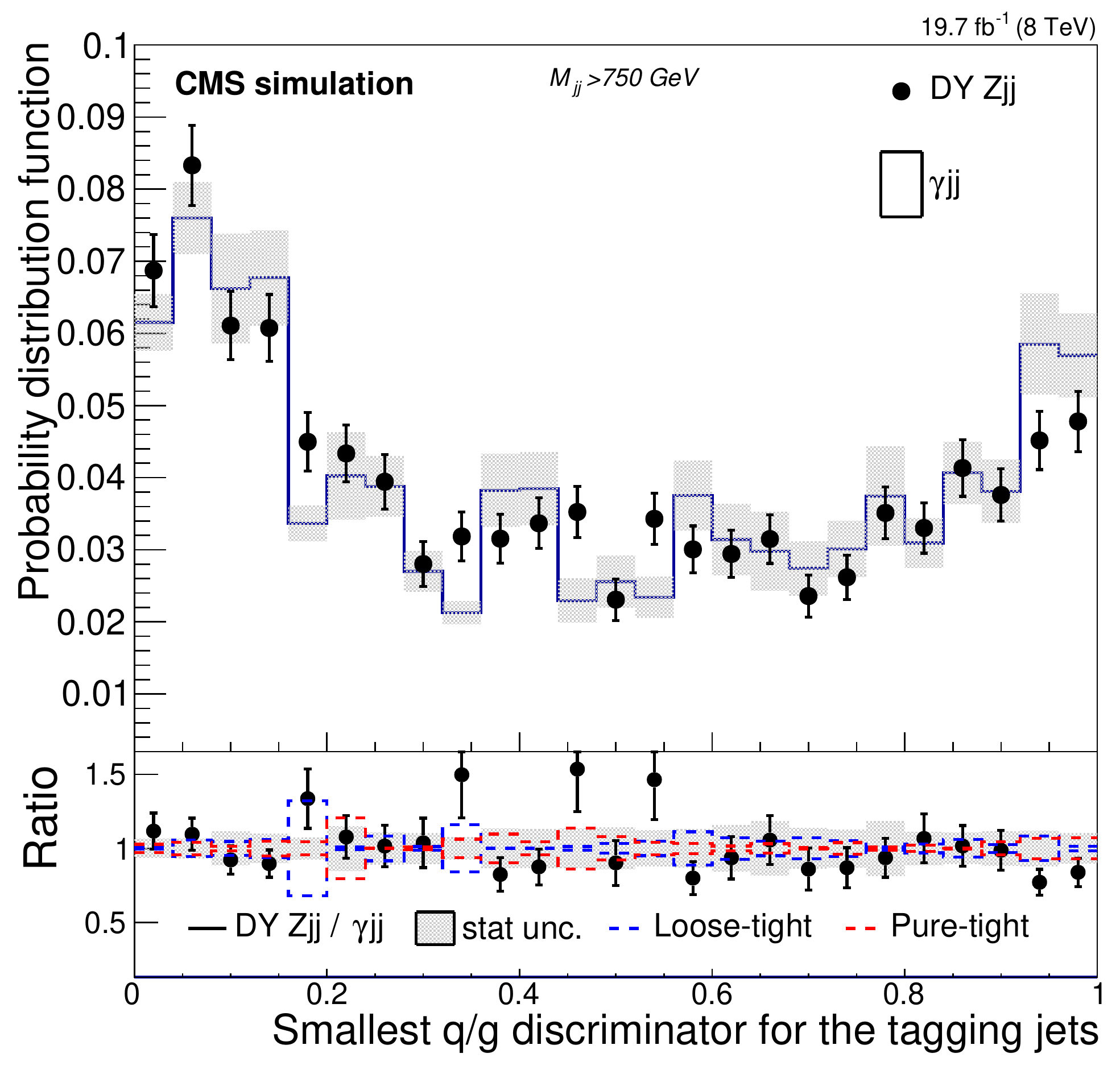}
\includegraphics[width=0.49\textwidth]{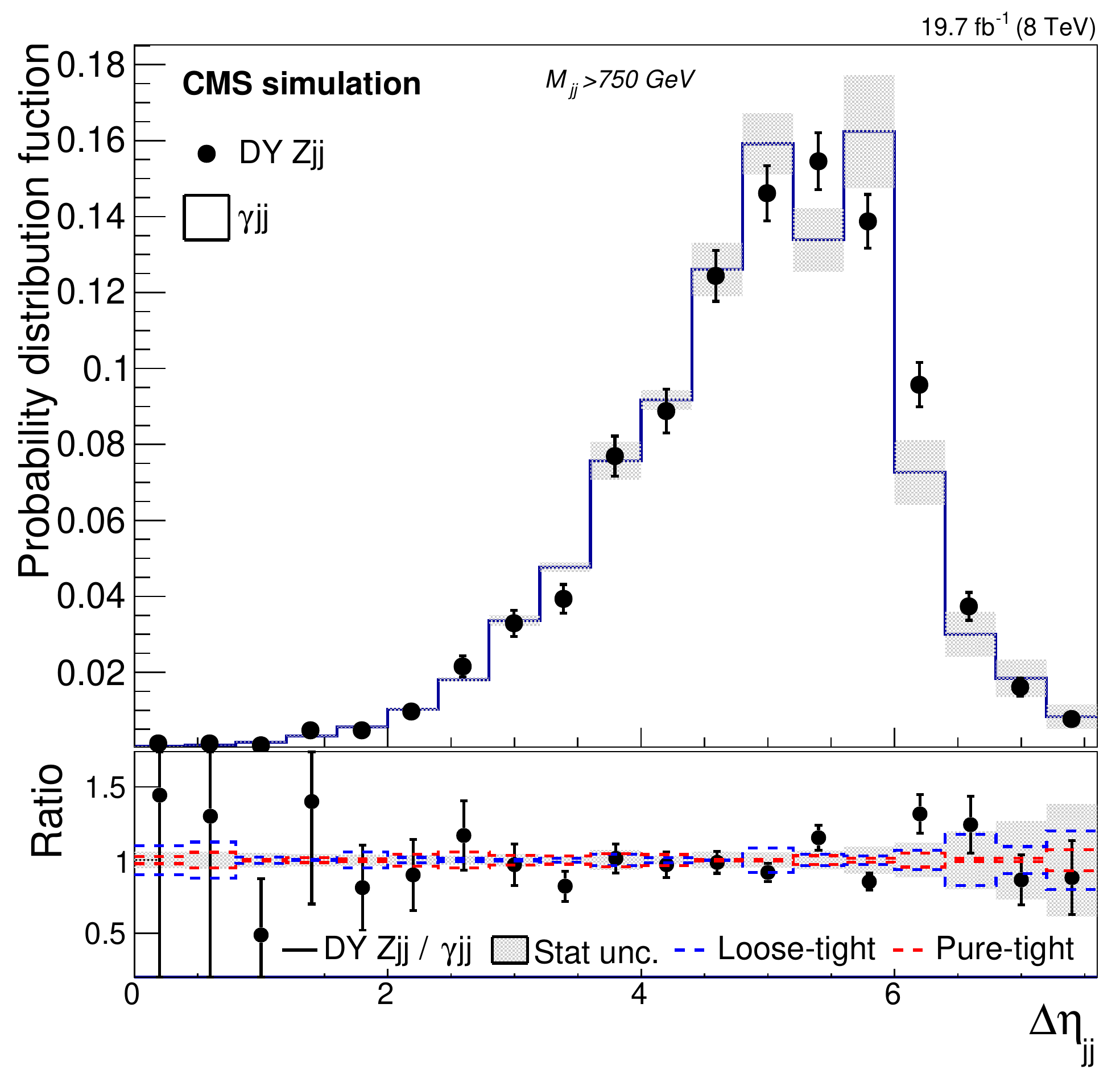}
\includegraphics[width=0.49\textwidth]{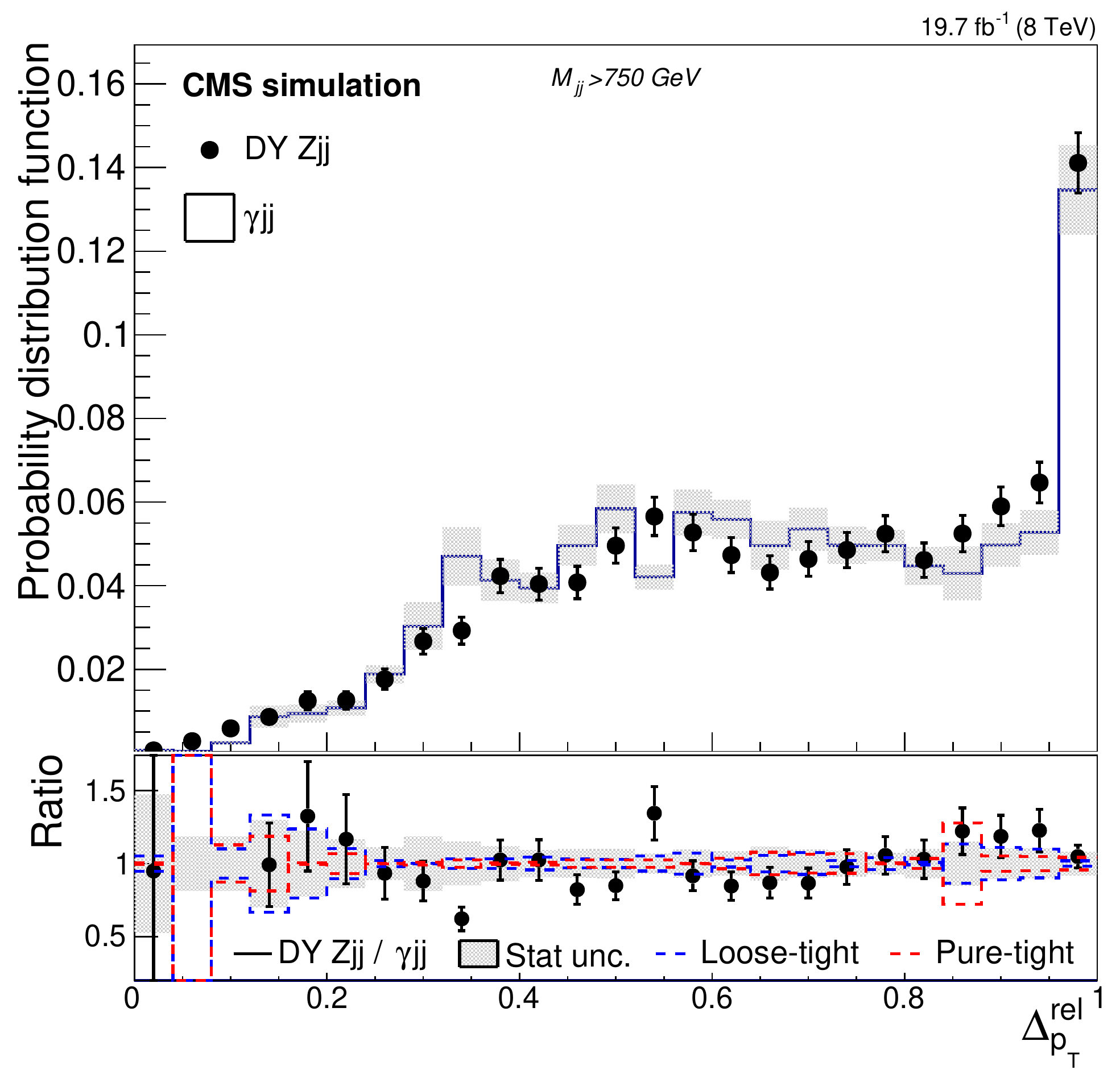}
\caption{
Comparison of the \dyzjj\ distributions with the prediction
from the photon control sample, for simulated events with $M_\mathrm{jj}>750\GeV$.
The upper left subfigure shows the distributions in the pseudorapidity $\eta$ of the most
forward tagging jet
and the upper right shows the smallest q/g discriminant of the two tagging jets.
The lower left shows the pseudorapidity separation $\Delta\eta_\mathrm{jj}$
and the lower right the relative \pt balance of the tagging jets $\Delta^\text{rel}_{\pt}$.
The DY \gjj\ distribution contains the contribution from prompt
and misidentified photons as estimated from simulation and it is compared to
the simulated \dyzjj\ sample in the top panel of each subfigure.
The bottom panels show the ratio between
the \dyzjj\ distribution and the photon-based prediction, and includes
the different sources of estimated total
uncertainty in the background shape from the photon control sample.
(See text for specification of impact of loose, tight and pure photons).
\label{fig:photonclosure}
}
\end{figure*}

The method is checked in simulation by characterising
the \dyzjj\ or direct photon events in different physical regions
defined according to the reconstructed $M_\mathrm{jj}$ and comparing both distributions.
Figure~\ref{fig:photonclosure} illustrates the
compatibility of simulated events with a high dijet invariant mass.
Good agreement is found for the  $\eta$ of the most forward jet, the $\Delta\eta_\mathrm{jj}$ variable
and the ratio between the \pt of the dijet system to the scalar sum of the tagging jets' \pt,
\begin{equation}
\Delta^\text{rel}_{\pt}=
\frac{\abs{\vec{p}_{\mathrm{T} \mathrm{j}_1}+\vec{p}_{\mathrm{T} \mathrm{j}_2}}}{\abs{\vec{p}_{\mathrm{T} \mathrm{j}_1}}+\abs{\vec{p}_{\mathrm{T} \mathrm{j}_2}}}.
\end{equation}

The smallest of the quark/gluon discriminant value among the tagging jets
is also found to be in agreement --- Fig.~\ref{fig:photonclosure} (top right).
In general, the kinematics of the tagging jets predicted from the photon sample
are found to be in agreement with those observed in DY \cPZ\ events also for
lower $M_\mathrm{jj}$ values.
A similar conclusion holds for other global event observables inspected
in the simulation, such as energy fluxes and angular correlations.

The result of the compatibility tests described above
have the potential to yield a correction factor
to be applied to the \dyzjj\ prediction from the photon data.
However due to the limited statistics in our simulation
and due to uncertainties in handling the simulation of residual
background from multijet events in data,
we have opted to use the simulation-based compatibility test results
to assign, instead, an uncertainty in the final shape.
We assign the difference in the compatibility tests relative
to a pure prompt-photon possibility as one of the systematic
uncertainties.
The changes observed in the compatibility test, obtained after varying the
PDF by its uncertainties synchronously in the two samples is also
assigned as a source of uncertainty.
In data, the difference between a ``tight'' and a ``loose'' photon selections
is, furthermore, assigned as an extra source of systematic
uncertainty.
The selection is tightened by applying stricter requirements on the
photon identification and isolation requirements.
This prescription is adopted to cover possible effects from the
contamination of multijet processes.

The final distributions for \dyzjj\ events are obtained after
subtracting a residual contamination from pure EW production of a
photon in association with two jets (\ewkgjj) \cite{Jager:2010aj}.
The diagrams for the latter process are similar to the ones of
Fig.~\ref{fig:sigdiagram}~(left) and (middle), where the $\cPZ/\gamma^*$
is now a real photon.
For a fiducial phase space defined by $M_\mathrm{jj} >120\GeV$,
$\ptj > 30\GeV$,
$\abs{\eta_\mathrm{j}}< 5$, $p_{\mathrm{T} \gamma}>50\GeV$ and
$\abs{\eta_\gamma}<1.5$, the production cross section of \ewkgjj\ process is expected to be 2.72\unit{pb},
based on the \MADGRAPH generator.
After event reconstruction and selection,
we estimate the ratio of the number of \ewkgjj\ candidate
events to the total number of photon events selected in data to be a factor of $\approx$5
times smaller than the ratio between the expected \ewkzjj\ and \dyzjj\ yields.
From simulations this ratio is expected to be independent of $M_\mathrm{jj} $.
In the subtraction procedure, a 30\%  normalisation
uncertainty is assigned to this residual process, which
corresponds to approximately twice the envelope of variations obtained for
the cross section at NLO with \textsc{vbfnlo},
after tightening the selection criteria and changing
the factorisation and renormalisation scales.

\begin{figure*}[htbp]
\centering
\includegraphics[width=0.4\textwidth]{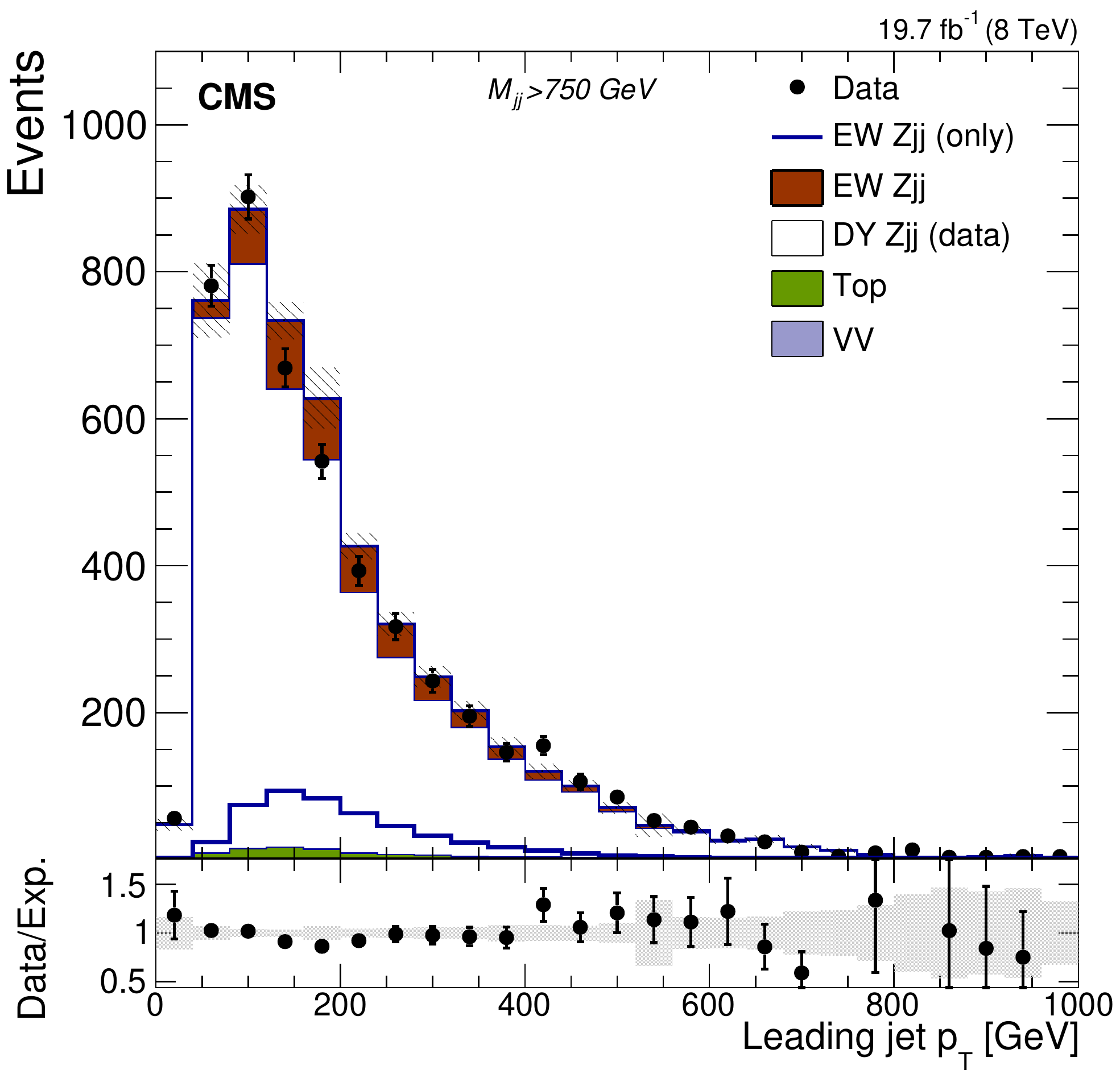}
\includegraphics[width=0.4\textwidth]{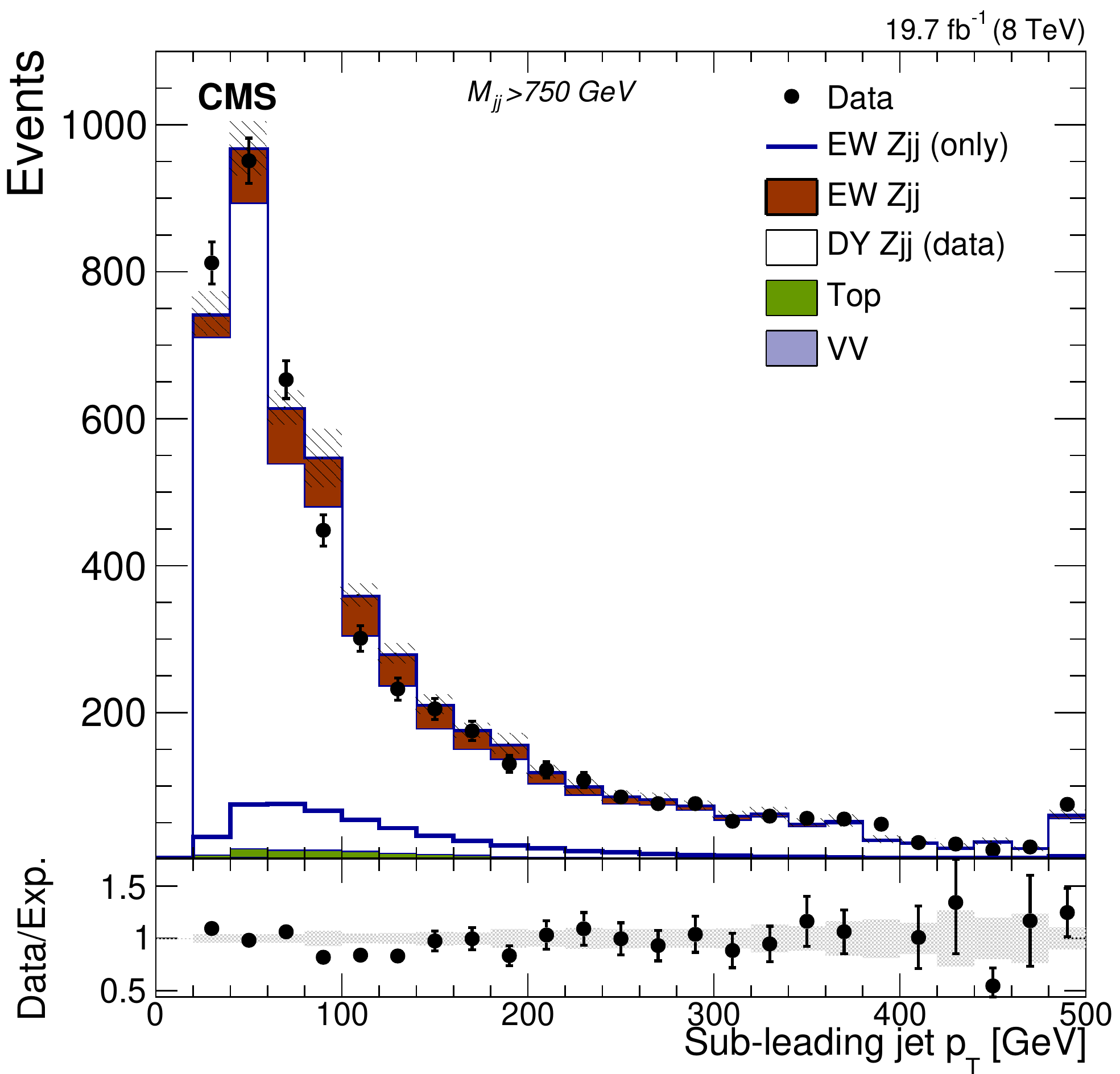}
\includegraphics[width=0.4\textwidth]{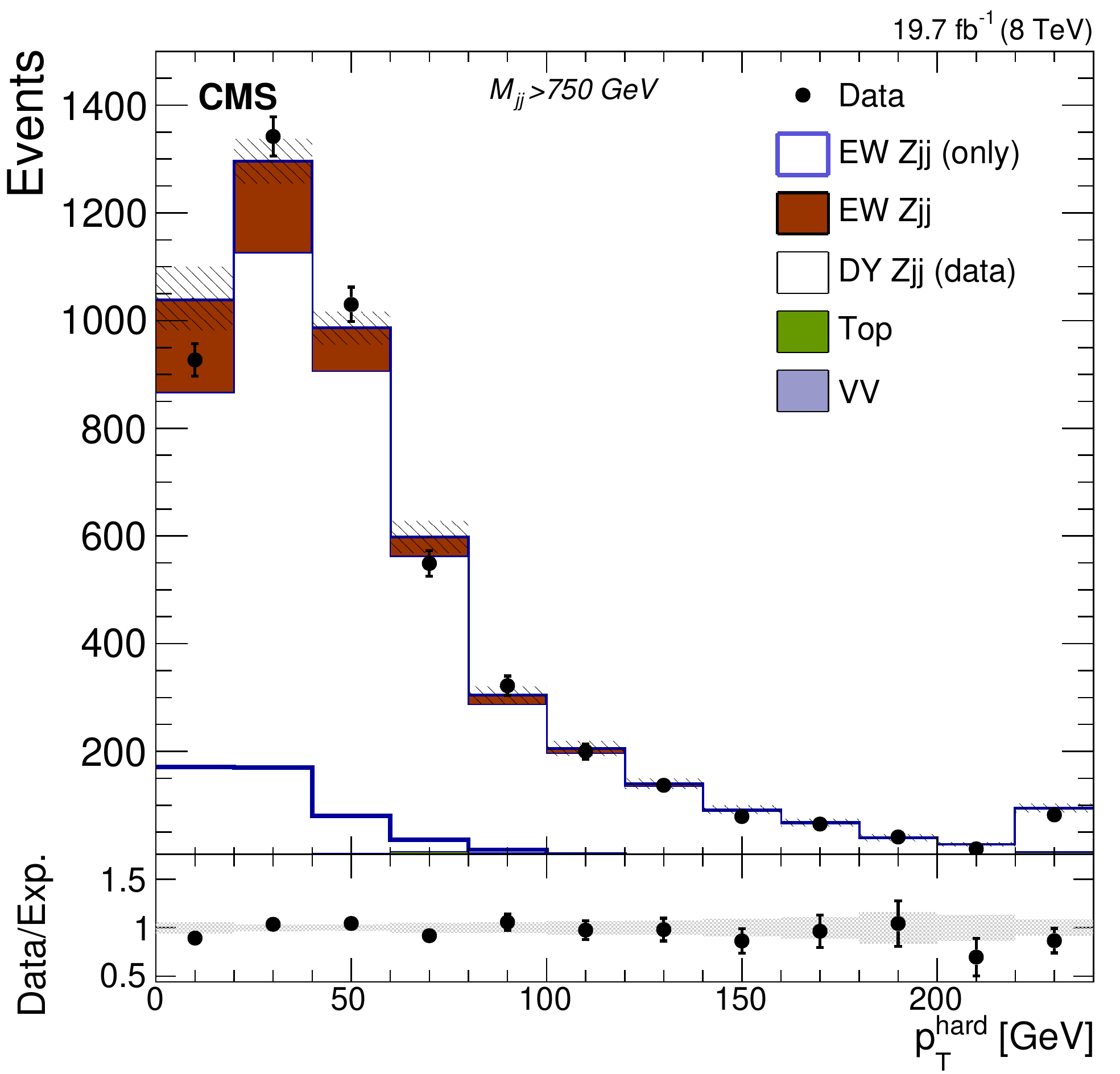}
\includegraphics[width=0.4\textwidth]{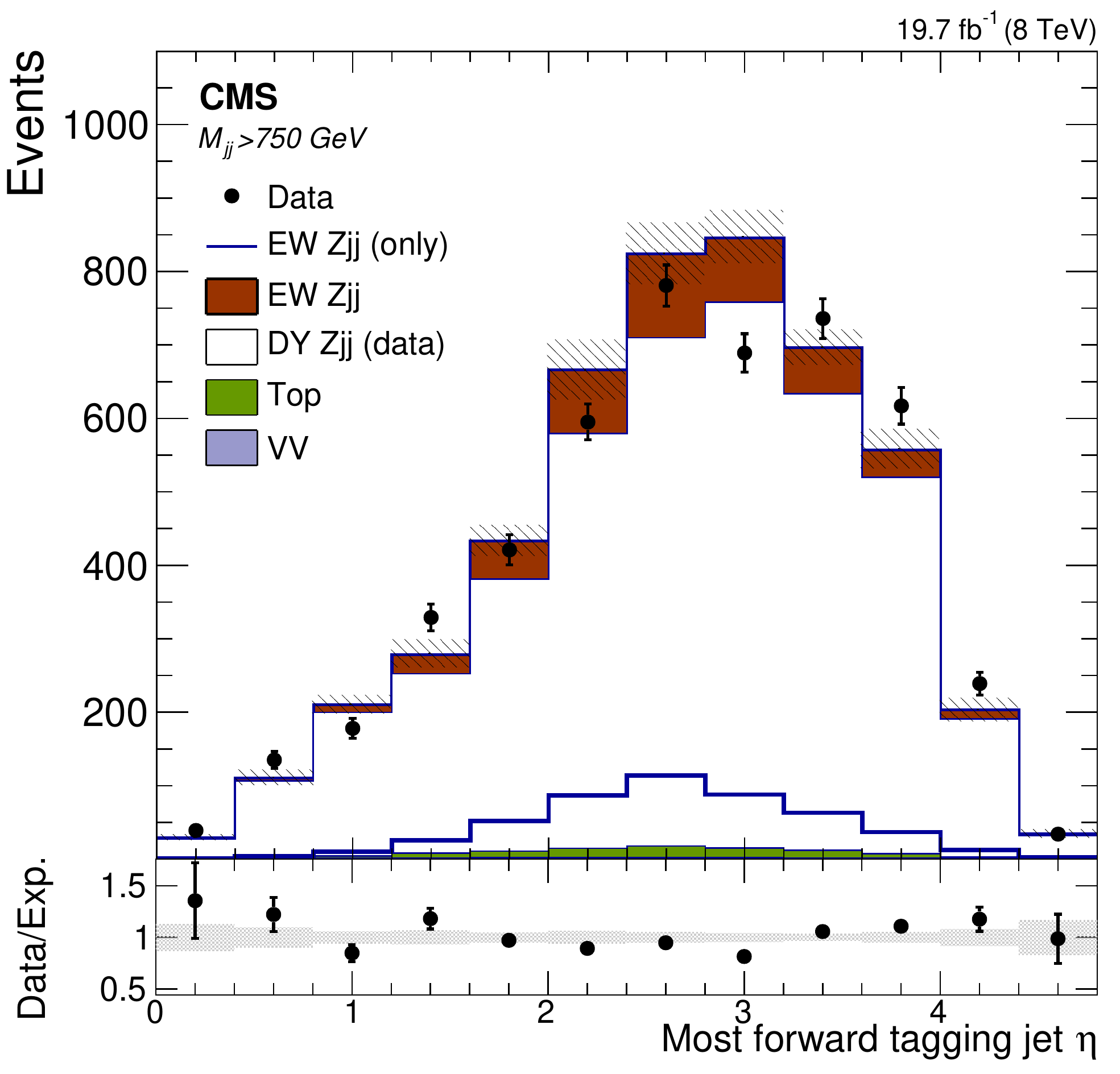}
\includegraphics[width=0.4\textwidth]{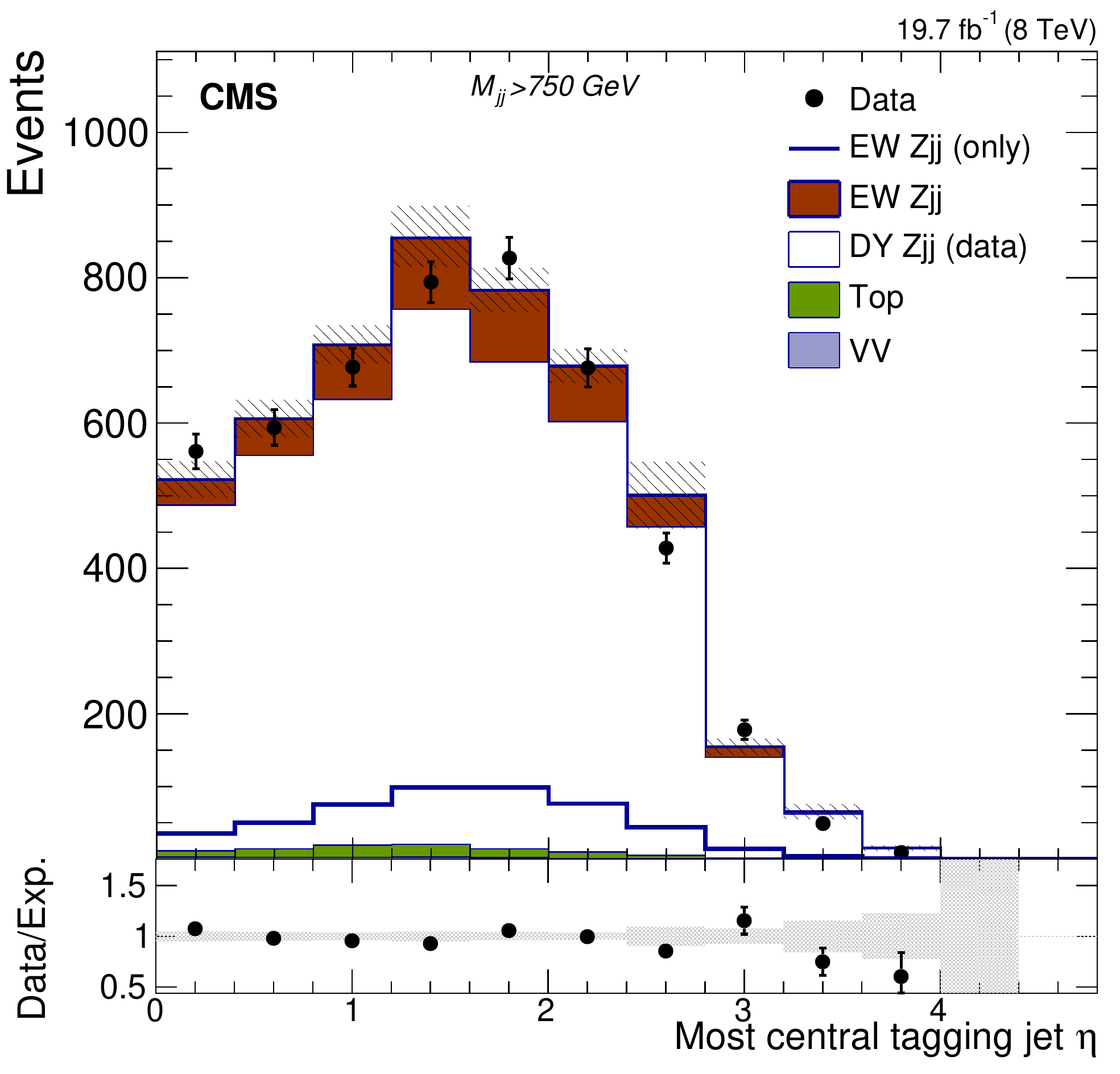}
\includegraphics[width=0.4\textwidth]{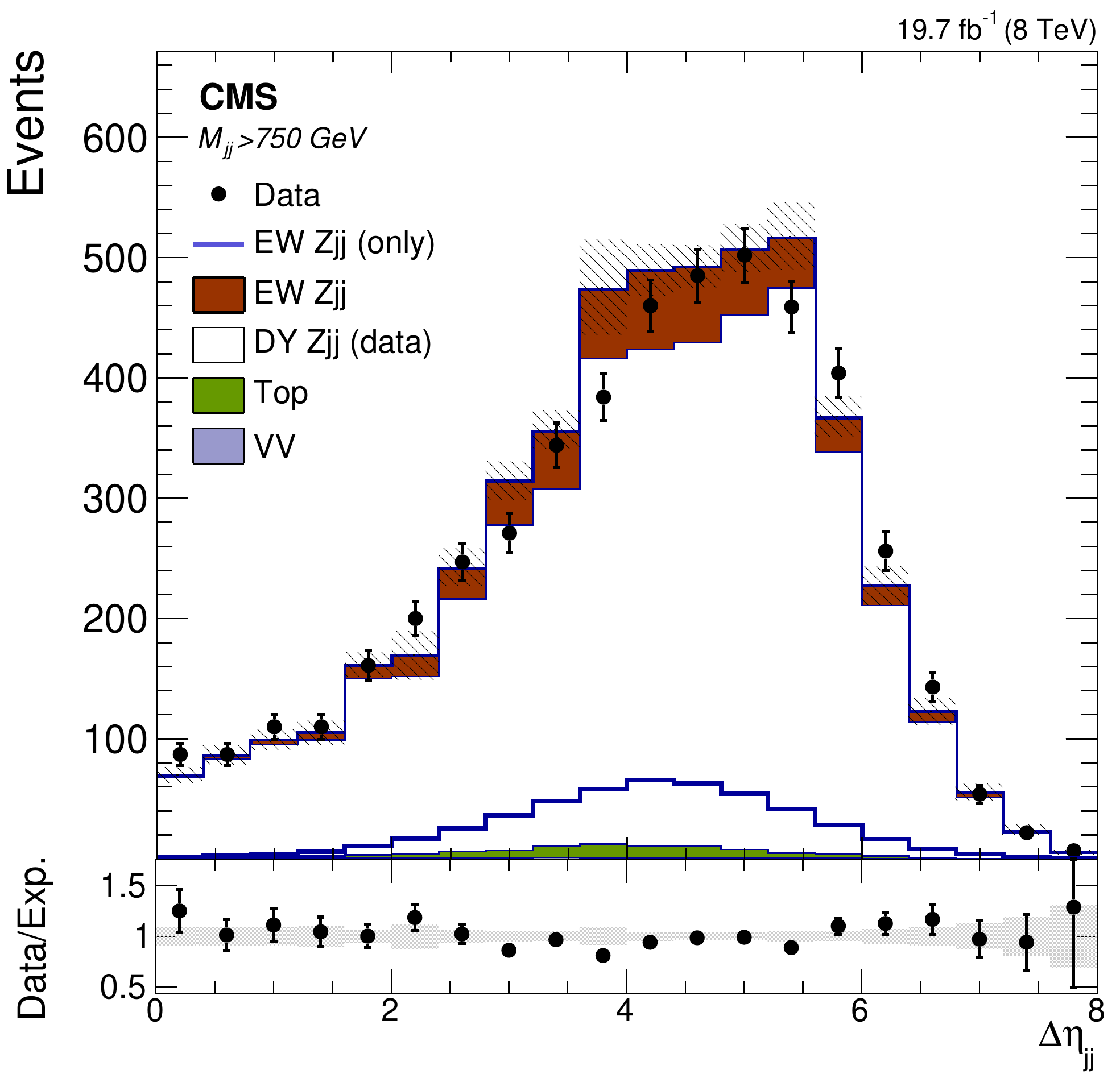}
\caption{
Distributions for the tagging jets for $M_\mathrm{jj}>750\GeV$ in the
combined dielectron and dimuon event sample:
(upper left) \pt of the leading jet, (upper right) \pt of the sub-leading jet,
(middle left) hard process \pt (dijet+$\cPZ$ system),
(middle right) $\eta$ of the most forward jet,
(lower left) $\eta$ of the most central jet and
(lower right) $\Delta\eta_\mathrm{jj}$ of the tagging jets.
In the top panels, the contributions from the different background
sources and the signal are shown stacked being data superimposed.
In all plots the signal shape is also superimposed separately as a thick line.
The bottom panels show the ratio between data and total prediction.
The total uncertainty assigned to the \dyzjj\ background estimate from \gjj\ control sample in data is shown
in all panels as a shaded grey band.
\label{fig:tagjetkin}
}
\end{figure*}

The results obtained when the
data-based prediction, used to characterise the \dyzjj\ contribution
to the reconstructed kinematics of the tagging jets in data,
show a good agreement for different dijet invariant mass categories.
Figure~\ref{fig:tagjetkin} illustrates the agreement
observed for $M_\mathrm{jj}>750\GeV$ in the distribution of different
variables:
(upper left) \pt of the leading jet, (upper right) \pt of the sub-leading jet,
(middle left) hard process \pt (dijet+$\cPZ$ system),
(middle right) $\eta$ of the most forward jet,
(lower left) $\eta$ of the most central jet and
(lower right) $\Delta\eta_\mathrm{jj}$ of the tagging jets.

\section{Signal discriminants and extraction procedure}
\label{sec:sigdisc}

We use a multivariate analysis technique
that provides separation of the \dyzjj\ and \ewkzjj\ components of the inclusive
\lljj\ spectrum.
As discussed previously, the \ewkzjj\ signal is characterised by a
large $\Delta\eta_\mathrm{jj}$ jet separation
that stems from the small-angle scattering of the two initial partons.
Owing to both the topological configuration and the large \pt
of the outgoing partons, the $M_\mathrm{jj}$ variable is also
expected to be large.
The evolution of $\Delta\eta_\mathrm{jj}$ with $M_\mathrm{jj}$ is expected
to be different in signal and background events and therefore
these characteristics are expected to yield the best separation power
between the \ewkzjj\ and the \dyzjj\ productions.
In addition, one can exploit the fact
that the $\cPZ$-boson candidate is expected to be produced centrally
in the rapidity region defined by the two tagging jets
and that the \Zjj\ 
system is approximately balanced in the transverse plane.
As a consequence, we expect the signal to be found with lower values of both $y^*$
and $\pt^\text{hard}$, compared to the DY background.
Other variables which can be used to enhance the separation
are related to the kinematics of the event
(\pt, rapidity, and distance between the jets and/or the $\cPZ$ boson)
or to the properties of the jets that are expected to be initiated by
quarks.
We combine these variables using three alternative
multivariate analyses with the goal of cross-checking the final
result.
All three analyses make use of boosted decision tree
(BDT) discriminators
implemented using \textsc{tmva} package~\cite{Hocker:2007ht}
to achieve the best expected separation between the
\ewkzjj\ and \dyzjj\ processes.

\begin{description}
\item[Analysis A] {\tolerance=1000 expands one of the procedures previously adopted for the
7\TeV measurement~\cite{Chatrchyan:2013jya}. It uses both dimuon and
dielectron final states and PF jet reconstruction. A multivariate
discriminator making use of the dijet and \cPZ\ boson kinematics is
built.  A choice is made for variables
which are robust against JES uncertainties.
Extra discrimination information,
related to the q/g nature of the jet, is included.
All processes are modelled from simulation, and
the description of each variable is verified by comparing data with
the simulation-based expectations in control regions.
\item[Analysis B] uses only the dimuon final state and the JPT jet reconstruction
approach. It builds a discriminator which tries to profit from the
full kinematics of the event including the tagging jets and the
\cPZ\ boson. Similarly to analysis A it expands one of the cross-check
procedures previously adopted for the 7\TeV
measurement~\cite{Chatrchyan:2013jya}
and relies on simulation-based prediction of the backgrounds.
\item[Analysis C]  uses solely dijet-related variables in the
 multivariate discriminator and selects both the  dimuon and dielectron final
states with PF jets. Lepton-related selection variables are not used as the main background is derived
from the photon control sample.
In this analysis events are split in four categories for $M_\mathrm{jj}$ values
in the intervals 450--550\GeV, 550--750\GeV, 750--1000\GeV, and above 1000\GeV,
which have been chosen to have similar numbers of expected signal
events.}
\end{description}

Table~\ref{tab:methodcomp} compares in more detail the three
independent analyses A, B and C.
From simulation, the statistical correlation between the analyses, if
performed with the same final state, is estimated to be $\approx$60\%.

\begin{table*}[htbp]
\centering
\topcaption{Comparison of the selections and variables used  in three different analyses.
The variables marked with the black circle are used in the discriminant of the indicated analysis.
}
\label{tab:methodcomp}
\begin{tabular}{l|ccc}
\hline
Analysis & \textit{A} & \textit{B} & \textit{C} \\\hline
\multirow{2}{*}{Channels} & \multirow{2}{*}{ee, $\mu\mu$} & \multirow{2}{*}{$\mu\mu$} & ee, $\mu\mu$ \\
& & & binned in $M_\mathrm{jj}$\\\hline
\multirow{4}{*}{Selection}
& \multicolumn{3}{c}{$p_{\mathrm{T} \mathrm{j}_1,\mathrm{j}_2}>50,30\GeV$}  \\
\cline{2-4}
& \multicolumn{2}{c|}{$R\pt^\text{hard}<0.14$} & $p_{\mathrm{T}\cPZ}>50\GeV$\\
& \multicolumn{2}{c|}{$\abs{y^{*}}<1.2$} & $\abs{y_\cPZ}<1.4442$\\
& \multicolumn{2}{c|}{$M_\mathrm{jj}>200\GeV$} & $M_\mathrm{jj}>450\GeV$\\\hline
Jets & PF & JPT & PF \\\hline
Variables used & & & \\
~~$M_\mathrm{jj}$ & \usedSign & \usedSign & \usedSign \\
~~$p_{\mathrm{T} \mathrm{j}_1},p_{\mathrm{T} \mathrm{j}_2}$ &   & \usedSign &  \usedSign  \\
~~$\eta_{\mathrm{j}_1},\eta_{\mathrm{j}_2}$ &  & &  \usedSign  \\
~~$\Delta_\text{rel}(\mathrm{jj})=\frac{\vert\vec{p}_{\mathrm{T} \mathrm{j}_1}+\vec{p}_{\mathrm{T} \mathrm{j}_2}\vert}{p_{\mathrm{T} \mathrm{j}_1}+p_{\mathrm{T} \mathrm{j}_2}}$ & & &  \usedSign  \\
~~$\Delta\eta_\mathrm{jj}$ & & \usedSign  & \\
~~$\vert\eta_{\mathrm{j}_1}\vert+\vert\eta_{\mathrm{j}_2}\vert$ & \usedSign & \usedSign & \usedSign  \\
~~$\Delta\phi_\mathrm{jj}$ &  & \usedSign & \usedSign  \\
~~$\Delta\phi_{\cPZ,{\mathrm{j}_1}}$ & & \usedSign &  \\
~~$y_\cPZ$ &  \usedSign & \usedSign & \\
~~$z^*_\cPZ$ & \usedSign & & \\
~~$p_{\mathrm{T} \cPZ}$& \usedSign & \usedSign & \\
~~$R\pt^\text{hard}$ & & \usedSign & \\
~~q/g discriminator & \usedSign & & \usedSign \\\hline
\dyzjj\ model & MC-based & MC-based & From data\\\hline
\end{tabular}
\end{table*}

\begin{figure*}[htp] {
\centering
\includegraphics[width=0.49\textwidth]{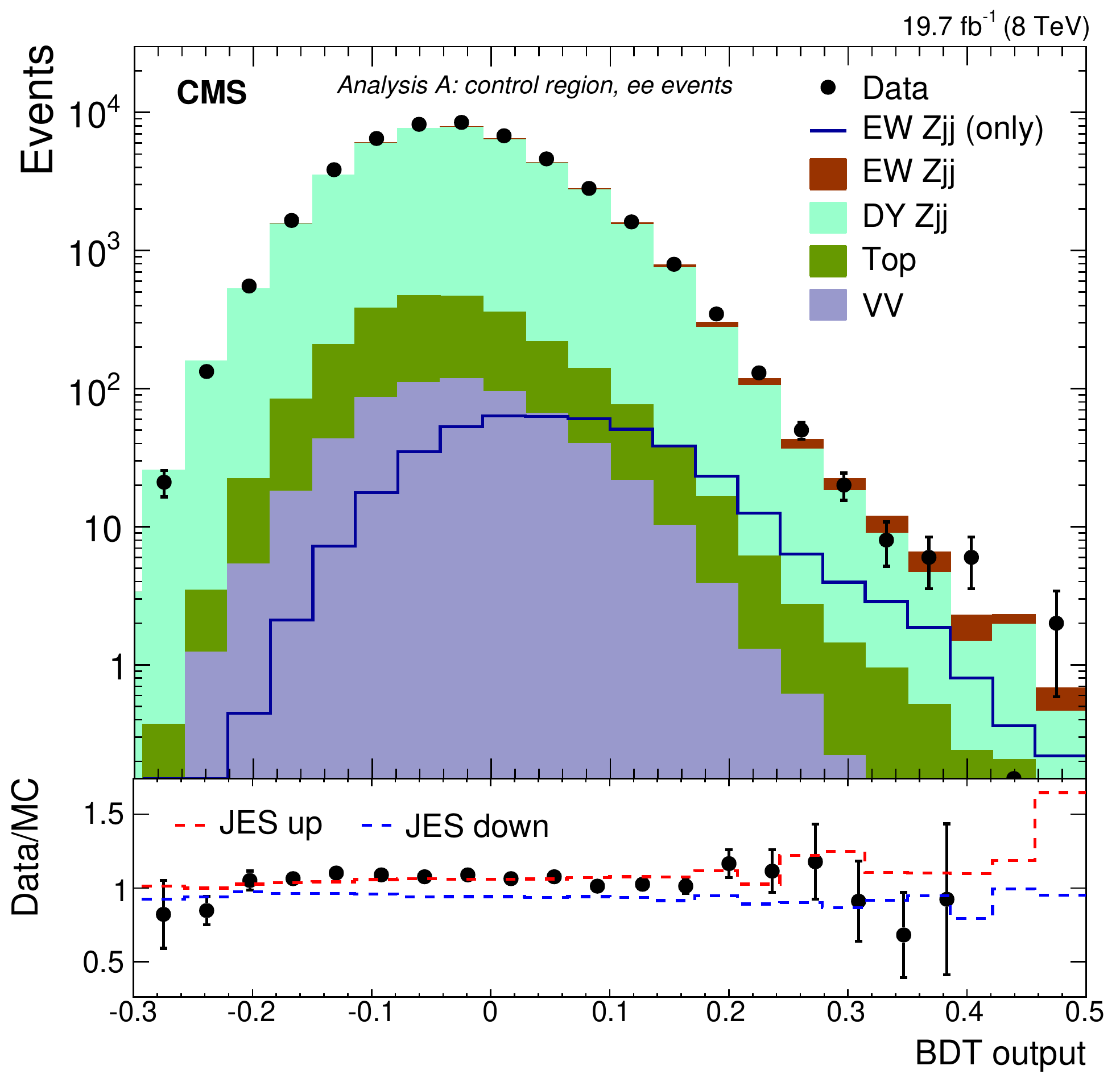}
\includegraphics[width=0.49\textwidth]{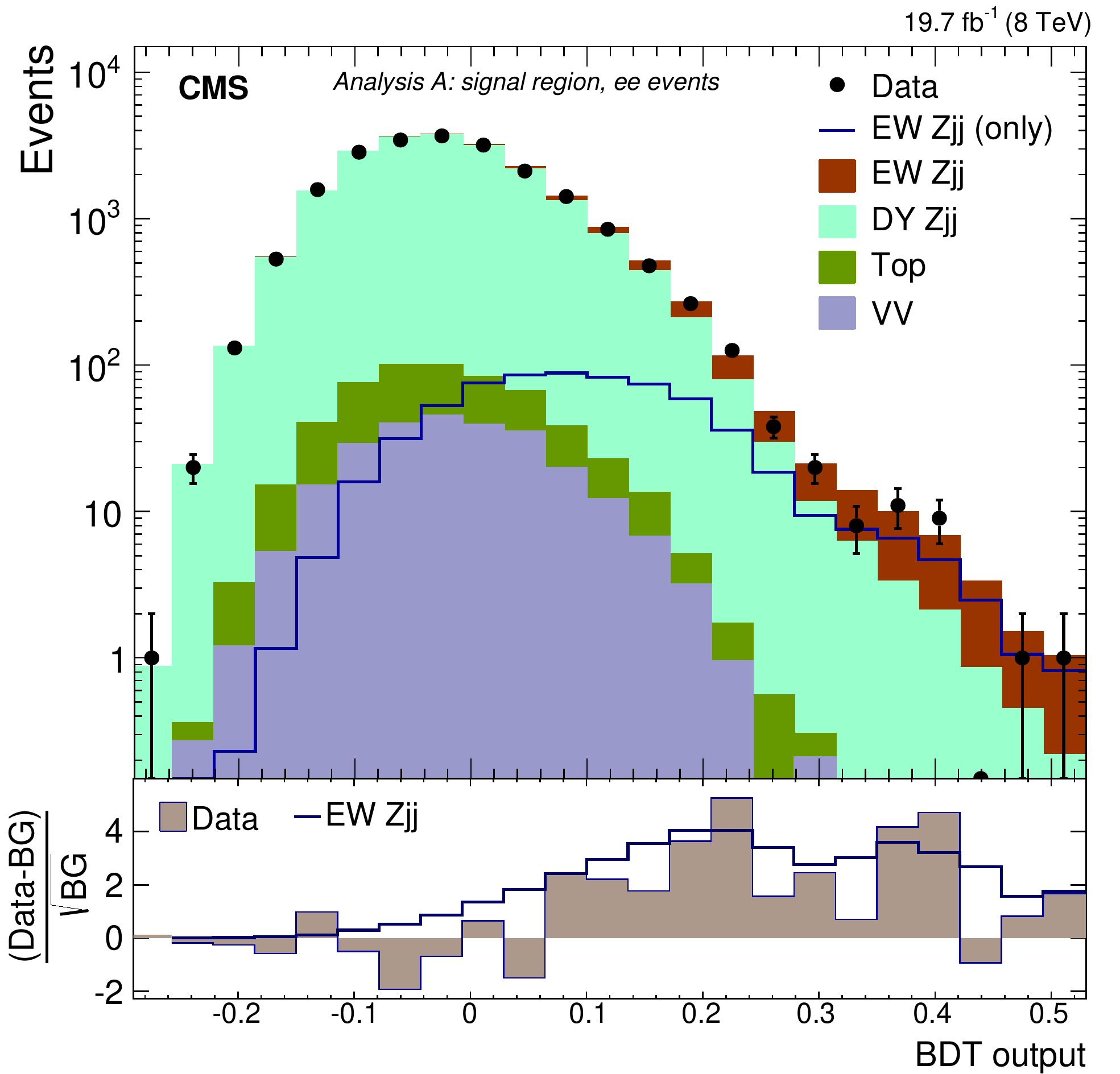}
\includegraphics[width=0.49\textwidth]{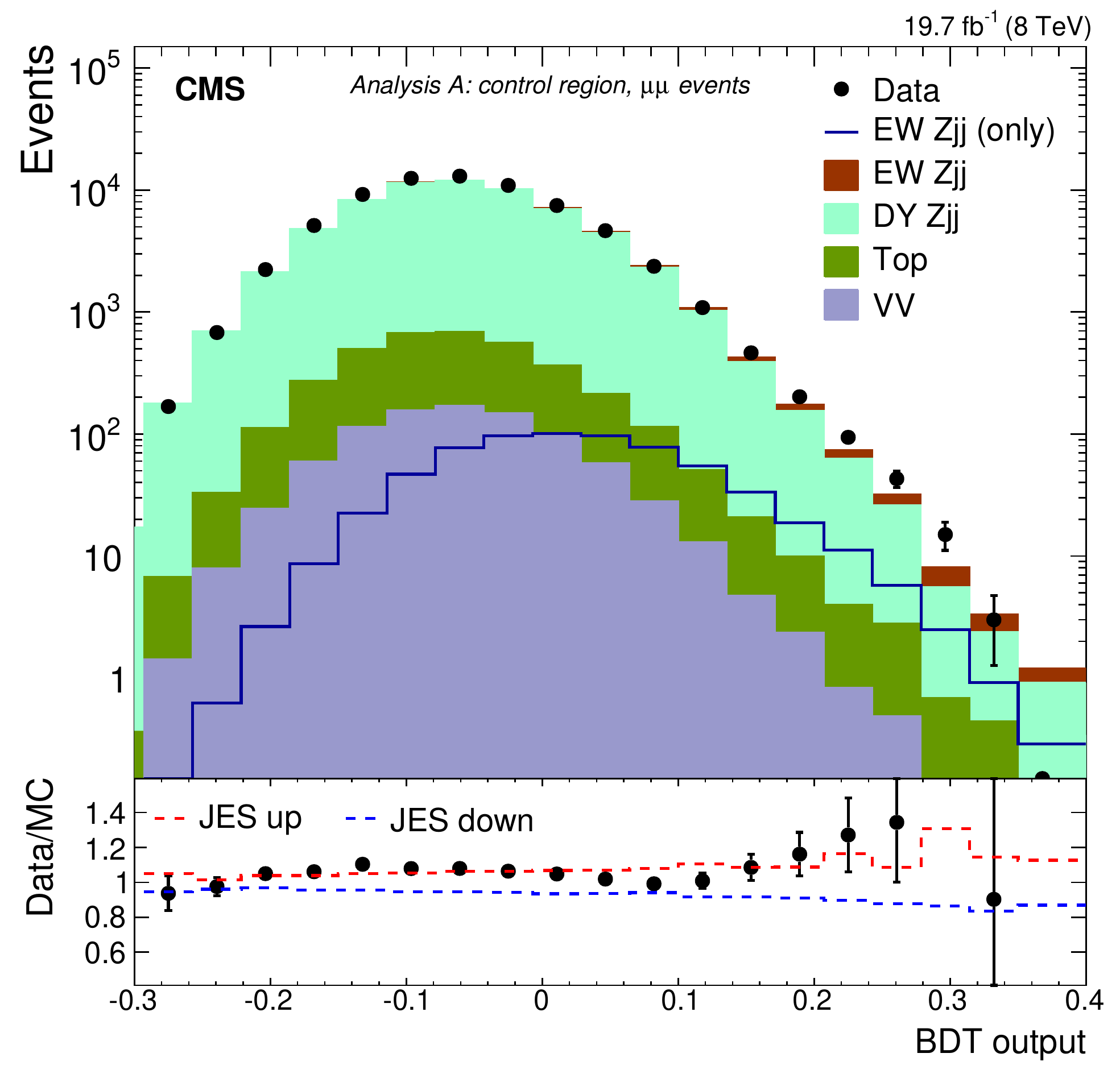}
\includegraphics[width=0.49\textwidth]{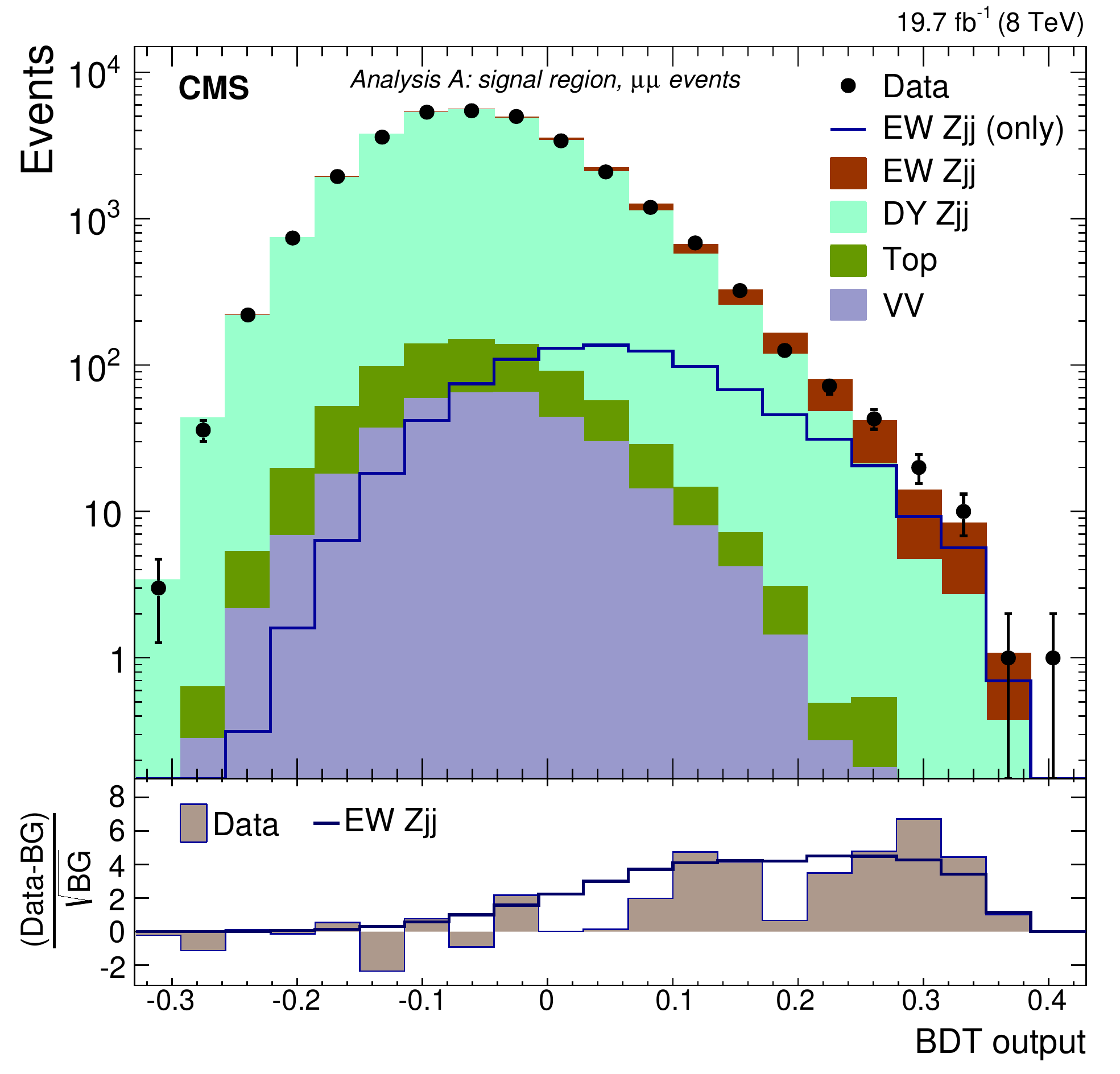}
\caption{
Distributions for the BDT discriminants in ee (top row) and $\mu\mu$
(bottom row) events, used by analysis A.
The distributions obtained in the control regions are shown at the
left while the ones obtained in the signal region are shown at the right.
The ratios for data to MC simulations are given in the bottom
panels in the left column, showing the impact of changes in JES by $\pm$1
SD.
The bottom panels of the right column show the differences between data or
the expected \ewkzjj\ contribution with respect to the background (BG).
}
\label{fig:discriminatorsa}
}
\end{figure*}

\begin{figure}[htb] {
\centering
\includegraphics[width=0.49\textwidth]{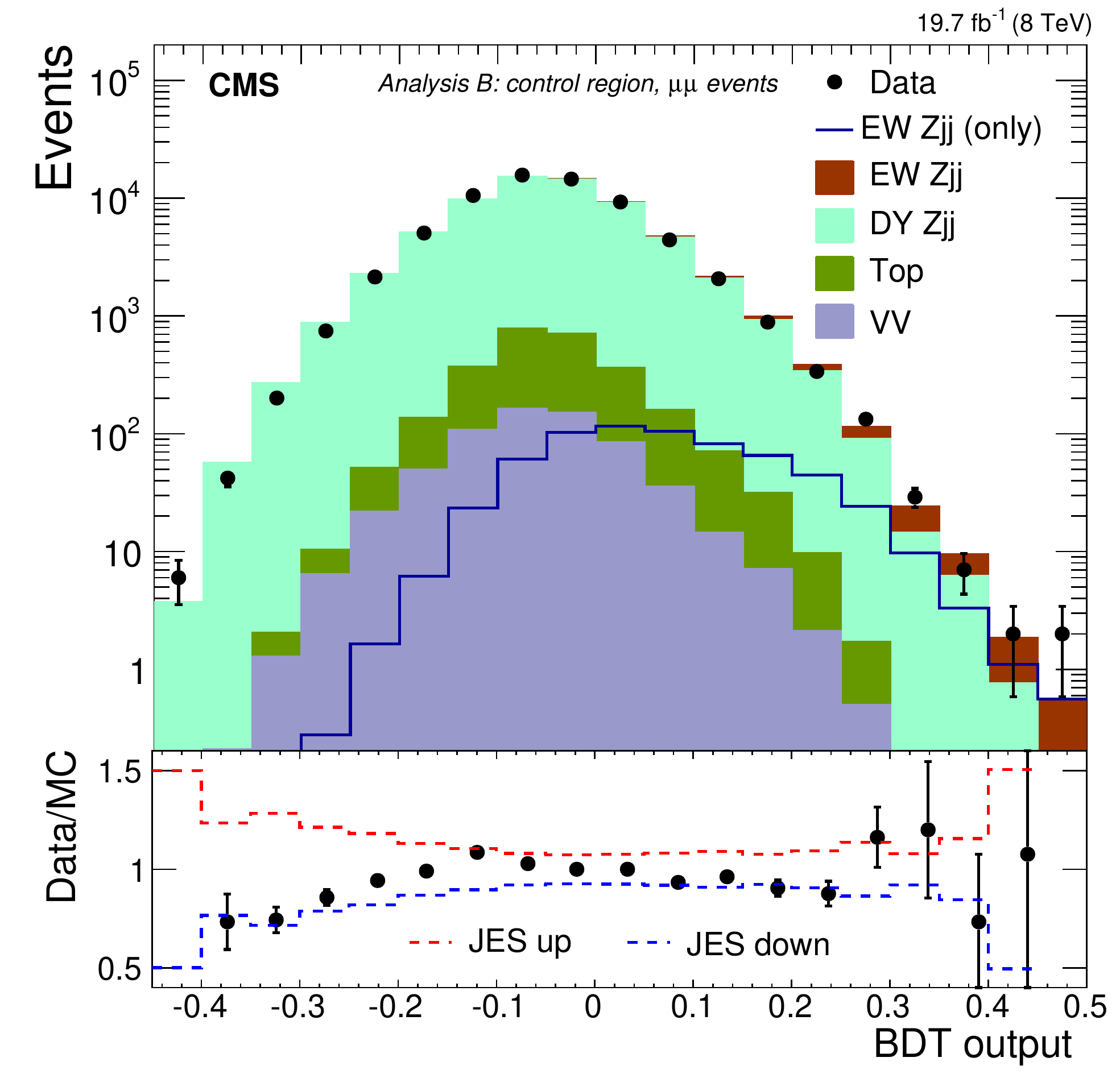}
\includegraphics[width=0.49\textwidth]{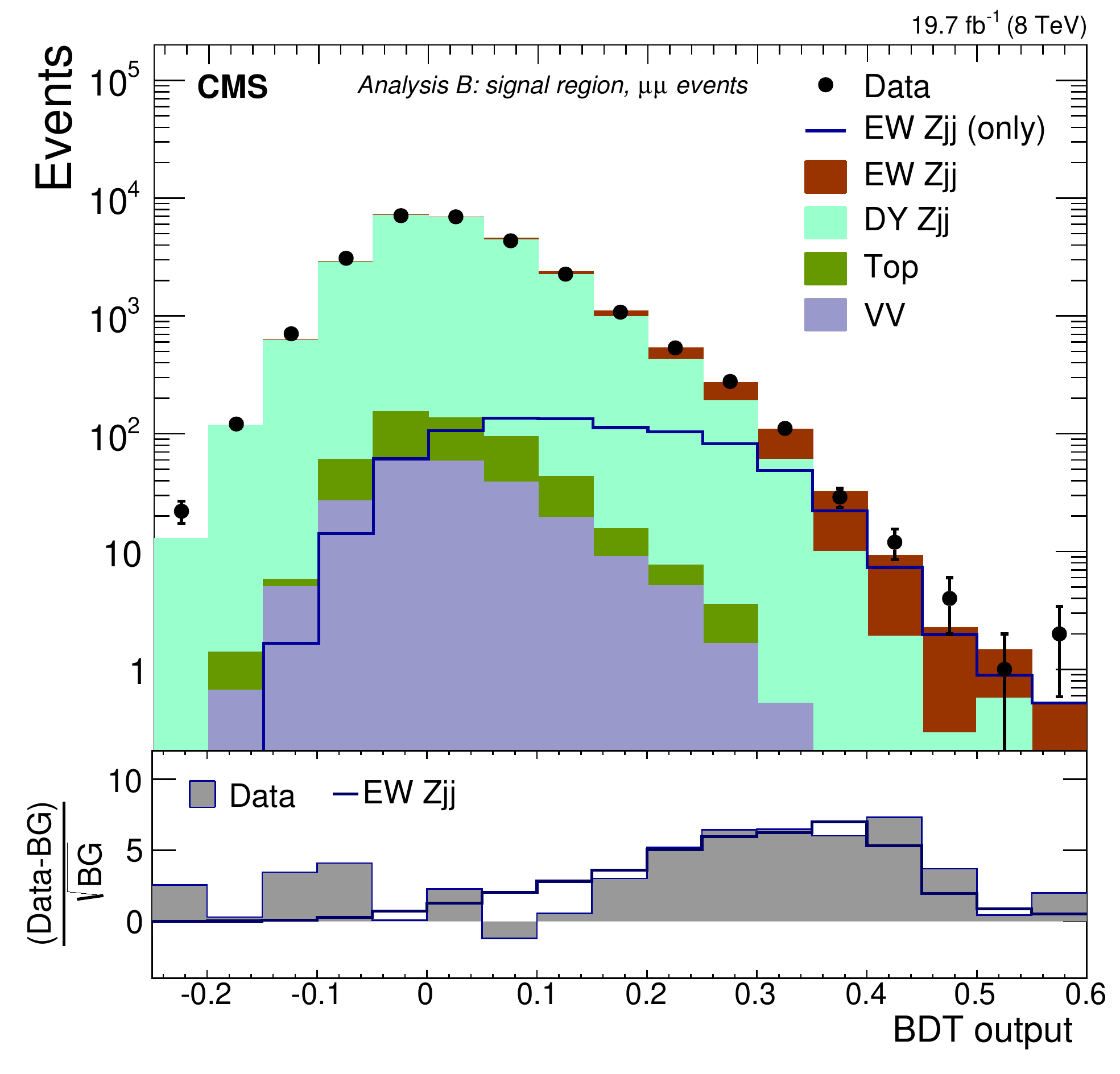}
\caption{
Distributions for the BDT discriminants in $\mu\mu$ events, for the
control region (\cmsLeft) and signal region (\cmsRight), used by analysis B.
The ratio for data to MC simulations is given in the bottom panel
on the left, showing the impact of changes in JES by $\pm$1
SD.
The bottom panel on the right shows the difference between data or
the expected \ewkzjj\ contribution with respect to the background (BG).
}
\label{fig:discriminatorsb}
}
\end{figure}

\begin{figure*}[htb] {
\centering
\includegraphics[width=0.4\textwidth]{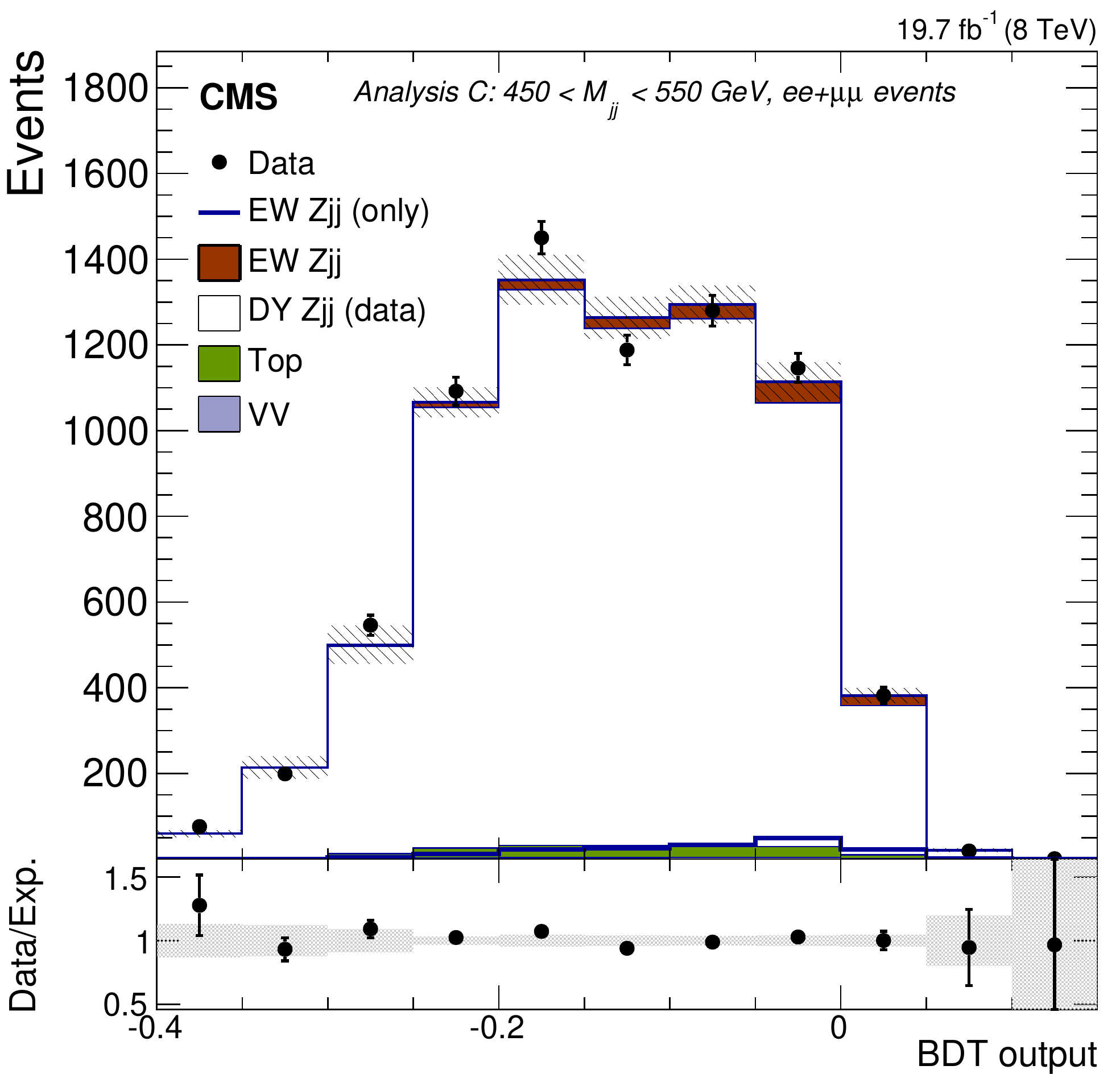}
\includegraphics[width=0.4\textwidth]{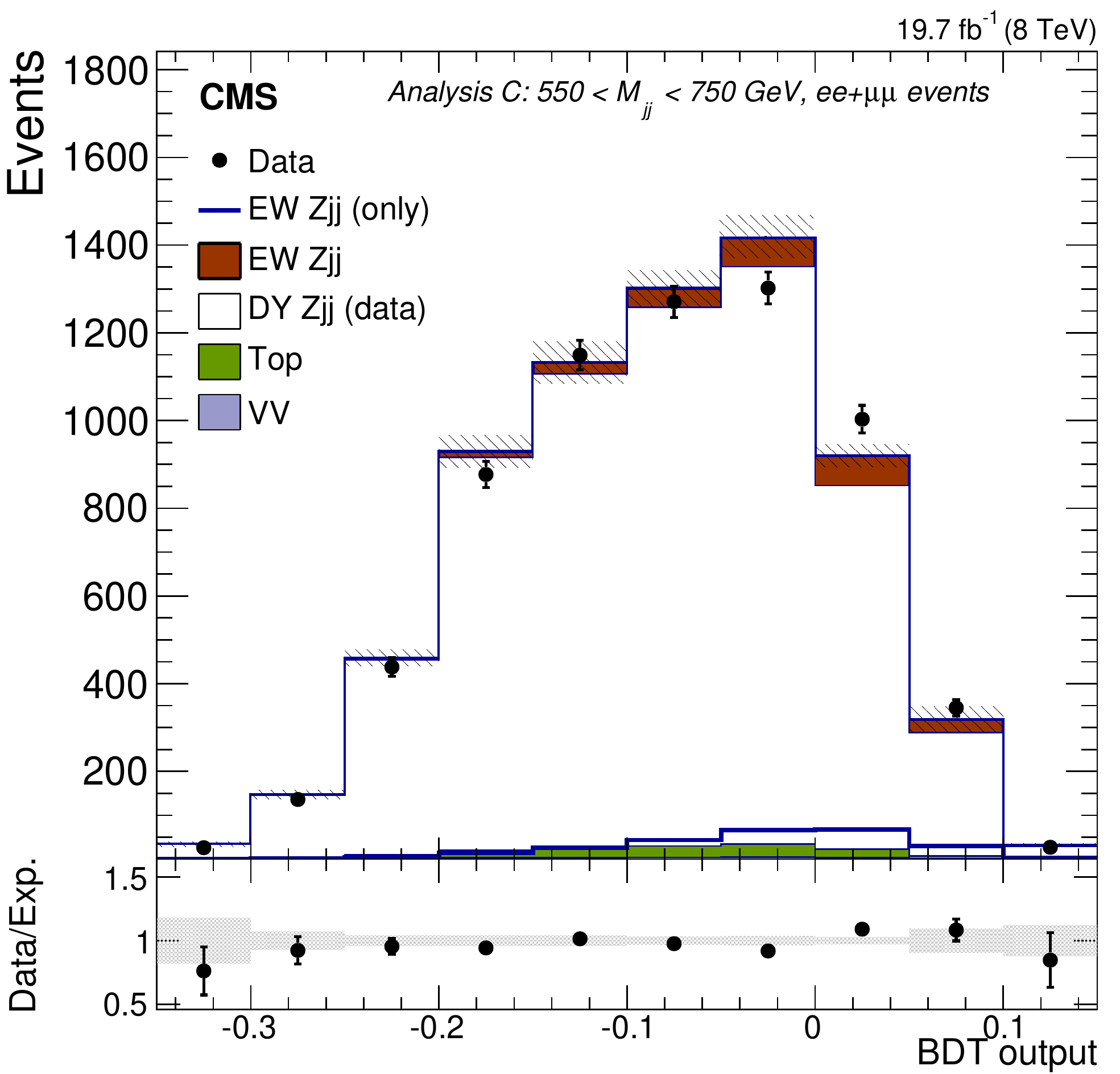}
\includegraphics[width=0.4\textwidth]{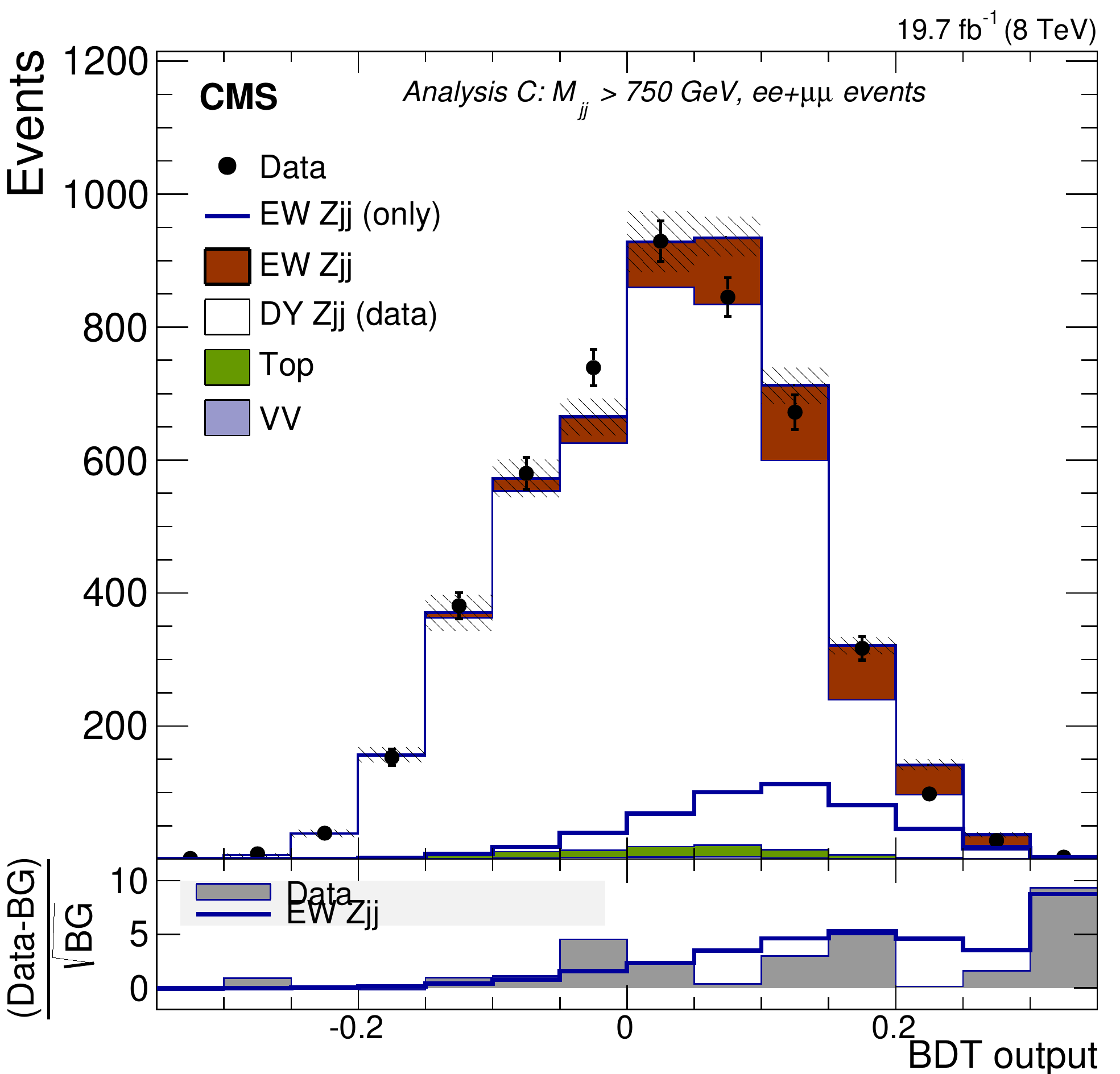}
\caption{
Distributions for the BDT discriminants in ee+$\mu\mu$ events for
different $M_\mathrm{jj}$ categories, used in analysis C.
The ratios at the bottom each subfigure of the top row gives the results of data to expectation
for the two control regions of $M_\mathrm{jj}$.
The lower panel of the bottom subfigure shows the difference between data or the
expected \ewkzjj\ contribution with respect to the background (BG).
}
\label{fig:discriminatorsc}
}
\end{figure*}

Figures~\ref{fig:discriminatorsa}, \ref{fig:discriminatorsb} and \ref{fig:discriminatorsc} show
the distributions of the discriminants for the three analyses.
Good agreement is observed overall in both the
signal and in the control regions which
are defined according to the value of the $R\pt^\text{hard}$ or $M_{\mathrm{jj}}$ variables (see Section~\ref{subsec:bckgmodel}).

{\tolerance=1500
Each analysis has a binned maximum likelihood formed from the expected
rates for each process, as function of the value of the discriminant,
which is used to fit simultaneously across the control and signal categories
the strength modifiers for the \ewkzjj\ and \dyzjj\ processes,
$\mu = \breakhere\sigma({\mathrm{EW}~\cPZ\mathrm{jj}}) / \sigma_\mathrm{LO}({\mathrm{EW}~\ell\ell\mathrm{jj}})$
and
$\upsilon = \sigma({\mathrm{DY}})/\sigma_\text{th}({\mathrm{DY}})$.
Nuisance parameters are added to modify the
expected rates and shapes according to the estimate of the systematic
uncertainties affecting the analysis and are mostly assumed to have a
log-normal distribution.
}

The interference between the \ewkzjj\ and
the \dyzjj\ processes is taken into account in the fitting procedure.
A parameterisation of the interference effects, as a function of
the parton-level $M_\mathrm{jj}$ variable, is derived from
the \MADGRAPH simulation described in Section~\ref{sec:simulation}.
The matrix elements for the \ewkzjj\ and \dyzjj\ processes provide
the total yields for the \lljj\ final state as
\begin{equation}
\hat{N}^{\ell\ell \mathrm{jj}}(\mu,\upsilon)= \mu N_{\mathrm{EW}~\cPZ \mathrm{jj}} + \sqrt{\mu\upsilon} N_\mathrm{I} + \upsilon N_{\mathrm{DY}~\cPZ \mathrm{jj}},
\label{eq:sigdef}
\end{equation}
where $N_{\mathrm{EW}~\cPZ\mathrm{jj}}$, $N_{\mathrm{DY}~\cPZ \mathrm{jj}}$ are the yields
for the \ewkzjj\ and \dyzjj\ processes, $N_\mathrm{I}$ is the
expected contribution from the interference to the total yield, and
$\mu$ and $\upsilon$ are the strength factors that modify the SM predictions.
In the absence of signal (or background) the contribution from the
interference term vanishes in Eq.~(\ref{eq:sigdef}).

The parameters of the model ($\mu$ and $\upsilon$) are determined
maximising a likelihood ($\mathcal{L}$).
Systematic uncertainties are incorporated in the fit
by scanning the profile likelihood ratio $\lambda$, defined as

\begin{equation}
\lambda(\mu,\nu) = \frac{\mathcal{L}(\mu,\nu,\hat{\hat\theta})}{\mathcal{L}(\hat{\mu},\hat{\nu},\hat\theta)},
\label{eq:profll}
\end{equation}

where the denominator has estimators $\hat{\mu}$,$\hat{\nu}$ and
$\hat\theta$ that maximise the likelihood, and
the numerator has estimators $\hat{\hat\theta}$ that maximise
the likelihood for the specified $\mu$ and $\nu$ strengths.
The statistical methodology used is similar to the one used in the CMS
Higgs analysis~\cite{Chatrchyan:2012ufa} using asymptotic formulas~\cite{Cowan:2010js}.
In this procedure some of the systematic uncertainties affecting the
measurement of the signal strength are partially constrained.
The \dyzjj\ strength is constrained by the uncertainties
in analyses A and B and is free to change in C.
In all cases the difference of the result relative to the one that
would have been obtained without taking the
interference term into account, is assigned as a systematic uncertainty
of the measurement.
This shall be discussed in more detail in the next section where the
systematic uncertainties affecting our analysis are summarised.

\section{Systematic uncertainties}
\label{sec:systunc}

The main systematic uncertainties affecting our measurement are classified into experimental and theoretical sources.

\subsection{Experimental uncertainties}
\label{subsec:expunc}

The following experimental uncertainties are considered:

\begin{description}

\item[Luminosity] --- A 2.6\% uncertainty is assigned to the value of the integrated luminosity~\cite{CMS-PAS-LUM-13-001}.

\item[Trigger and selection efficiencies] --- We assign total 2\%
  and 3\% uncertainties on the total trigger and selection efficiencies in
  the ee and $\mu\mu$ channels, respectively. These uncertainties have been
  estimated by comparing the lepton efficiencies expected in
  simulation and measured in data with
  a ``tag-and-probe'' method~\cite{inclusWZ3pb}.

\item[Jet energy scale and resolution] --- The energy of the jets enters
  in our analysis not only at the selection level but also in the
  computation of the kinematic variables used in forming discriminants.
  The uncertainty on JES affects therefore both the
  expected event yields, through the migration of events to different
  bins, and the final distributions.
  In addition to the standard JES uncertainty,
  the residual difference in the response
  observed in the balancing of a $\cPZ$ or $\gamma$
  candidate with a jet, discussed in Section~\ref{sec:controlreg}, is
  assigned as a systematic uncertainty.
  The effect of the JES uncertainty is studied by
rescaling up and down the reconstructed jet energy
  by a \pt- and $\eta$-dependent scale factor~\cite{Chatrchyan:2011ds}.
  An analogous approach is used for the JER.
  In both cases the uncertainties are derived separately of PF and JPT jets.

\item[q/g discriminator] --- The uncertainty on the performance of the
  q/g discriminator has been measured
using independent \cPZ+jet and dijet data,
after comparing with the corresponding simulation predictions~\cite{CMS-PAS-JME-13-002}.
The pa\-ram\-e\-tri\-za\-tion of the estimated uncertainty is used on an
event-per-event basis to derive alternative predictions
for the signal and background which are profiled in the fit for the signal.

\item[Pileup] --- Pileup is not expected to affect the identification and
  isolation of the leptons or the corrected energy of the jets. When
  the jet clustering algorithm is run, pileup can,
  however, induce a distortion of the reconstructed dijet system due to the contamination
  of tracks and calorimetric deposits. We evaluate this uncertainty by generating two
  alternative distributions after changing the number of pileup interactions by $\pm$5\%,
  according to the uncertainty on the inelastic pp cross section at $\sqrt{s}=8~\TeV$.

\item[Statistics of simulation] --- For signal and backgrounds which are estimated from
  simulation we form envelopes for the distributions by shifting all bin
  contents simultaneously up or down by its statistical uncertainty.
  This generates two alternatives to the nominal shape to be analysed.
  However, when a bin has an uncertainty which is $>10\%$,
  we assign an additional, independent uncertainty to it in the fit in
  order to avoid overconstraining a specific background from a single bin in the fit.

\end{description}

\subsection{Theoretical uncertainties}
\label{subsec:thunc}

We have considered the following theoretical uncertainties in the analysis:

\begin{description}

\item[PDF] --- The PDF uncertainties are evaluated by considering the \textsc{pdf4lhc} prescription~\cite{Alekhin:2011sk,Botje:2011sn,Ball:2010de,Pumplin:2002vw,Martin:2009iq},
  where for each source a new weight is extracted event-by-event and used to
  generate an alternative signal distribution. The up and down changes
  relative to the nominal prediction for each independent variable and are added in
  quadrature to estimate the final uncertainty.

\item[Factorisation and renormalisation scales] ---
  In contrast to the main background,
  the two signal process partons originate from
  electroweak vertices. Changing the QCD factorisation and
  renormalisation scales is therefore not
  expected to have a large impact on the final cross section.
  The renormalisation scale, in particular, is not expected to have
  any impact at LO.
  Changing the values of $\mu_F$ and $\mu_R$ from their defaults
  by 2 or 1/2
  we find a variation of $\approx 4\%$ in \MADGRAPH and in \textsc{vbfnlo}.
  As the change in the scales can also affect the expected
  kinematics, we use the
  altered $\mu_R/\mu_F$ samples to extract a weight that is applied
  at the generator level on an event-by-event basis. The parameterisation
  is done as function of the dilepton \pt. The changes induced
  in the form of the discriminant at the reconstruction level are assigned as
  systematic uncertainties.

\item[DY~Zjj prediction] --- For the modelling of the \dyzjj\
  background from simulation, as we indicated previously, we consider the full difference
  between the Born-level \MADGRAPH prediction and the NLO prediction
  based on \MCFM as a systematic uncertainty.
  The differences are particularly noticeable at very large $M_\mathrm{jj}$ and at large $y^*$.
  For the data-based modelling of \dyzjj\ we consider the effect
  induced on the discriminant functions from five distinct sources.
  Not all are of theoretical nature, nevertheless, we list them here for simplicity.
  We consider not only the statistical size of the photon sample
  but also the difference observed in data selected with a loose-photon selection relative
  to the data selected with a tight-photon selection.
  From simulation, the expected difference, between the
  tight-photon selection and a pure photon sample is also considered,
  and added in quadrature to the previous.
  Furthermore, we consider the envelope
  of the PDF changes induced in the simulated compatibility tests, and the
  contamination from residual \ewkgjj\ events in the photon sample.
  For the latter, we assign a 30\% uncertainty to the \ewkgjj\ contribution,
which is added in quadrature to the statistical uncertainty
  in the simulated events for this process.

\item[Normalisation of residual backgrounds] --- Diboson and top-quark
  processes are modelled with a MC simulation.
Thus, we assign an intrinsic uncertainty in their normalisation
according to their uncertainty which arises from the PDF and
factorisation/renormalisation scales.
The  uncertainties are assigned based on~\cite{Campbell:2010ff,Czakon:2013goa,Kidonakis:2012db}.

\item[Interference between \ewkzjj\ and \dyzjj\ ] --- The difference observed in the fit when the
interference term is neglected relative to the nominal result
is used to estimate the uncertainty due to the interference of
the signal and the background.

\end{description}

\subsection{Summary of systematic uncertainties}
\label{subsec:uncsummary}

Table~\ref{tab:systssummary} summarises the systematic uncertainties
described above. We give their magnitudes at the input level, and whether they are
treated as normalisation uncertainties or uncertainties in the distributions used to
fit the data.
The uncertainties are organised according to their experimental or
theoretical nature.

\begin{table*}[htbH]
\centering
\topcaption{
        Summary of the relative variation of uncertainty sources (in \%)
        considered for the evaluation of the systematic uncertainties in the different analyses.
        A filled or open circle signals whether that uncertainty affects the distribution or the absolute rate of a process in the fit, respectively.
        For some of the uncertainty sources ``variable'' is used to signal that the range is not unambiguously quantifiable by a range, as it depends
        on the value of the discriminants, event category and may also have a statistical component.
}
\label{tab:systssummary}
\begin{tabular}{llccc}
\hline
 \multicolumn{2}{c}{Source} & Shape & Methods A,B & Method C \\
\hline
\multirow{7}{*}{\rotatebox[origin=c]{90}{Experimental}}
& Luminosity & $\circ$ & \multicolumn{2}{c}{2.6}\\
& Trigger/selection & $\circ$ & \multicolumn{2}{c}{2--3}\\
& JES and residual jet response & \usedSign & \multicolumn{2}{c}{1--10}\\
& JER & \usedSign & \multicolumn{2}{c}{6--15}\\
& Pileup   & \usedSign & \multicolumn{2}{c}{6}\\
& Simulation statistics & \usedSign & \multicolumn{2}{c}{variable}\\
& \dyzjj\ distribution (data) & \usedSign & --- & variable\\\hline
\multirow{6}{*}{\rotatebox[origin=c]{90}{Theoretical}}
& PDF & \usedSign & \multicolumn{2}{c}{variable}\\
& $\mu_R/\mu_F$ (signal) & \usedSign & \multicolumn{2}{c}{variable} \\
& \dyzjj\ shape (MC) & \usedSign & variable & ---\\
& \dyzjj\ shape (PDF and \ewkgjj\ contribution)  & \usedSign & --- & variable \\
& Interference & \usedSign & \multicolumn{2}{c}{100} \\
& Normalisation of top-quark and diboson processes &$\circ$ & \multicolumn{2}{c}{7--10} \\
\hline
\end{tabular}
\end{table*}

\section{Measurement of the \texorpdfstring{\ewkzjj}{(EW) Zjj} production cross section}
\label{sec:results}

Table~\ref{tab:finalyieldsI} reports the expected and observed
event yields after imposing a minimum value for the discriminators
used in methods A and B such that ${S/B}>10\%$.
Table~\ref{tab:finalyieldsII} reports the event yields
obtained in each category for method C.
Fair agreement is observed between data and expectations for the sum of
signal and background, for both methods, in all categories.

\begin{table*}[htbp]
\centering
\topcaption{Event yields expected after fits to background and signal
  processes in methods A or B, using the initial selections
(summarised in Table~\ref{tab:methodcomp}),
and requiring ${S/B}>10\%$. The yields are
  compared to the data observed in the different channels and categories.
  The total uncertainties quoted for signal, \dyzjj,
dibosons (VV), and processes with top
  quarks (\ttbar and single top quarks)
are dominated by JES uncertainties
and include other sources, \eg, the statistical fluctuations in the MC samples . \label{tab:finalyieldsI}}
\reBox{
\begin{tabular}{lccccccc}\hline
 Selection & Channel & VV & Top quark & \dyzjj & Total backgrounds & \ewkzjj & Data\\
\hline
\multirow{3}{*}{Initial}
 &$ \Pe\Pe$ (A) & $255{\pm} 14$ & $314{\pm} 15$ & $20083{\pm} 857$ & \boldmath $20652{\pm} 857$ & $659{\pm} 16$ &\boldmath $20752$\\
 &$ \mu\mu$ (A) & $355{\pm} 15$ & $456{\pm} 16 $& $30042{\pm} 1230$ & \boldmath $30853 {\pm} 1230$ & $925{\pm} 22$ &\boldmath $30306$\\
 &$ \mu\mu$ (B) & $226{\pm} 13$ & $295{\pm} 12 $& $25505{\pm} 1735$ & \boldmath $26026 {\pm} 1735$ & $833{\pm} 14$ &\boldmath $26651$\\
\hline
 BDT$>$0.05 & $ \Pe\Pe$ (A) & $56{\pm} 6$ & $50{\pm}7$& $3541{\pm} 169 $ & \boldmath $3647{\pm} 169$ & $427{\pm}12$&\boldmath $3979$\\

 BDT$>$0.05 & $ \mu\mu$ (A) &$38{\pm} 5$ & $36{\pm}5$ & $2867{\pm} 135$ &\boldmath $ 2941 {\pm} 135$ & $459{\pm}14$&\boldmath $3182$\\

 BDT$>$0.1 & $ \mu\mu$ (B) &$36{\pm} 3$ & $35{\pm}3$ & $3871{\pm} 273$ &\boldmath $ 3942 {\pm} 273$ & $514{\pm}12$&\boldmath $4312$\\
\hline

\end{tabular}
}
\end{table*}

\begin{table*}[htbp]
\centering
\topcaption{Event yields expected before the fit to background and signal processes in method C.
  The yields are compared to the data observed in the different channels and categories.
  The total systematic uncertainty assigned to the normalisation of the processes is shown. \label{tab:finalyieldsII}}
\reBox{
\begin{tabular}{cccccccc}\hline
$M_\mathrm{jj}$ (\GeVns{}) & Channel & VV & Top quark & \dyzjj & Total backgrounds & \ewkzjj & Data\\
\hline
\multirow{2}{*}{450--550}
& $\Pe\Pe$ &$20{\pm}2$&$68{\pm}4$&$5438{\pm}731$&\boldmath $5526{\pm}731$&$94{\pm}6$&\boldmath $5809$\\
&$\mu\mu$&$27{\pm}2$&$96{\pm}4$&$7325{\pm}983$&\boldmath $7448{\pm}983$&$128{\pm}8$&\boldmath $8391$\\
\hline
\multirow{2}{*}{550--750}
& $\Pe\Pe$ &$16{\pm}1$&$56{\pm}3$&$3802{\pm}496$&\boldmath $3874{\pm}664$&$112{\pm}7$&\boldmath $4139$\\
&$\mu\mu$&$30{\pm}2$&$69{\pm}4$&$5234{\pm}683$&\boldmath $5333{\pm}896$&$155{\pm}10$&\boldmath $5652$\\
\hline
\multirow{2}{*}{750--1000}
& $\Pe\Pe$ &$5.4{\pm}0.5$&$20{\pm}2$&$1300{\pm}188$&\boldmath $1325{\pm}236$&$73{\pm}5$&\boldmath $1384$\\
&$\mu\mu$&$7.5{\pm}0.6$&$26{\pm}2$&$1846{\pm}262$&\boldmath $1880{\pm}313$&$98{\pm}6$&\boldmath $1927$\\
\hline
\multirow{2}{*}{$>$1000}
& $\Pe\Pe$ &$2.7{\pm}0.4$&$10.2{\pm}0.8$&$600{\pm}84$&\boldmath $613{\pm}90$&$84{\pm}6$&\boldmath $684$\\
&$\mu\mu$&$4.2{\pm}0.4$&$13{\pm}1$&$913{\pm}127$&\boldmath$930{\pm}122$&$114{\pm}8$&\boldmath $923$\\
\hline
\end{tabular}
}
\end{table*}

The signal strength is extracted from the fit to the discriminator
shapes as discussed in Section~\ref{sec:sigdisc}.
Table~\ref{tab:fitresults} summarises the results obtained for the fits
to the signal strengths in each method.
The results obtained are compatible among the dilepton
channels and different methods, and in agreement with the SM prediction of unity.
Methods A and B are dominated by the systematic uncertainty stemming
from the modelling of the \dyzjj\ background and the interference
with the \ewkzjj\ signal. Method C is dominated by the statistical
uncertainty in the fit and, due to tighter selection criteria, is expected
to be less affected by the modelling of the
interference. In method C, the \dyzjj\ modelling
uncertainty is partially due to the statistics of the photon sample.
With the exception of jet energy resolution, which has a larger impact
in method C due to its tighter $M_\mathrm{jj}$ selection, all other
uncertainties are of similar magnitude for the different methods.

\begin{table*}[htbp]
\centering
\topcaption{Fitted signal strengths in the different analyses and channels
including the statistical and systematic uncertainties.
For method C, only events with $M_\mathrm{jj}>450\GeV$ are used.
The breakup of the systematic components of the uncertainty is given in detail
in the listings. \label{tab:fitresults}}
\begin{tabular}{l|ccc|c|ccc}\hline
& \multicolumn{3}{c|}{Analysis A} & Analysis B & \multicolumn{3}{c}{Analysis C} \\
& $\Pe\Pe$ & $\mu\mu$ & $\Pe\Pe+\mu\mu$ & $\mu\mu$ & $\Pe\Pe$  & $\mu\mu$  & $\Pe\Pe+\mu\mu$ \\
\hline
~~~Luminosity                            & 0.03 & 0.03 & 0.03 & 0.03 & 0.03 & 0.03 & 0.03\\
~~~Trigger/lepton selection        & 0.04 & 0.04 & 0.04 & 0.04 & 0.04 & 0.04 & 0.04\\
~~~JES+residual response            & 0.06 & 0.05 & 0.05 & 0.04 & 0.06 & 0.05 & 0.05\\
~~~JER                               & 0.02 & 0.02 & 0.02 & 0.02 & 0.04 & 0.04 & 0.03\\
~~~Pileup                           & 0.01 & 0.02 & 0.02 & 0.01 & 0.01 & 0.01 & 0.01\\
~~~\dyzjj                            & 0.07 & 0.05 & 0.07 & 0.08 & 0.14 & 0.12 & 0.13\\
~~~q/g discriminator                    & $<$0.01 & $<$0.01& $<$0.01 & --- & $<$0.01 &$<$0.01 & $<$0.01\\
~~~Top, dibosons                         & 0.01 & 0.01 & 0.01 & 0.01 & $<$0.01 & $<$0.01 & $<$0.01\\
~~~Signal acceptance                   & 0.03 & 0.04 & 0.04 & 0.04 & 0.06 & 0.06 & 0.06\\
~~~DY/EW \cPZ jj interference           & 0.14 & 0.14 & 0.14 & 0.13 & 0.06 & 0.08 & 0.08\\\hline
Systematic uncertainty                &0.19 & 0.18 & 0.19 & 0.17 & 0.17 & 0.17 & 0.18\\\hline
Statistical uncertainty                 & 0.11 & 0.10 & 0.07 & 0.09 & 0.24 & 0.21 & 0.16 \\\hline
$\mu=\sigma/\sigma_\text{th}$   & 0.82 & 0.86 & 0.84 & 0.89 & 0.91 & 0.85 & 0.88 \\\hline
\end{tabular}
\end{table*}

\begin{figure}[htp] {
\centering
\includegraphics[width=0.49\textwidth]{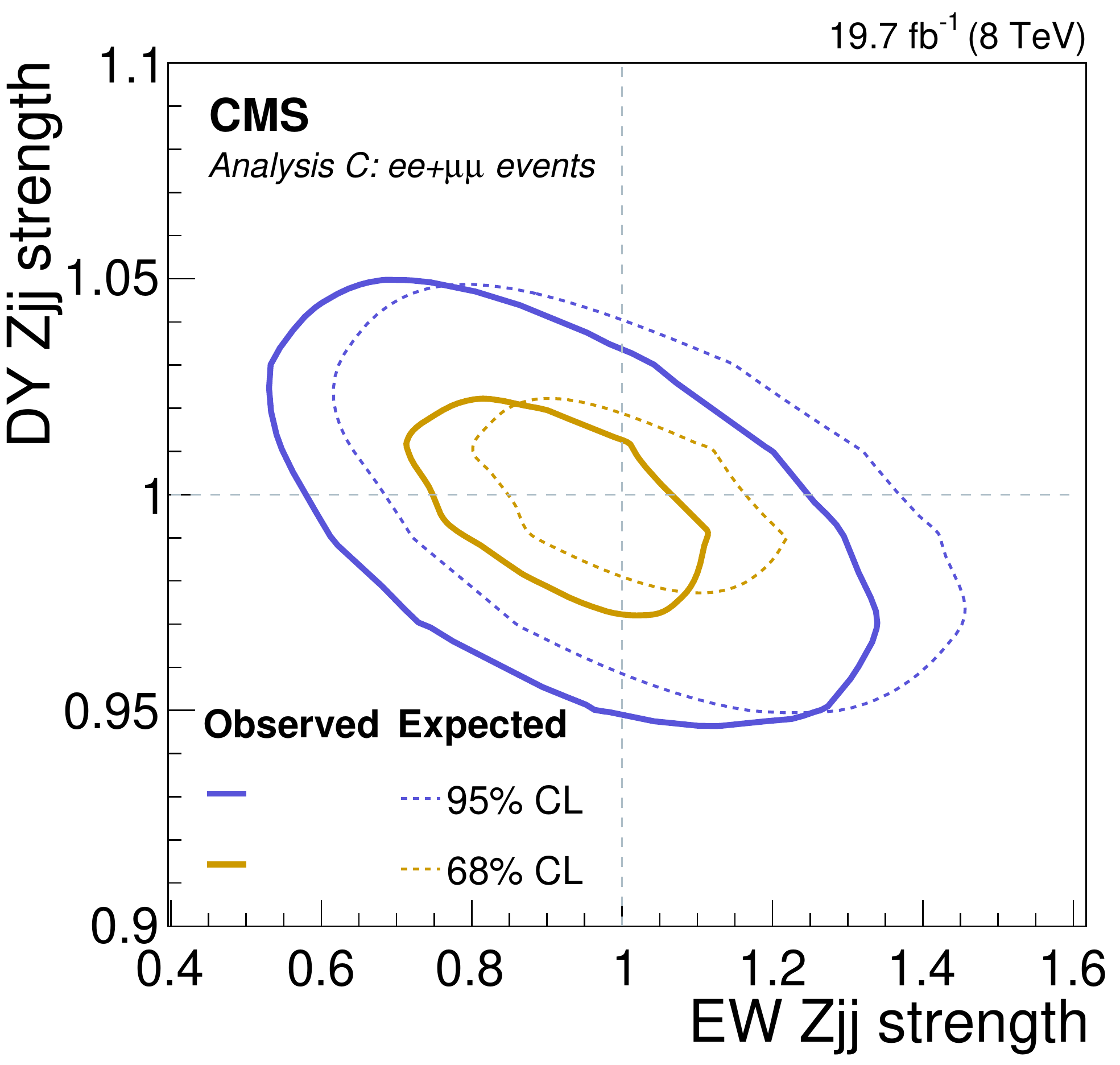}
\caption{
Expected and observed contours for the 68\% and 95\% CL intervals on
the \ewkzjj\ and DY signal strengths, obtained with method C after
combination of the ee and $\mu\mu$ channels.
}
\label{fig:munucont}
}
\end{figure}

For the results from method C, the 68\% and 95\% confidence levels (CL)
obtained for the combined fit of the \ewkzjj\ and \dyzjj\ strengths
are shown in Fig.~\ref{fig:munucont}.
Good agreement is found with the SM prediction for both components,
as well as with the expected magnitude of the CL intervals.
The \dyzjj\ strength is measured to be
$0.978\pm0.013\stat\pm0.036\syst$ in the ee channel,
$1.016\pm0.011\stat\pm0.034\syst$ in the $\mu\mu$ channel,
and $0.996\pm0.008\stat\pm0.025\syst$ after the combination of the
previous two.

From the combined fit of the two channels in analysis A we obtain the signal strength
\begin{equation*}
\mu=0.84\pm 0.07\stat \pm0.19\syst=0.84\pm0.20\,\text{(total)},
\end{equation*}
corresponding to a measured signal cross section
\ifthenelse{\boolean{cms@external}}{
\begin{multline*}
\sigma({\mathrm{EW}~\ell\ell\mathrm{jj}})=\\
174\pm 15\stat \pm 40\syst\unit{fb}=174\pm42\,\text{(total)}\unit{fb},
\end{multline*}
}{
\begin{equation*}
\sigma({\mathrm{EW}~\ell\ell\mathrm{jj}})=174\pm 15\stat \pm 40\syst\unit{fb}=174\pm42\,\text{(total)}\unit{fb},
\end{equation*}
}
in agreement with the SM prediction
$\sigma_\mathrm{LO}(\mathrm{EW}\,\ell\ell\mathrm{jj})=208\pm 18\unit{fb}$.
Using the same statistical methodology, as described in Section~\ref{sec:sigdisc},
the background-only hypothesis is excluded with a significance
greater than 5$\sigma$.

\section{Study of the hadronic and jet activity in \cPZ+jet events}
\label{sec:hadactivity}

After establishing the signal, we examine the properties of the hadronic
activity in the selected events.
Radiation patterns and
the profile of the charged hadronic activity as a function of several
kinematic variables
are explored in a region dominated by the main background, \dyzjj;
these studies are presented in Sections~\ref{subsec:jetrad} and~\ref{subsec:soft}.
The production of additional jets in a region with a
larger contribution of \ewkzjj\ processes
is furthermore pursued in Section~\ref{subsec:highpur}.
We expect a significant suppression of the hadronic activity
in signal events because the final-state objects have origin in
purely electroweak interactions, in contrast with the radiative QCD
production of jets in \dyzjj\ events.
The reconstructed distributions are compared directly to the
prediction obtained with a full simulation of the CMS detector
(see Section~\ref{sec:simulation}) and extends the studies reported in~\cite{Chatrchyan:2011ne}
to the phase space region of interest for the study of the \ewkzjj\ process.

\subsection{Jet radiation patterns}
\label{subsec:jetrad}

For the \cPZ+jets events,
the observables referred to as ``radiation patterns'' correspond to:
(i) the number of jets, $N_\mathrm{j}$,
(ii) the total scalar sum of the transverse momenta of jets
reconstructed within $\abs{\eta}<4.7$,  $\HT$,
(iii) $\Delta\eta_\mathrm{jj}$ between the two jets with $\pt>40\GeV$
    which span the largest pseudorapidity gap in the event
   (not required to be the two leading-\pt jets), and
(iv) the cosine of the azimuthal angle difference, $\cos \abs{\phi_{\mathrm{j}_1} - \phi_{\mathrm{j}_2}} = \cos \Delta\phi_\mathrm{jj}$,
     for the two jets with criterion (iii).
These observables are measured using
events that are required to satisfy the $\cPZ\to\mu\mu$ and
$\cPZ\to \Pe\Pe$ selection criteria of analyses A and B.
These observables are investigated following the prescriptions and
suggestions from Ref.~\cite{Binoth:2010ra},
where the model dependence is estimated by comparing
different generators.

\begin{figure*}[hbtp]
\centering
\includegraphics[width=0.45\textwidth]{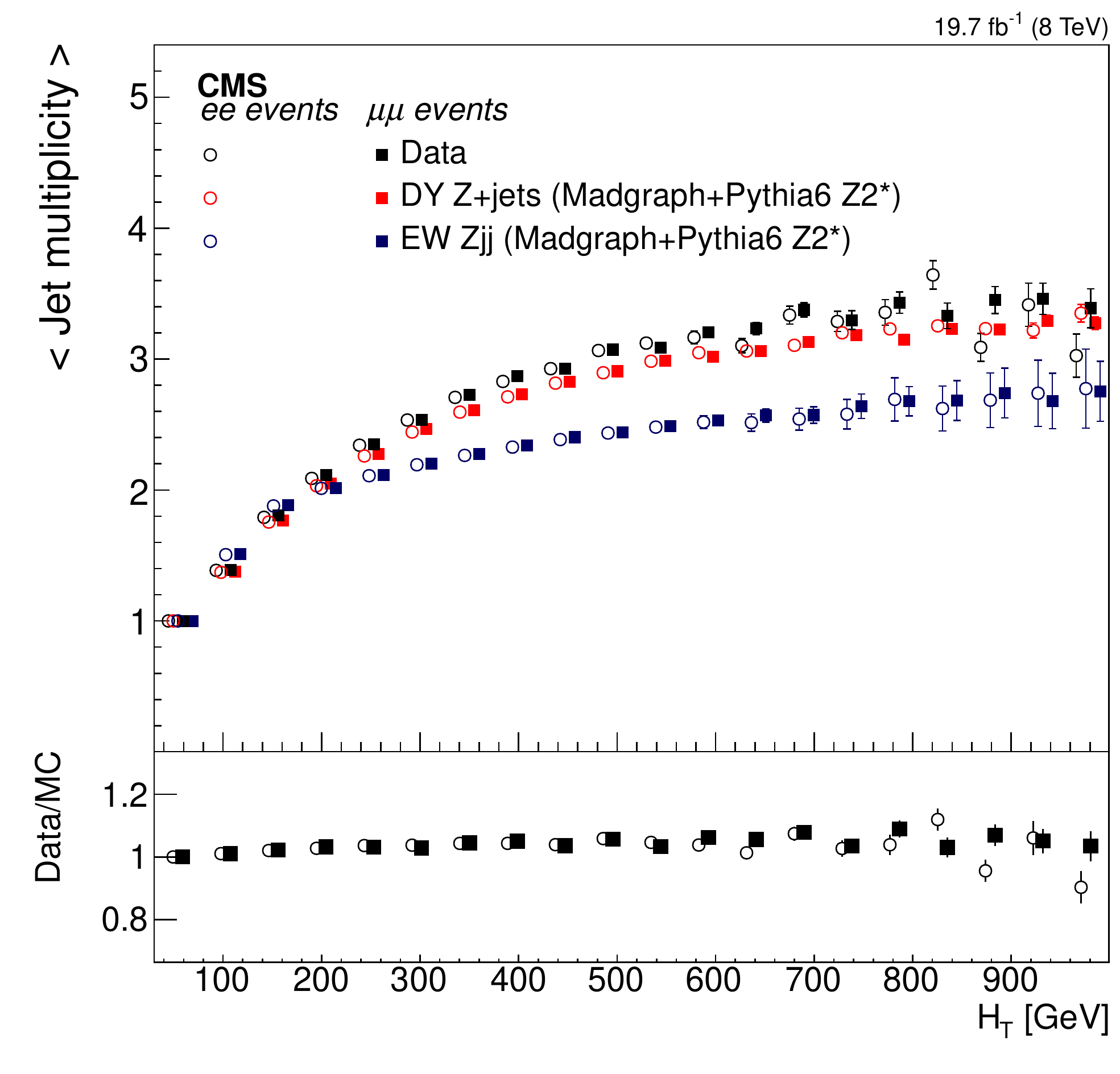}
\includegraphics[width=0.45\textwidth]{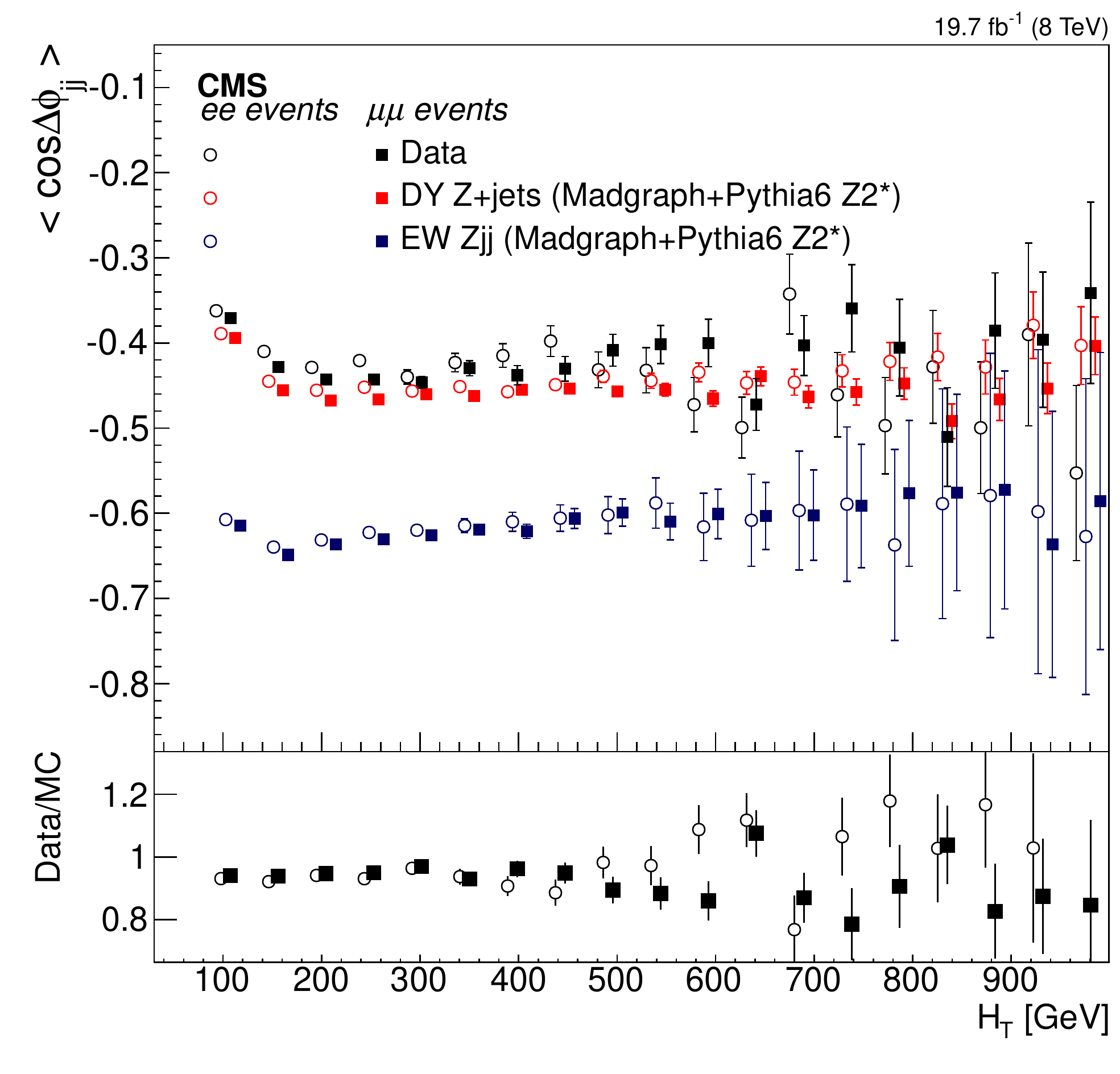}
\caption{(left) The average number of jets with $\pt > 40\GeV$ as a function of the total $\HT$
in events containing a \cPZ\ and at least one jet,
and (right) average $\cos\Delta\phi_\mathrm{jj}$ as a function of the total $\HT$
in events containing a \cPZ\ and at least two jets.
The ratios of data to expectation are given below the main panels.
At each ordinate, the entries are separated for clarity.
The expectations for \ewkzjj\ are shown separately.
The data and simulation points are shown with their statistical uncertainties.}
\label{fig:Ht_vs}
\end{figure*}

\begin{figure*}[hbtp]
\centering
\includegraphics[width=0.45\textwidth]{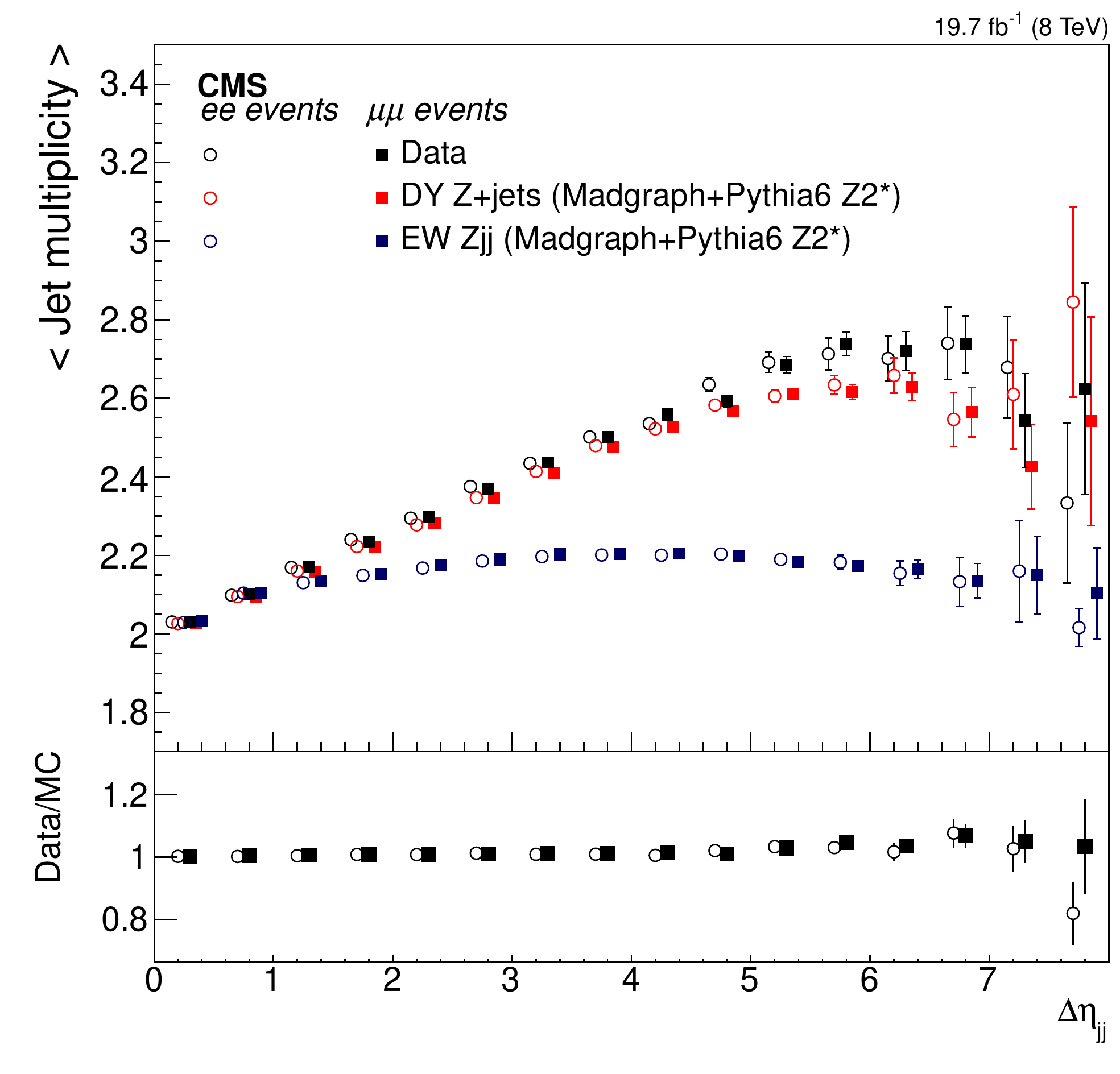}
\includegraphics[width=0.45\textwidth]{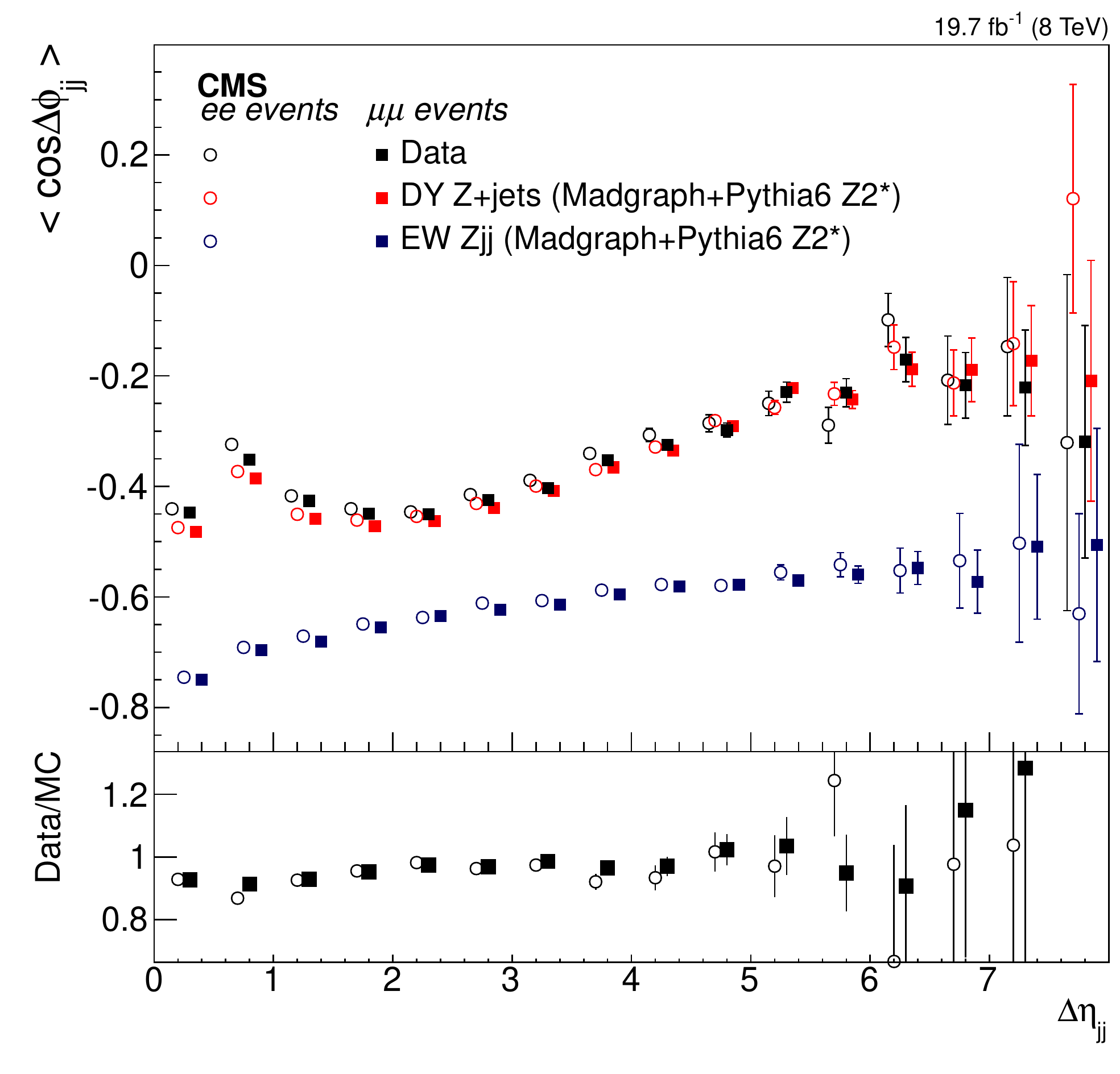}
\caption{(left) The average number of jets with $\pt > 40\GeV$ as a function of
the pseudorapidity distance between the dijet with
largest $\Delta\eta$, and
(right) average $\cos\Delta\phi_\mathrm{jj}$ as a
function of $\Delta\eta_\mathrm{jj}$ between the dijet with largest $\Delta\eta$.
In both cases events containing a \cPZ\ and at least two jets are used.
The ratios of data to expectation are given below the main panels.
At each ordinate, the entries are separated for clarity.
The expectations for \ewkzjj\ are shown separately.
The data and simulation points are shown with their statistical
uncertainties.}
\label{fig:deltaeta_vs}
\end{figure*}

Figures~\ref{fig:Ht_vs} and~\ref{fig:deltaeta_vs}
show  the average number of jets and the average $\cos\Delta\phi_\mathrm{jj}$
as a function of the total $\HT$ and $\Delta\eta_\mathrm{jj}$. The \MADGRAPH + \PYTHIA (ME-PS)
predictions are in good agreement with the data, even in the regions
of largest $\HT$ and $\Delta\eta_\mathrm{jj}$.
In both cases we estimate that the contribution from \ewkzjj\ is $<1\%$.
Jet multiplicity increases both as function of $\HT$ and $\Delta\eta_\mathrm{jj}$.
The increase of $\HT$ and $\Delta\eta_\mathrm{jj}$ induces, in average,
an increase of jet multiplicity and leads to different dijet
configurations in the azimuthal plane.
In average the two selected jets are separated by $120^0\deg$,
independently of $\HT$.
This separation tends to decrease for larger $\Delta\eta_\mathrm{jj}$ separation.
The behavior observed for $\cos\Delta\phi_\mathrm{jj}$ when
$\Delta\eta_\mathrm{jj}<0.5$
is related to the jet distance parameter used in the reconstruction (R=0.5).
In data, the separation of the jets in the $\cos\Delta\phi_\mathrm{jj}$
variable, is observed to be $<$5\% smaller with respect to the simulation.

\subsection{Study of the charged hadronic activity}
\label{subsec:soft}

For this study, a collection  is formed
of high-purity tracks \cite{CMS-PAS-TRK-10-005} with $\pt > 0.3\GeV$,
uniquely associated with the main PV in the event.
Tracks associated with the two leptons or with the tagging jets are
excluded from the selection.
The association between the selected tracks and the reconstructed PVs
is carried out by minimising the longitudinal
impact parameter which is defined
as the $z$-distance
between the PV and the point of closest approach of
the track helix to the PV, labeled $d_z^\mathrm{PV}$.
The association is required to satisfy the conditions $d_z^\mathrm{PV}<2\unit{mm}$ and
 $d_z^\mathrm{PV}<3\delta d_z^\mathrm{PV}$, where $\delta d_z^\mathrm{PV}$ is the uncertainty on $d_z^\mathrm{PV}$.

A collection of ``soft track-jets'' is defined
by clustering the selected tracks using the anti-\kt clustering algorithm~\cite{Cacciari:2008gp}
with a distance parameter of $R=0.5$. The use of track jets represents a
clean and well-understood
method~\cite{CMS-PAS-JME-10-006} to reconstruct jets with energy
as low as a few \GeVns{}.
These jets are not affected by pileup, because of the association of
their tracks with the hard-scattering vertex~\cite{CMS-PAS-JME-08-001}.

To study the central hadronic activity between the tagging jets,
only track jets of low \pt, and within
$\eta^\text{tag jet}_\text{min}+0.5 < \eta < \eta^\text{tag jet}_\text{max}-0.5 $ are
considered. For each event, we compute the scalar sum of the \pt of up to
three leading-\pt soft-track jets, and define it as
the soft $\HT$ variable.
This variable  is chosen to monitor
the hadronic activity in the rapidity interval between the two jets.

\begin{figure}[hbtp]
\centering
\includegraphics[width=0.45\textwidth]{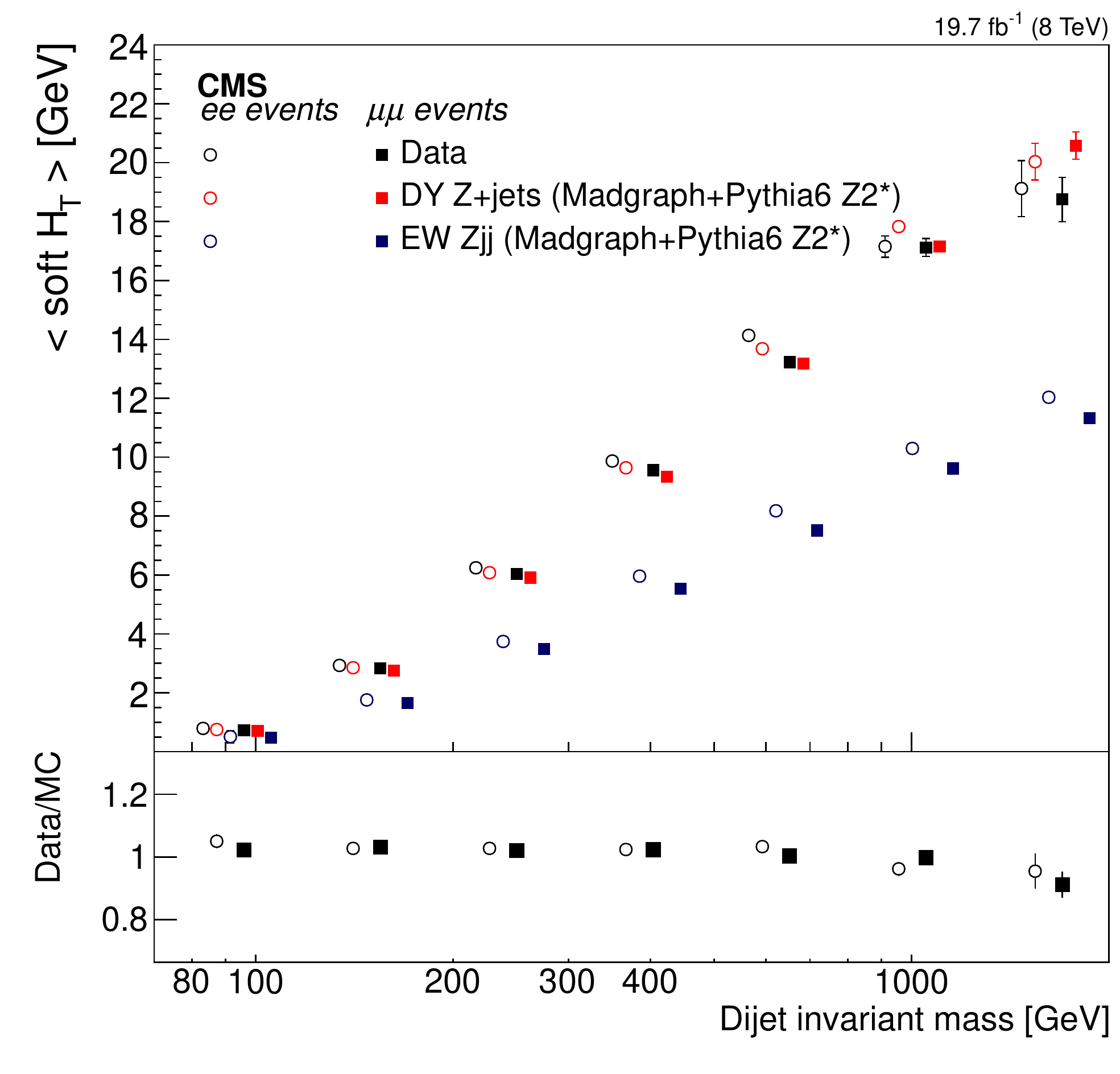}
\includegraphics[width=0.45\textwidth]{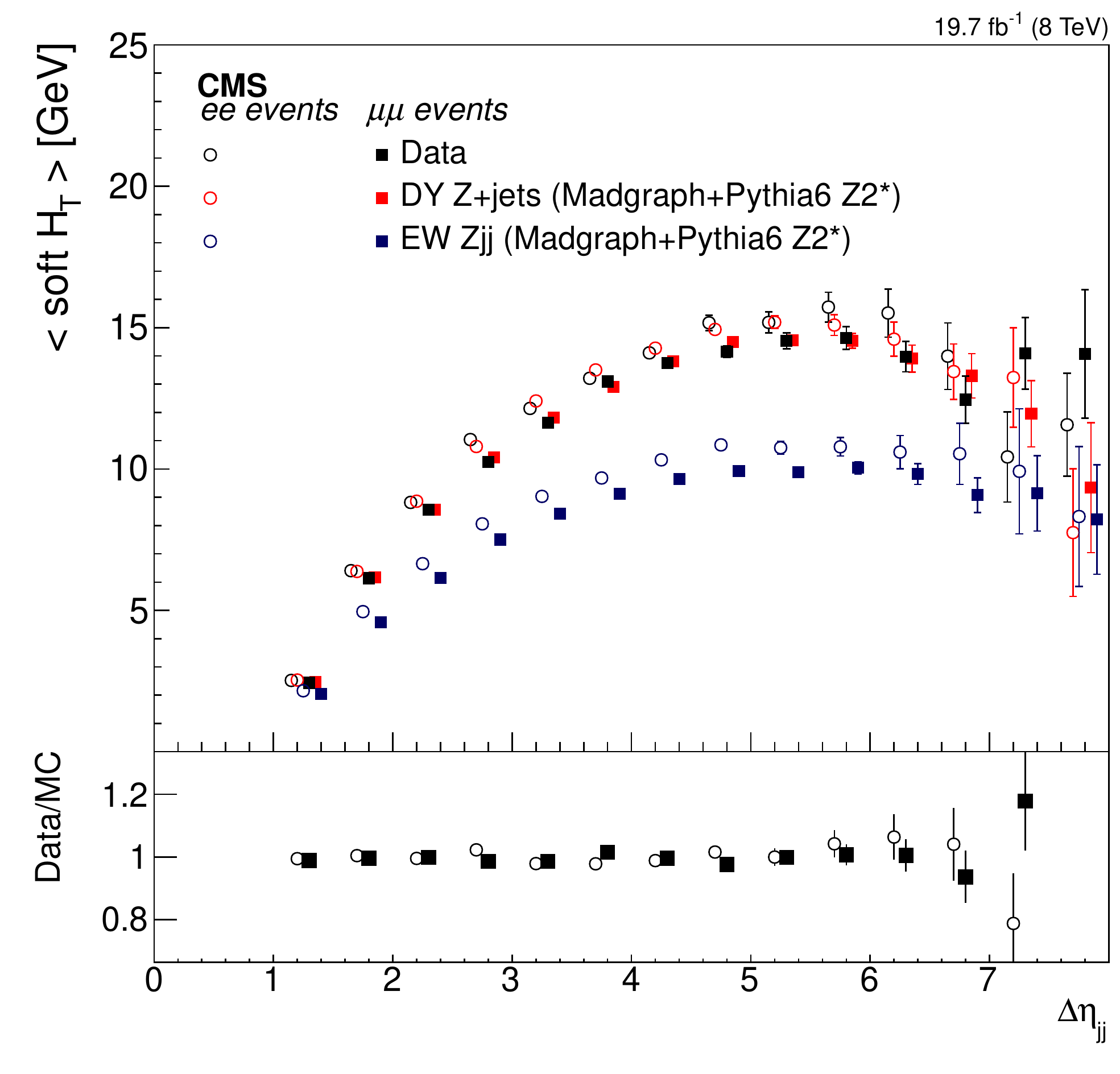}
\caption{Average soft $\HT$ computed using the three leading
soft-track jets reconstructed in the $\Delta\eta_\mathrm{jj}$
pseudorapidity interval between the tagging jets
that have $\pt>50\GeV$ and $\pt>30\GeV$.
The average soft $\HT$ is shown as function of:
(\cmsLeft) $M_\mathrm{jj}$ and (\cmsRight) $\Delta\eta_\mathrm{jj}$
for both the dielectron and dimuon channels.
The ratios of data to expectation are given below the main panels.
At each ordinate, the entries are separated for clarity.
The expectations for \ewkzjj\ are shown separately.
The data and simulation points are shown with their statistical
uncertainties.
\label{fig:softHT_vs}}
\end{figure}

The dependence of the average soft $\HT$ for the \Zjj\ events
as a function of $M_\mathrm{jj}$ and $\Delta\eta_\mathrm{jj}$ is shown in Fig.~\ref{fig:softHT_vs}.
Inclusively, the contribution from \ewkzjj\ is estimated to be at the level of 1\%,
but it is expected to evolve as function of the different variables, being 5\% (20\%)
for $\vert\Delta\eta_\mathrm{jj}\vert>4$ ($M_\mathrm{jj}>1\TeV$).
Overall, good agreement is observed between data and the simulation.
The average value of the soft $\HT$ is observed to increase linearly
with $M_\mathrm{jj}$, and to saturate its value for $\Delta\eta_\mathrm{jj}>5$, as a
consequence of the limited acceptance of the CMS tracker.

\subsection{Jet activity studies in a high-purity region}
\label{subsec:highpur}

The evidence for EW production of $\ell\ell \mathrm{jj}$ final states
can also be supported through a study of the emission of a third and
other extra jets in a region of high signal purity, \ie for large $M_{jj}$.
In this study, we compare two regions, one with $M_\mathrm{jj}>750\GeV$
and another with $M_\mathrm{jj}>1250\GeV$.
Aside from the two tagging jets used in the preselection, we use all PF-based jets with a
$\pt>15\GeV$ found within the $\Delta\eta_\mathrm{jj}$ of the tagging jets.
The background is modelled from the photon control sample (analysis C), and
uses the normalisations obtained from the fit discussed in
Section~\ref{sec:results}.
Where relevant we also compare the results using the MC-based modelling
of the background.

\begin{figure*}[htp]
\centering
\includegraphics[width=0.49\textwidth]{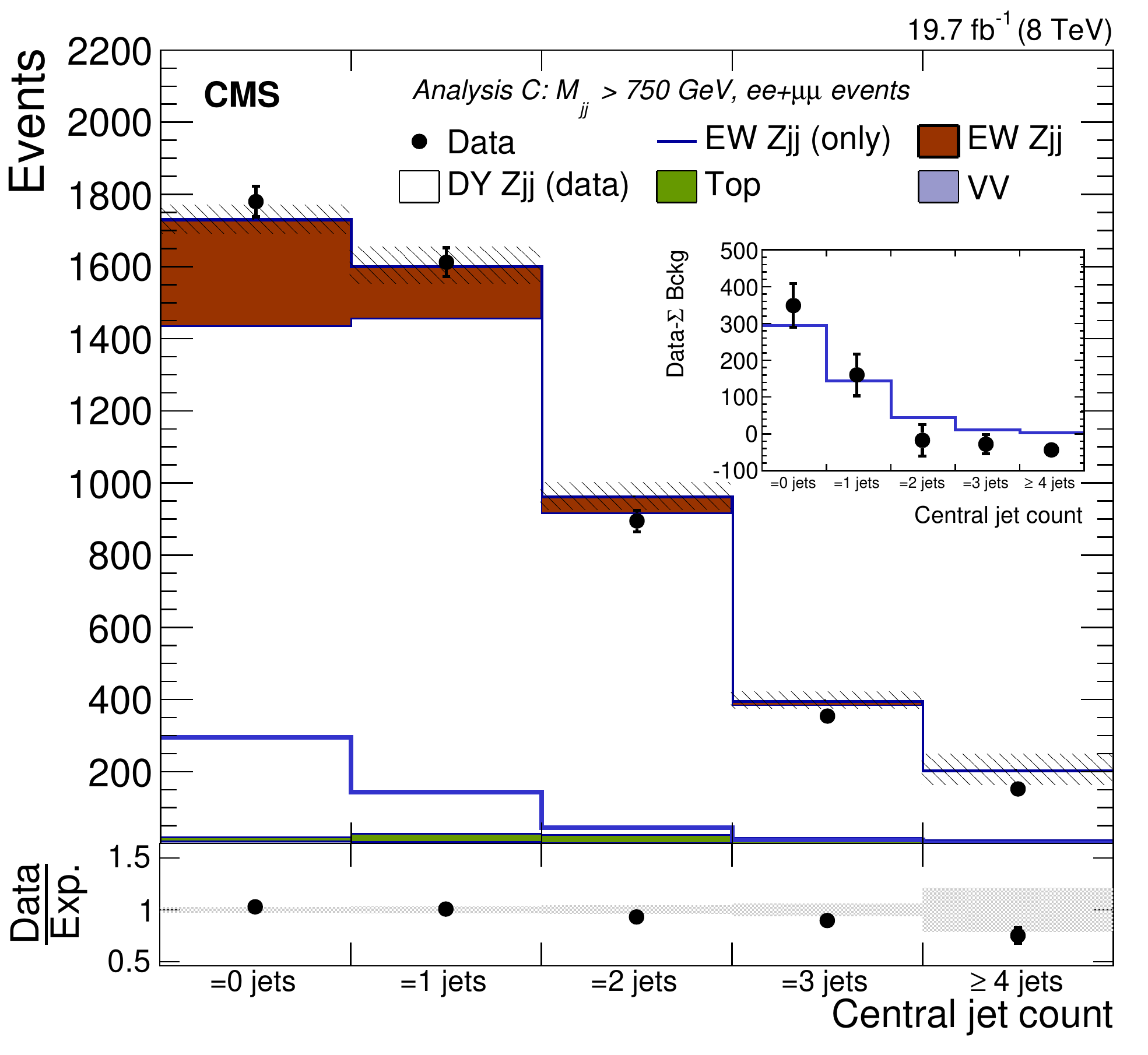}
\includegraphics[width=0.49\textwidth]{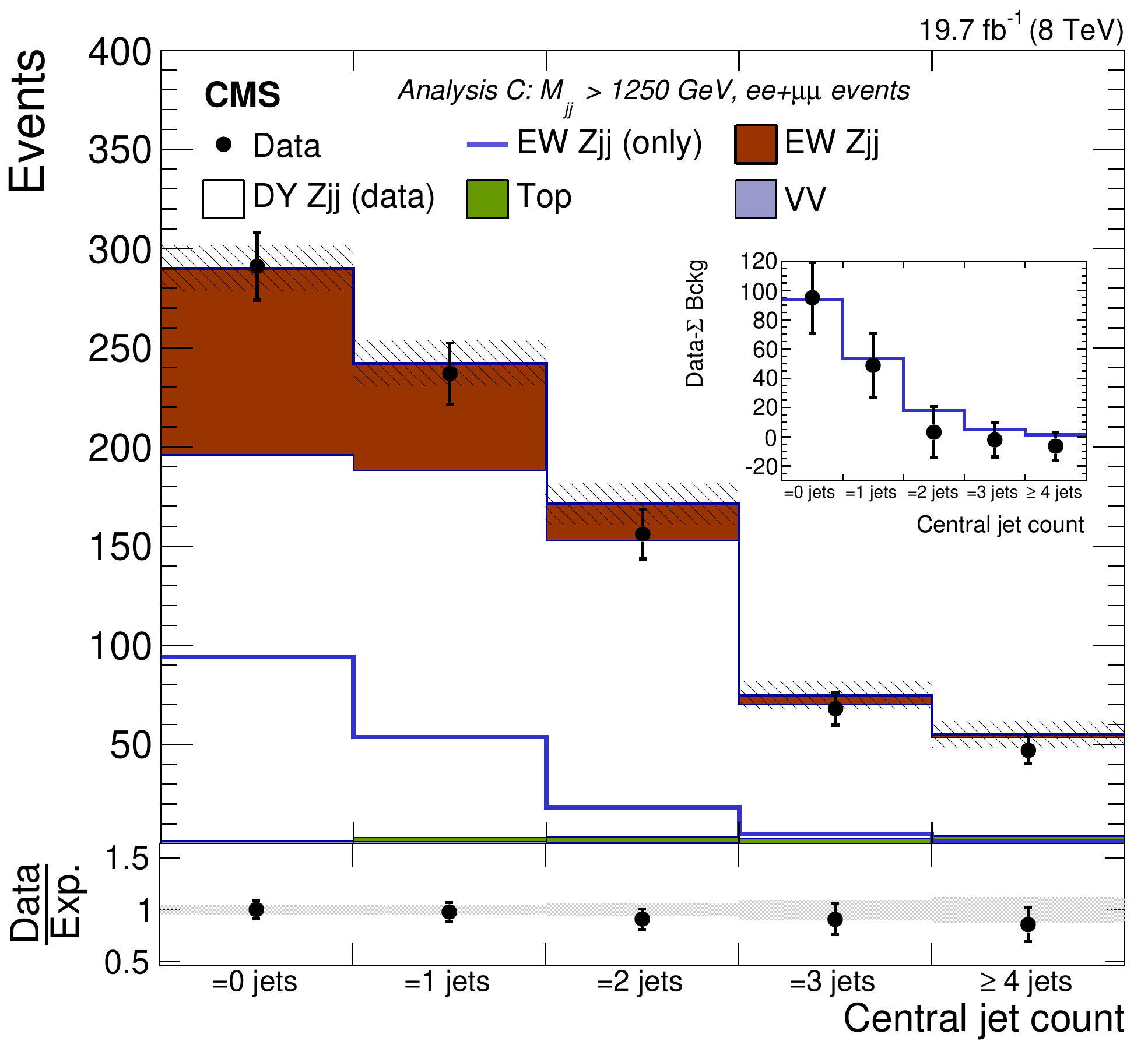}
\includegraphics[width=0.49\textwidth]{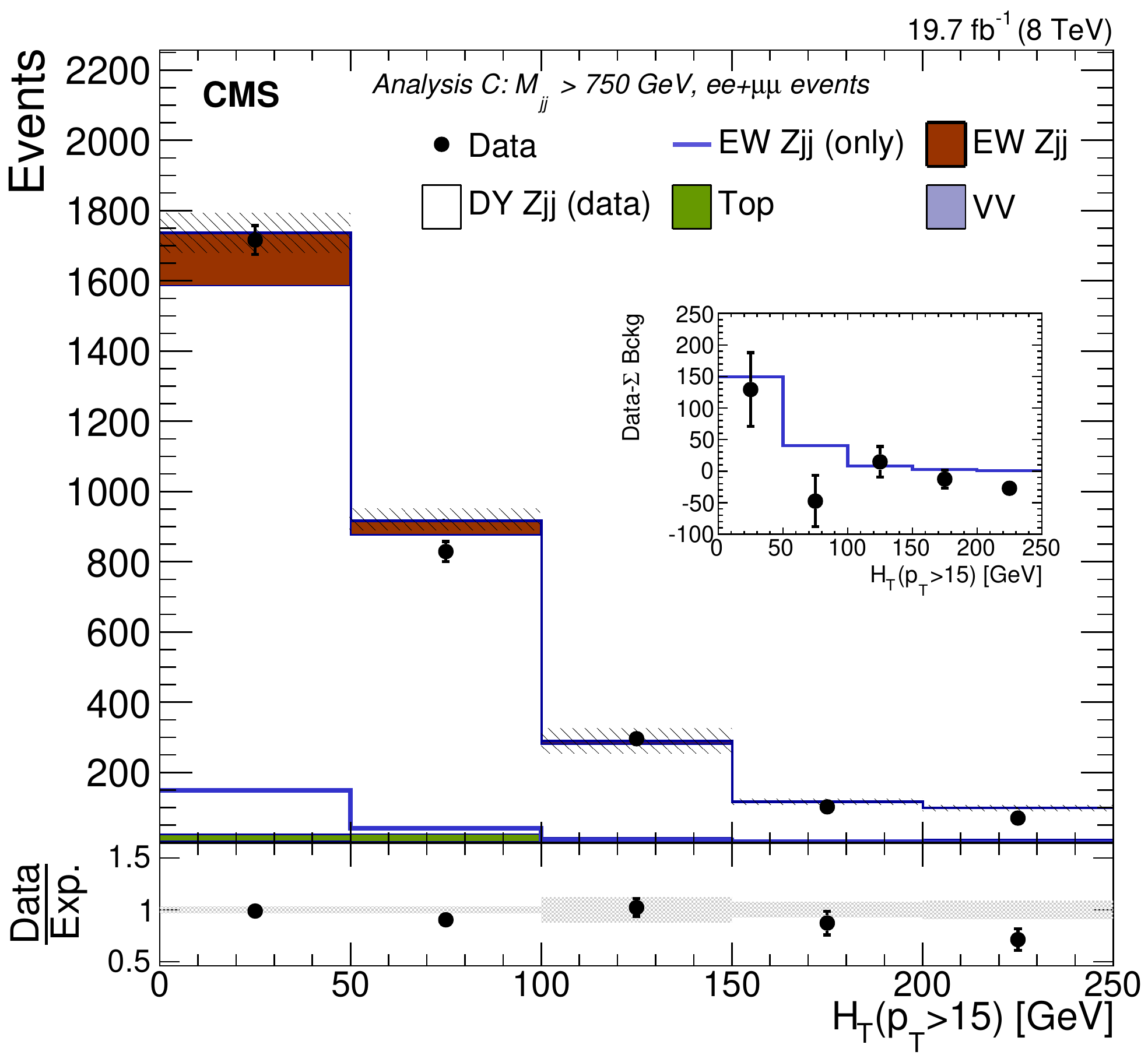}
\includegraphics[width=0.49\textwidth]{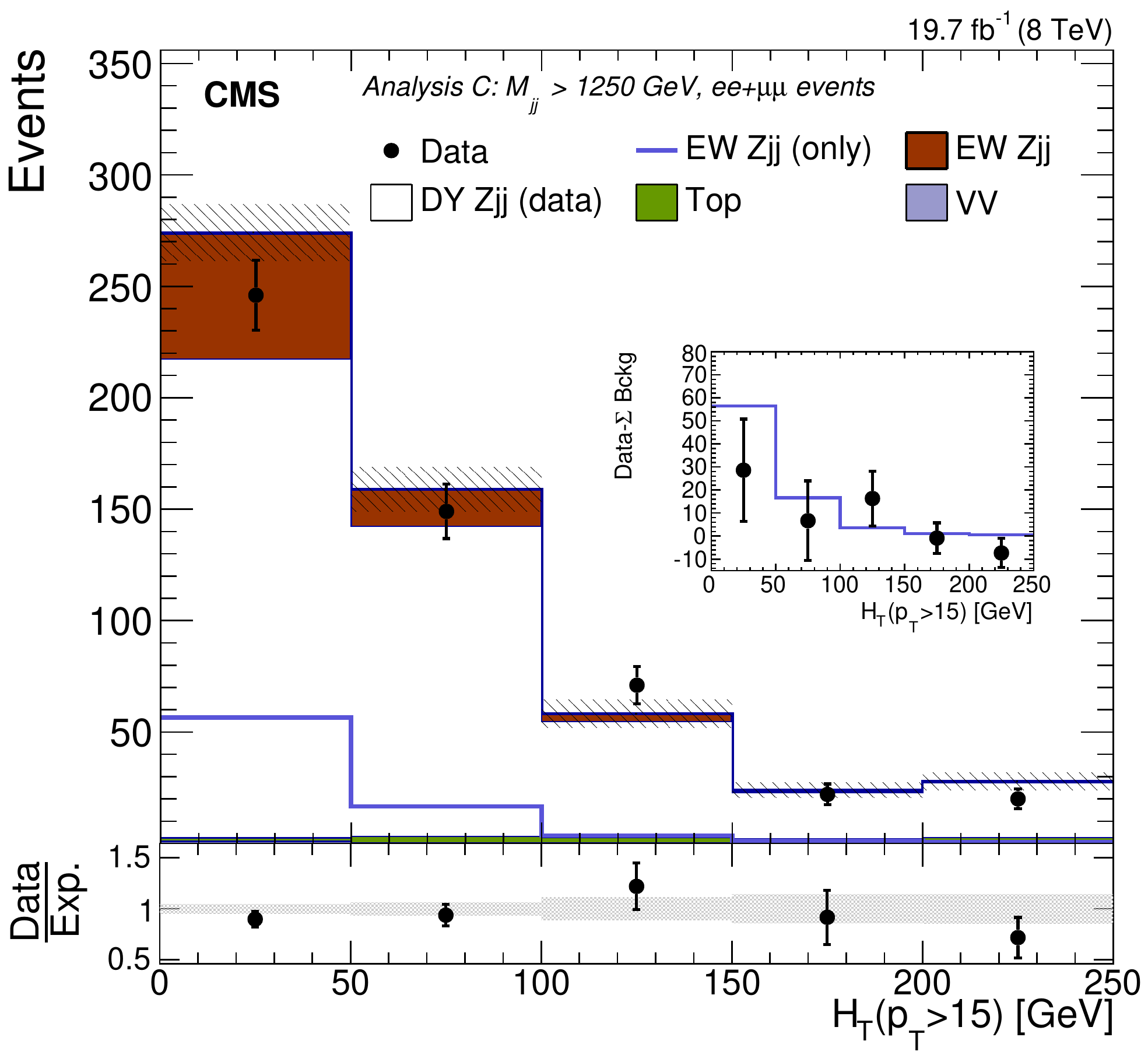}
\caption{
Additional jet multiplicity (top row), and corresponding $\HT$ (bottom row)
within the $\Delta\eta_{\mathrm{jj}}$ of the two tagging jets
in events with $M_\mathrm{jj}>750\GeV$ (left column) or $M_\mathrm{jj}>1250\GeV$ (right column).
In the main panels the expected contributions from \ewkzjj, \dyzjj, and residual
backgrounds are shown stacked, and compared to the observed data.
The signal-only contribution is superimposed separately
and it is also compared to the residual data after the subtraction
of the expected backgrounds in the insets.
The ratio of data to expectation is represented by point markers in the bottom panels.
The total uncertainties assigned to the expectations are represented as shaded bands.}
\label{fig:hadingap}
\end{figure*}

The number of extra jets,
as well as their scalar \pt sum ($\HT$),
are shown in Fig.~\ref{fig:hadingap}.
Data and expectations are generally in good agreement for both
distributions in the two $M_\mathrm{jj}$ regions.
A clear suppression of the emission of a third jet is observed in
data, when we take into account the background-only predictions.
After subtraction of the background,
which is shown as an inset in the different figures,
we observe that slightly less extra jets tend to be counted in data with respect to
the simulated signal.
Notice that in the simulation of the signal, the extra jets have
their origin in a parton-shower approach (see Section~\ref{sec:simulation}).

\begin{figure*}[htp]
\centering
\includegraphics[width=0.49\textwidth]{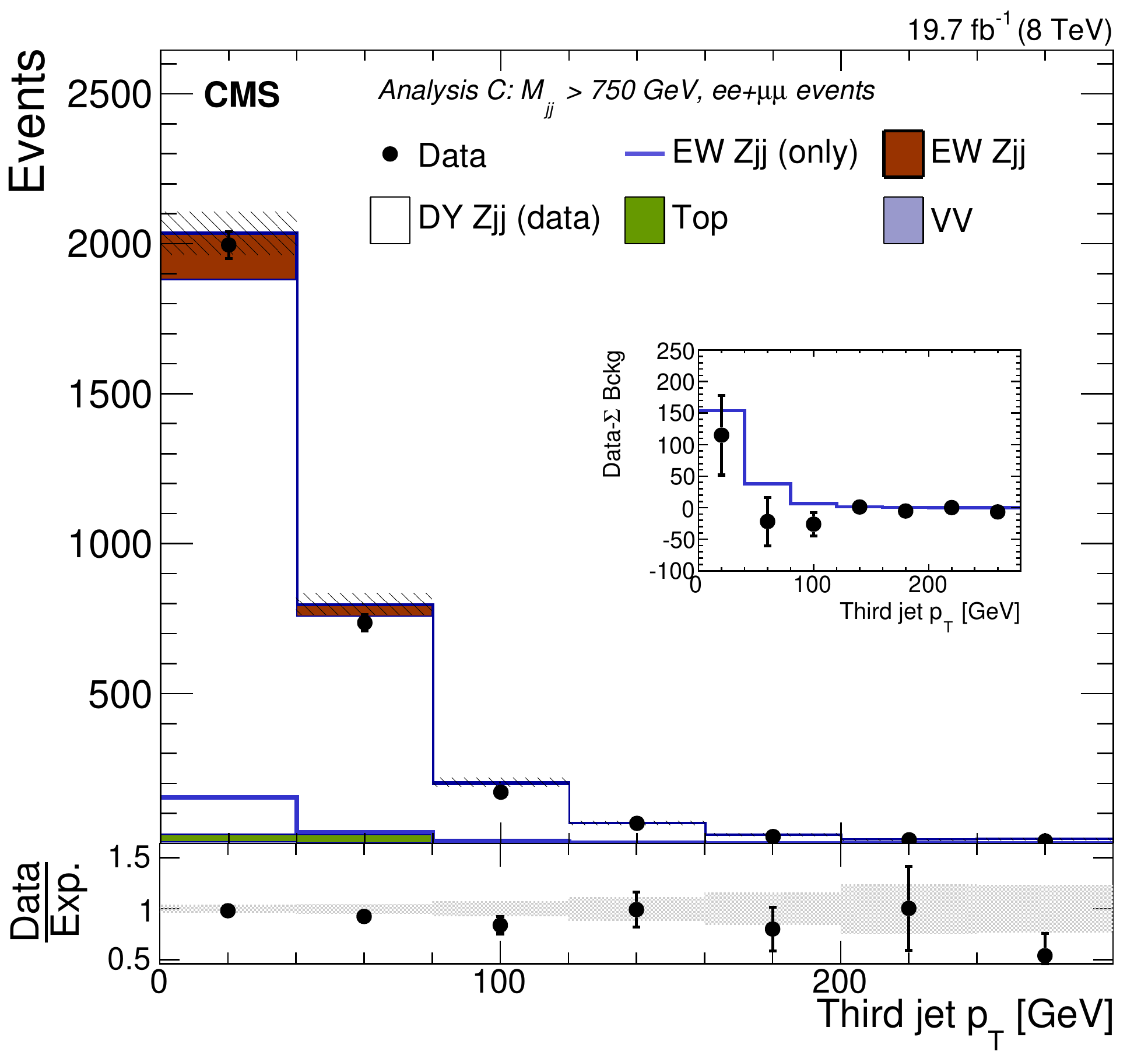}
\includegraphics[width=0.49\textwidth]{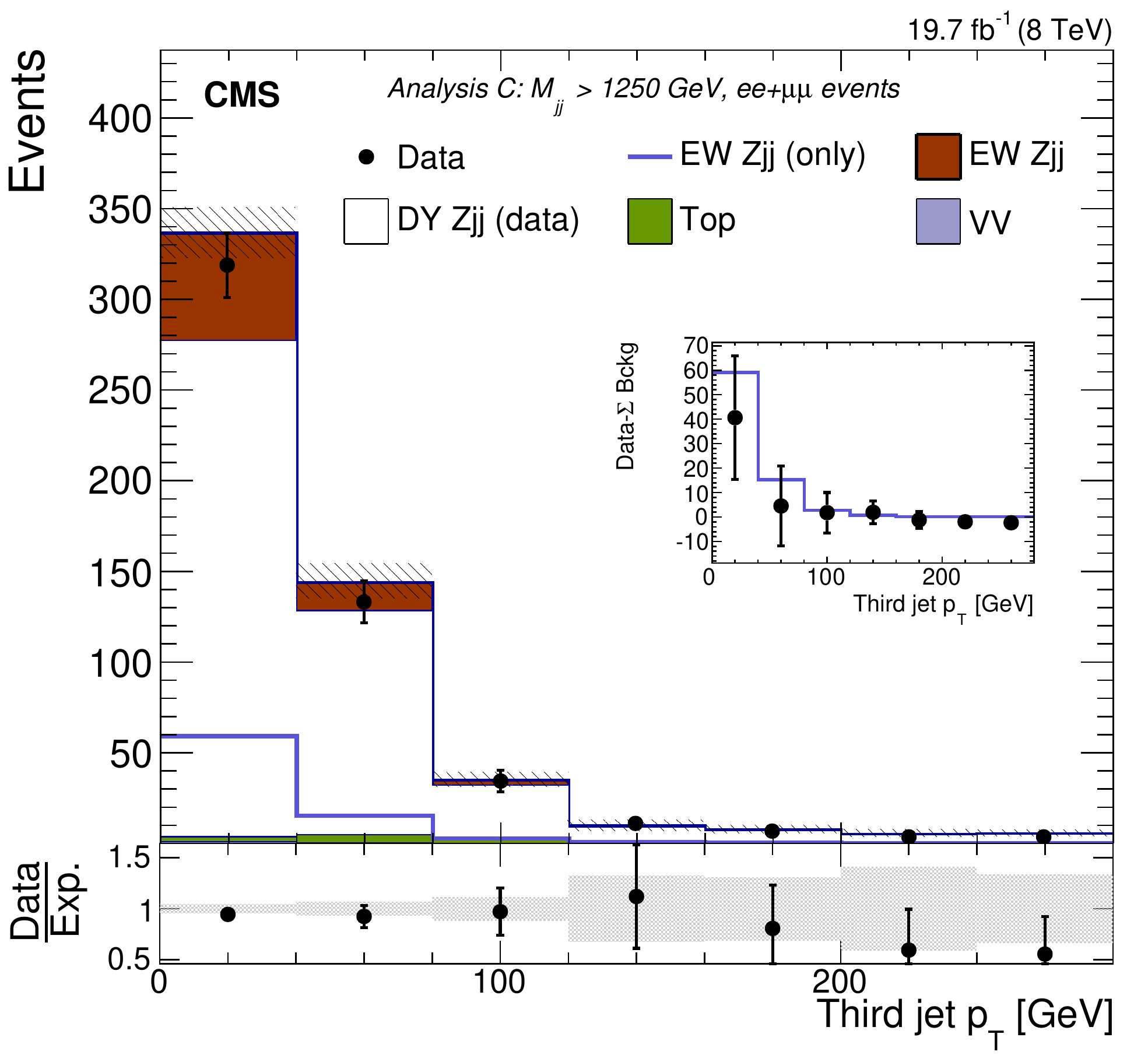}
\includegraphics[width=0.49\textwidth]{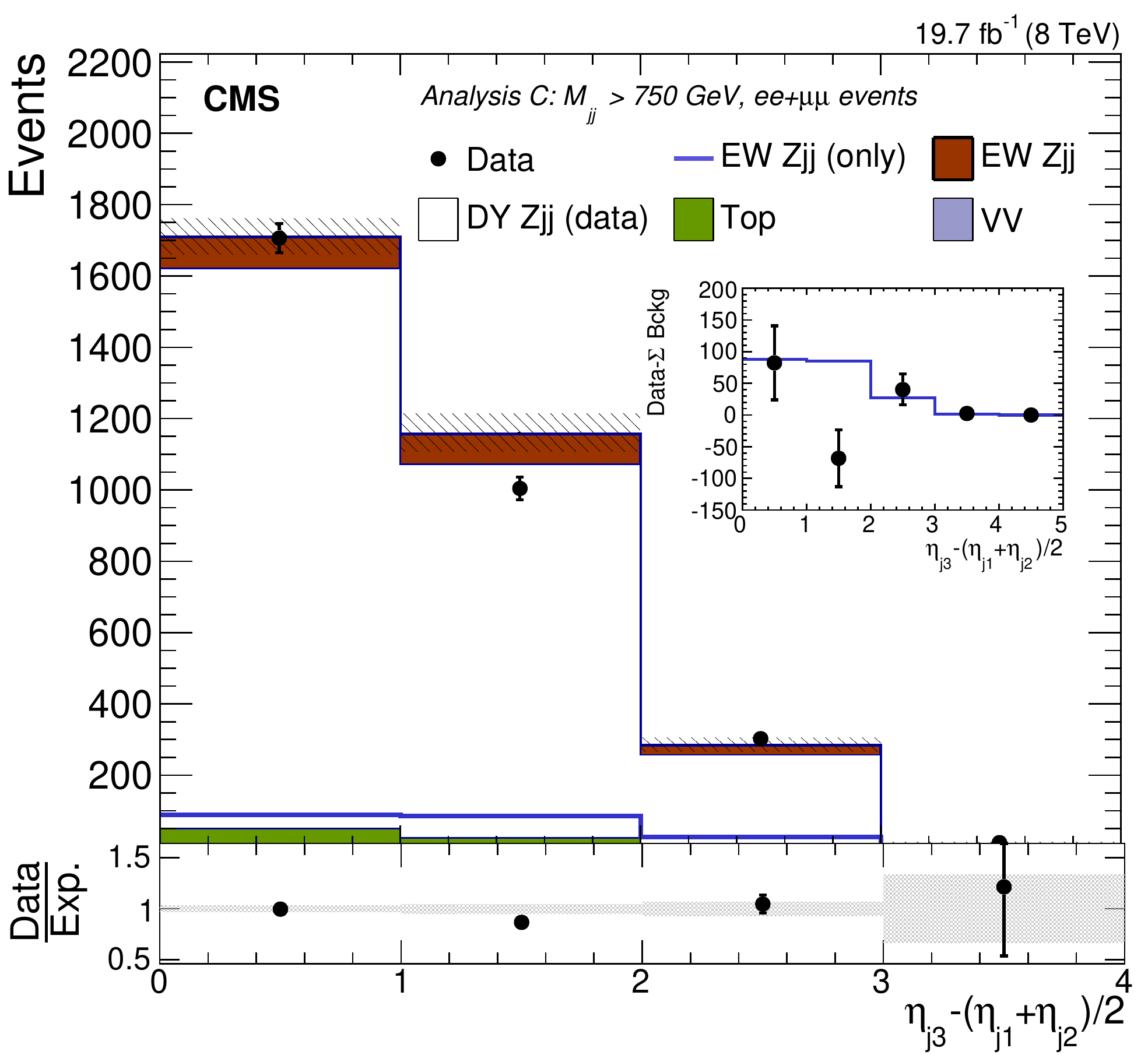}
\includegraphics[width=0.49\textwidth]{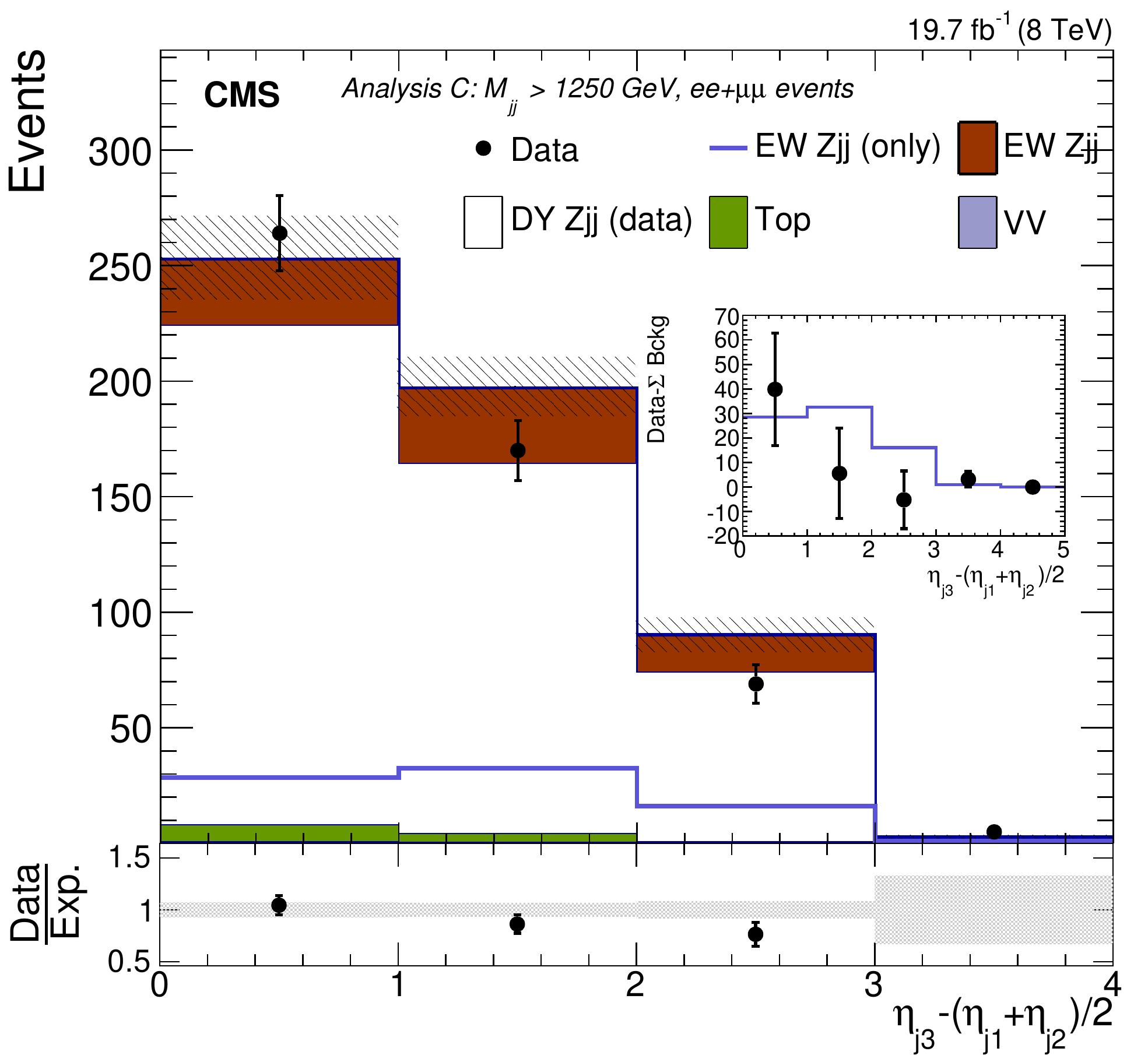}
\caption{
(top row) \pt and  (bottom row) $\eta^*_{\mathrm{j}3}$
of the leading additional jet
within the $\Delta\eta_{\mathrm{jj}}$ of the two tagging jets
in events with $M_\mathrm{jj}>750\GeV$ (left column) or $M_\mathrm{jj}>1250\GeV$ (right column).
The explanation of the plots is similar to Fig.~\ref{fig:hadingap}. \label{fig:thirdjet}}
\end{figure*}

The \pt values and the pseudorapidities relative to the average of the two tagging jets, \ie
$\eta^*_{\mathrm{j}3}=\eta_{\mathrm{j}3}-(\eta_{\mathrm{j}1}+\eta_{\mathrm{j}2})/2$,
of the third leading-\pt jet in the event,
are shown in Fig.~\ref{fig:thirdjet}.
There are some deviations of the data observed relative to the
predictions.
In particular, the third jet is observed to be slightly more central
than expected.
The poor statistical and other uncertainties prevent us,
however, from drawing further conclusions.

\begin{figure*}[htp]
\centering
\includegraphics[width=0.49\textwidth]{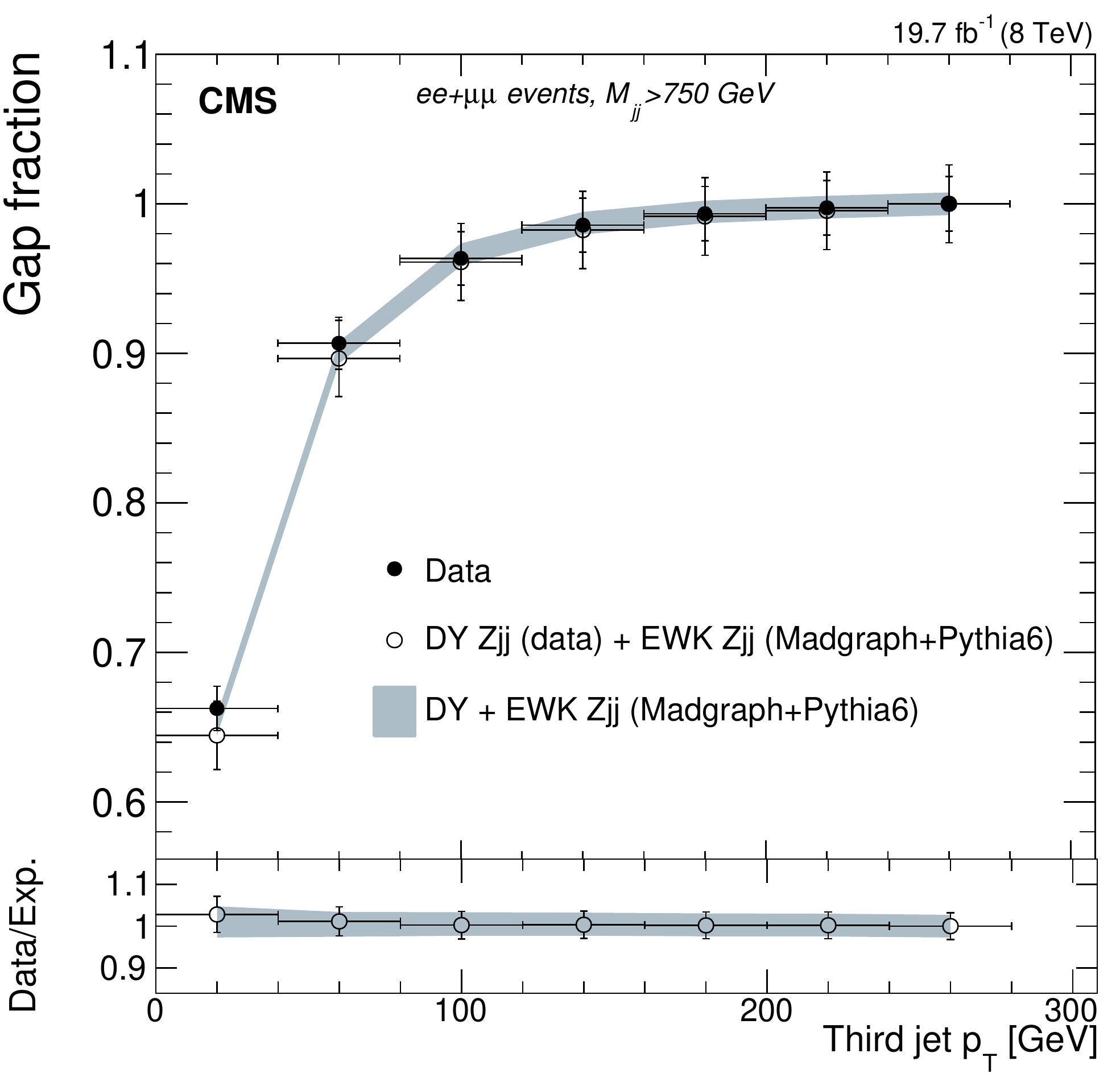}
\includegraphics[width=0.49\textwidth]{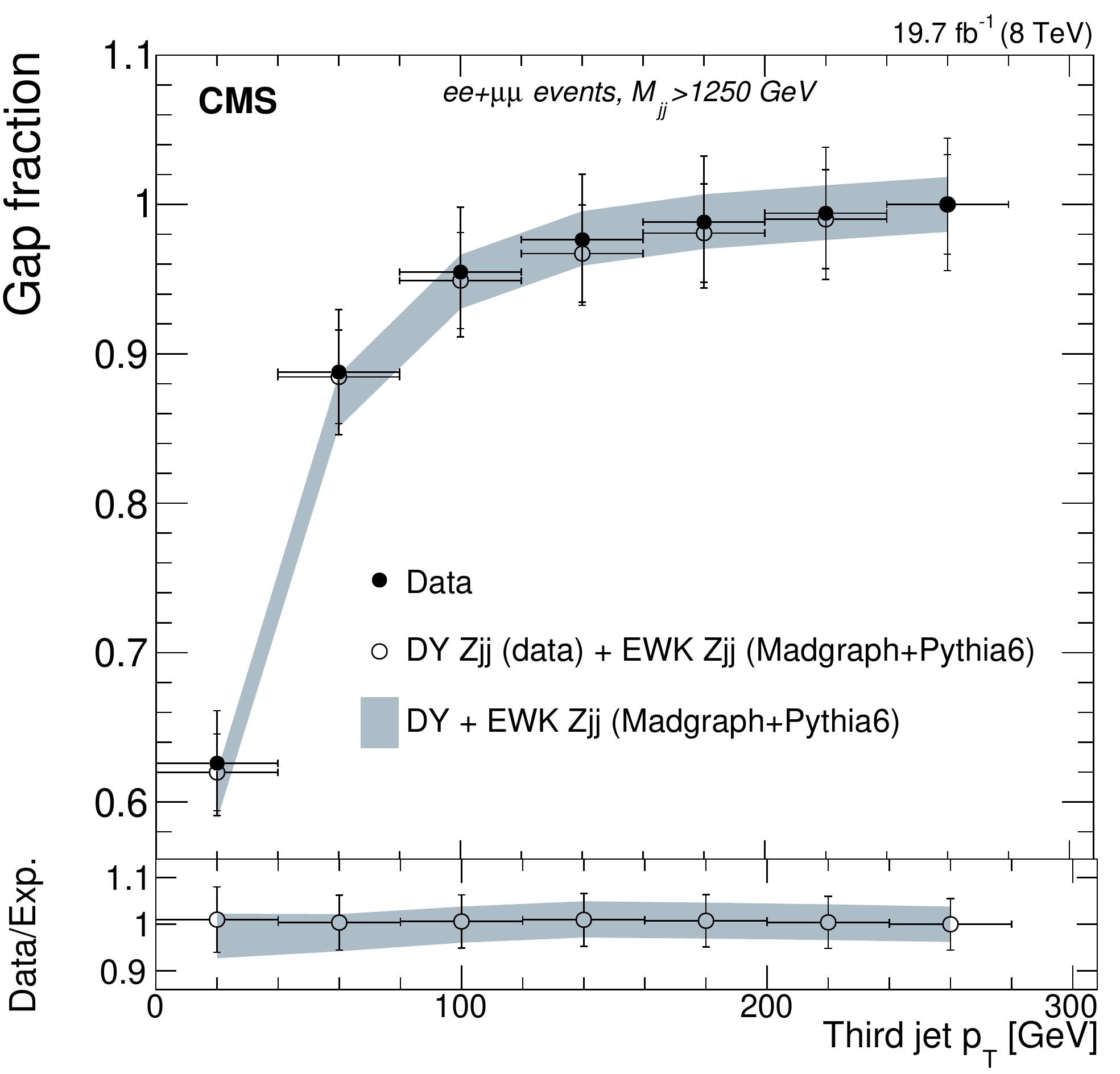}
\includegraphics[width=0.49\textwidth]{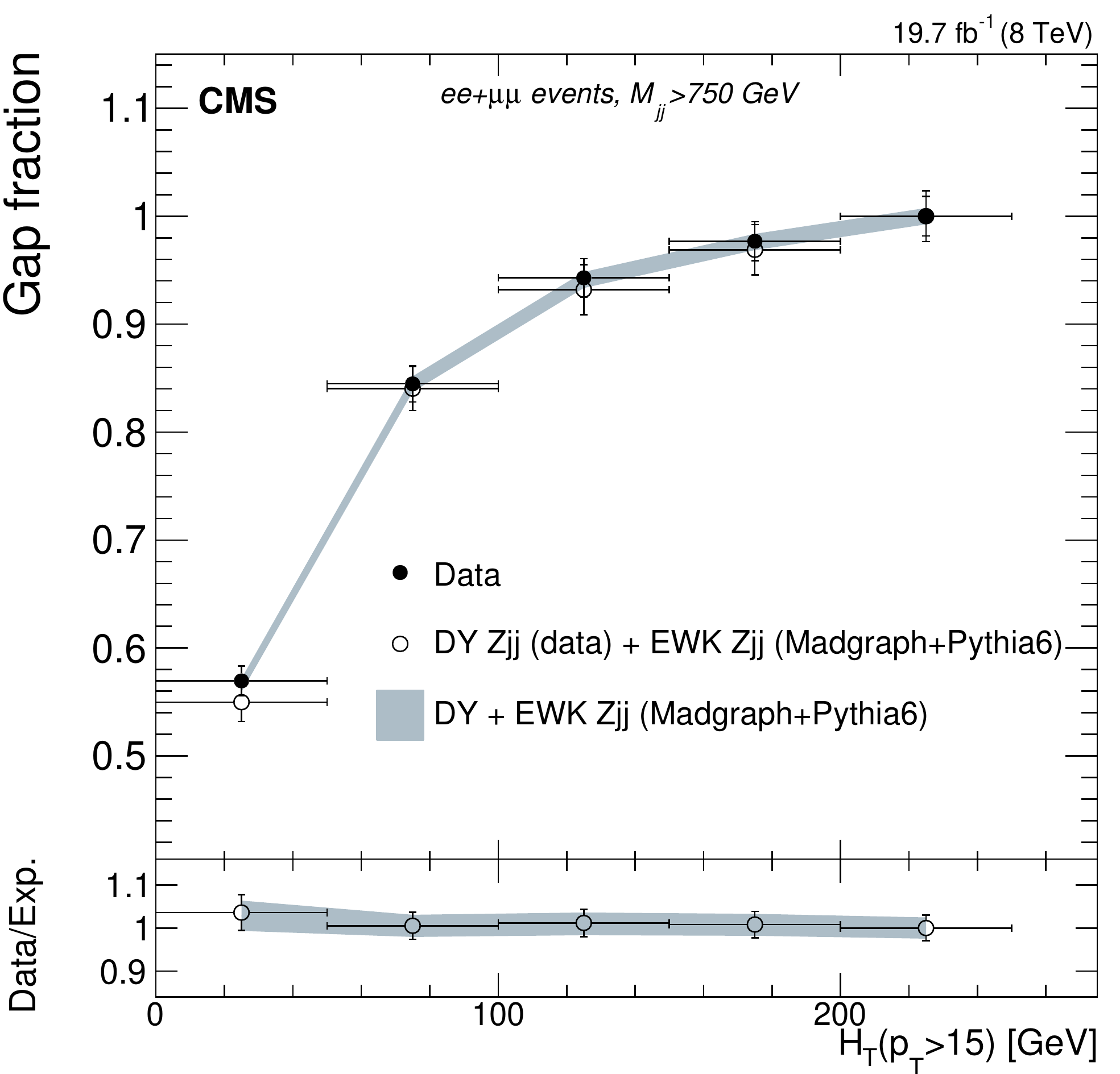}
\includegraphics[width=0.49\textwidth]{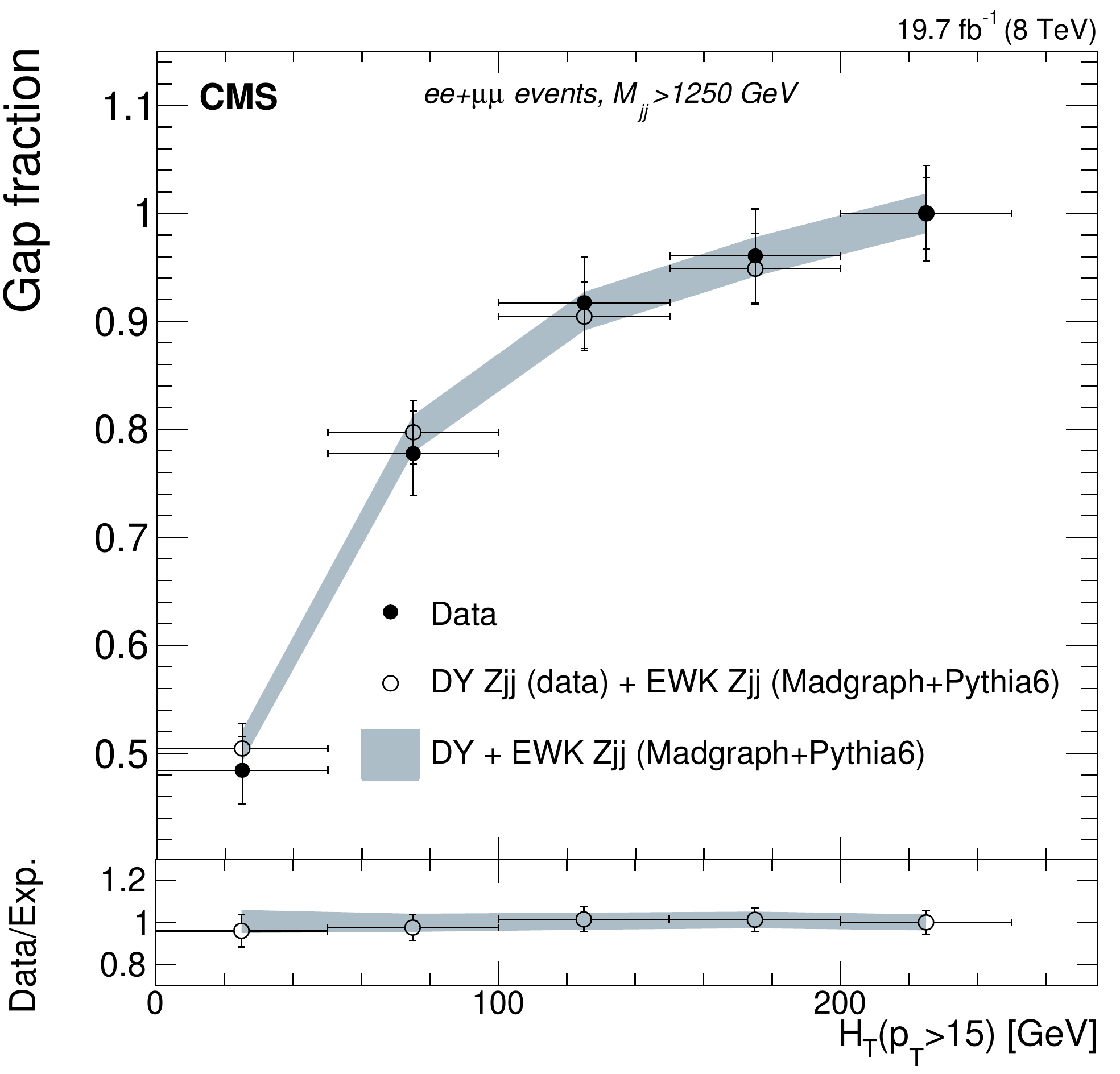}
\caption{
Gap fractions for:
(top row) \pt of leading additional jet,
(bottom row) the $\HT$ variable
within the $\Delta\eta_{\mathrm{jj}}$ of the two tagging jets
in events with $M_\mathrm{jj}>750\GeV$ (left) or $M_\mathrm{jj}>1250\GeV$ (right).
The observed gap fractions in data
are compared to two different signal plus background predictions
where \dyzjj\ is modelled either from \gjj\ data or from simulation.
The bottom panels show the ratio between the observed data
and different predictions.
}
\label{fig:rapgapveto}
\end{figure*}

The above distributions can be used to compute gap fractions.
We define a gap fraction as the fraction of events which do not
have reconstructed kinematics above a given threshold.
The most interesting gap fractions can be computed for the \pt of the
leading additional jet, and the $\HT$ variable.
These gap fractions are, in practice, measurements of the efficiency of
extra jet veto in VBF-like topologies.
By comparing different expectations with the observed data
we can quantify how reliable is the modelling of the extra jet
activity,
in particular in a signal-enriched region.
Figure~\ref{fig:rapgapveto} shows the gap fractions expected and
observed in data.
Two expectations are compared: the one using a full MC approach
and the one where the \dyzjj\ background is predicted from the \gjj\ data.
Both predictions are found to be in agreement with the data
for the \pt of the leading additional jet and the soft $\HT$ variable.

\section{Summary}
\label{sec:summary}

The cross section for the purely electroweak production of a Z boson
in association with two jets
in the \lljj\ final state, in  proton-proton collisions at
$\sqrt{s}=8\TeV$ has been measured to be
\begin{equation*}
\sigma({\mathrm{EW}~\ell\ell\mathrm{jj}})=174\pm 15\stat \pm 40\syst\unit{fb},
\end{equation*}
in agreement with the SM prediction.
Aside from the two analyses previously used to determine the cross
section of this process at 7\TeV~\cite{Chatrchyan:2013jya}, a new
analysis has been implemented
using a data-based model for the main background.
The increased integrated luminosity recorded at 8\TeV, an improved
selection method, and more precise modelling of signal and background processes
have allowed us to obtain
a more precise measurement of the \ewkzjj\ process relative to the 7\TeV result.

Studies of the jet activity in the selected events
show generally good agreement with the \MADGRAPH{}+\PYTHIA predictions.
In events with high signal purity, the additional hadron activity
has also been characterised, as well as the gap fractions.
Good agreement has been found between data and QCD predictions.

\begin{acknowledgments}
We congratulate our colleagues in the CERN accelerator departments for the excellent performance of the LHC and thank the technical and administrative staffs at CERN and at other CMS institutes for their contributions to the success of the CMS effort. In addition, we gratefully acknowledge the computing centres and personnel of the Worldwide LHC Computing Grid for delivering so effectively the computing infrastructure essential to our analyses. Finally, we acknowledge the enduring support for the construction and operation of the LHC and the CMS detector provided by the following funding agencies: BMWFW and FWF (Austria); FNRS and FWO (Belgium); CNPq, CAPES, FAPERJ, and FAPESP (Brazil); MES (Bulgaria); CERN; CAS, MoST, and NSFC (China); COLCIENCIAS (Colombia); MSES and CSF (Croatia); RPF (Cyprus); MoER, ERC IUT and ERDF (Estonia); Academy of Finland, MEC, and HIP (Finland); CEA and CNRS/IN2P3 (France); BMBF, DFG, and HGF (Germany); GSRT (Greece); OTKA and NIH (Hungary); DAE and DST (India); IPM (Iran); SFI (Ireland); INFN (Italy); NRF and WCU (Republic of Korea); LAS (Lithuania); MOE and UM (Malaysia); CINVESTAV, CONACYT, SEP, and UASLP-FAI (Mexico); MBIE (New Zealand); PAEC (Pakistan); MSHE and NSC (Poland); FCT (Portugal); JINR (Dubna); MON, RosAtom, RAS and RFBR (Russia); MESTD (Serbia); SEIDI and CPAN (Spain); Swiss Funding Agencies (Switzerland); MST (Taipei); ThEPCenter, IPST, STAR and NSTDA (Thailand); TUBITAK and TAEK (Turkey); NASU and SFFR (Ukraine); STFC (United Kingdom); DOE and NSF (USA).

Individuals have received support from the Marie-Curie programme and the European Research Council and EPLANET (European Union); the Leventis Foundation; the A. P. Sloan Foundation; the Alexander von Humboldt Foundation; the Belgian Federal Science Policy Office; the Fonds pour la Formation \`a la Recherche dans l'Industrie et dans l'Agriculture (FRIA-Belgium); the Agentschap voor Innovatie door Wetenschap en Technologie (IWT-Belgium); the Ministry of Education, Youth and Sports (MEYS) of the Czech Republic; the Council of Science and Industrial Research, India; the HOMING PLUS programme of Foundation for Polish Science, cofinanced from European Union, Regional Development Fund; the Compagnia di San Paolo (Torino); the Consorzio per la Fisica (Trieste); MIUR project 20108T4XTM (Italy); the Thalis and Aristeia programmes cofinanced by EU-ESF and the Greek NSRF; and the National Priorities Research Program by Qatar National Research Fund.
\end{acknowledgments}
\vspace*{2.5ex}

\bibliography{auto_generated}

\cleardoublepage \appendix\section{The CMS Collaboration \label{app:collab}}\begin{sloppypar}\hyphenpenalty=5000\widowpenalty=500\clubpenalty=5000\input{FSQ-12-035-authorlist.tex}\end{sloppypar}
\end{document}

%% file: FSQ-12-035-authorlist.tex
\textbf{Yerevan Physics Institute,  Yerevan,  Armenia}\\*[0pt]
V.~Khachatryan, A.M.~Sirunyan, A.~Tumasyan
\vskip\cmsinstskip
\textbf{Institut f\"{u}r Hochenergiephysik der OeAW,  Wien,  Austria}\\*[0pt]
W.~Adam, T.~Bergauer, M.~Dragicevic, J.~Er\"{o}, C.~Fabjan\cmsAuthorMark{1}, M.~Friedl, R.~Fr\"{u}hwirth\cmsAuthorMark{1}, V.M.~Ghete, C.~Hartl, N.~H\"{o}rmann, J.~Hrubec, M.~Jeitler\cmsAuthorMark{1}, W.~Kiesenhofer, V.~Kn\"{u}nz, M.~Krammer\cmsAuthorMark{1}, I.~Kr\"{a}tschmer, D.~Liko, I.~Mikulec, D.~Rabady\cmsAuthorMark{2}, B.~Rahbaran, H.~Rohringer, R.~Sch\"{o}fbeck, J.~Strauss, A.~Taurok, W.~Treberer-Treberspurg, W.~Waltenberger, C.-E.~Wulz\cmsAuthorMark{1}
% ...
\vskip\cmsinstskip
% ...
1:~~Also at Vienna University of Technology, Vienna, Austria\\
2:~~Also at CERN, European Organization for Nuclear Research, Geneva, Switzerland\\
% ...
